\documentclass[sort&compress,11pt]{article}
\pdfoutput=1

\usepackage{amssymb,amsmath,amsthm, mathtools}
\usepackage[toc,page]{appendix}
\usepackage{graphicx}
\usepackage{slashed}
\usepackage[]{hyperref}
\usepackage[utf8]{inputenc}
\usepackage{color}
\usepackage{cite}
\usepackage{subcaption}

\newcommand{\change}[1]{{\color{red} #1}} 
\def\change{\relax}

\newcommand{\mydstrut}{\rule[-1.2ex]{0ex}{1.8ex}}
\newcommand{\mystrut}{\rule[-1.4ex]{0ex}{4.2ex}}

\makeatletter
\newcommand\footnoteref[1]{\protected@xdef\@thefnmark{\ref{#1}}\@footnotemark}
\makeatother

\if@twoside \oddsidemargin -17pt \evensidemargin 00pt
\else \oddsidemargin 00pt \evensidemargin 00pt
\fi
\expandafter\ifx\csname mathrm\endcsname\relax\def\mathrm#1{{\rm #1}}\fi

\makeatother

\def\reffi#1{\mbox{Fig.~\ref{#1}}}
\def\reffis#1{\mbox{Figs.~\ref{#1}}}
\def\refta#1{\mbox{Table~\ref{#1}}}
\def\reftas#1{\mbox{Tables~\ref{#1}}}
\def\refse#1{\mbox{Section~\ref{#1}}}

\def\refapp#1{\mbox{App.~\ref{#1}}}

\def\citere#1{\mbox{Ref.~\cite{#1}}}
\def\citeres#1{\mbox{Refs.~\cite{#1}}}

\def\ie{i.e.\ }
\def\eg{e.g.\ }

\allowdisplaybreaks[3]   

\newcommand{\cmm}[1]{\ensuremath{#1}\ifmmode\else{}\fi}
\newcommand{\nmc}[2]{\newcommand{#1}{\cmm{#2}}}

\newcommand{\FortranNinety}{{\sc Fortran95}}
\newcommand{\PythonTwo}{{\sc Python 2.7}}
\newcommand{\PythonThree}{{\sc Python 3.x}}
\newcommand{\Recola}{{\sc RECOLA}}
\newcommand{\RecolaTwo}{{\sc RECOLA2}}
\newcommand{\FeynArts}{{\sc FeynArts}}
\newcommand{\QGS}{{\tt{QGS}}}
\newcommand{\QGraf}{{\tt{QGRAF}}}
\newcommand{\GraphShot}{{\tt{GraphShot}}}
\newcommand{\FormCalc}{{\sc FormCalc}}
\newcommand{\Reptil}{{\sc REPT1L}}
\newcommand{\Collier}{{\sc COLLIER}}
\newcommand{\HAWKTwo}{{\sc \mbox{HAWK 2.0}}}
\newcommand{\UFO}{{\sc UFO}}
\newcommand{\Sympy}{{\sc SymPy}}
\newcommand{\FORM}{{\sc FORM}}
\newcommand{\Feynrules}{{\sc Feynrules}}
\newcommand{\SARAH}{{\sc SARAH}}
\newcommand{\OpenLoops}{{\sc OpenLoops}}
\newcommand{\MG}{{\sc MadGraph5\_aMC@NLO}}
\newcommand{\GoSam}{{\sc GoSam}}

\newcommand{\GeV}{\unskip\,\mathrm{GeV}}

\newcommand{\fba}{\unskip\,\mathrm{fb}}

\newcommand{\pba}{\unskip\,\mathrm{pb}}

\nmc{\Pp}{\mathrm{p}}
\nmc{\PW}{\mathrm{W}}
\nmc{\PZ}{\mathrm{Z}}
\nmc{\PV}{\mathrm{V}}
\nmc{\PH}{\mathrm{H}}
\nmc{\Pl}{\mathrm{l}}
\nmc{\Pm}{\mathrm{\mu}}
\nmc{\Pv}{\mathrm{v}}
\nmc{\Pu}{\mathrm{u}}
\nmc{\Pd}{\mathrm{d}}
\nmc{\Pc}{\mathrm{c}}
\nmc{\Ps}{\mathrm{s}}
\nmc{\Pe}{\mathrm{e}}

\nmc{\g}{g}
\nmc{\gy}{{g^\prime}}
\nmc{\Gf}{G_\mathrm{F}}
\nmc{\slo}{\sigma_\mathrm{LO}}
\nmc{\EW}{\mathrm{EW}}
\nmc{\QCD}{\mathrm{QCD}}
\nmc{\SUSY}{\mathrm{SUSY}}
\nmc{\QED}{\mathrm{QED}}
\nmc{\LO}{\mathrm{LO}}
\nmc{\NLO}{\mathrm{NLO}}
\nmc{\snlo}{\sigma_\mathrm{NLO}}
\nmc{\snloew}{\sigma^\mathrm{EW}_\mathrm{NLO}}
\nmc{\dew}{\delta_\EW}
\nmc{\dewm}{\dew^{\msbar}}
\nmc{\dewb}{\dew^{\mathrm{BFM}}}
\nmc{\dewps}{\dew^{p^*}}
\nmc{\BGR}{\mathrm{BGR}}

\nmc{\B}{{\rm B}}
\nmc{\R}{{\rm R}}

\nmc{\ii}{\mathrm{i}}

\nmc{\rd}{\mathrm{d}}

\nmc{\ft}{\mathrm{F.T.}}

\nmc{\Az}{A_0}

\nmc{\h}{h}
\nmc{\Hl}{H_\mathrm{l}}
\nmc{\Hs}{H_\mathrm{s}}
\nmc{\Hbl}{\hat{H}_\mathrm{l}}
\nmc{\Hh}{H_\mathrm{h}}
\nmc{\Hbh}{\hat{H}_\mathrm{h}}
\nmc{\Hnov}{\widetilde{H}}
\nmc{\GZ}{G_0}
\nmc{\GbZ}{\hat{G}_0}
\nmc{\Ha}{H_\mathrm{a}}
\nmc{\Hba}{\hat{H}_\mathrm{a}}
\nmc{\Hp}{H^+}
\nmc{\Hm}{H^-}
\nmc{\Hpm}{H^\pm}
\nmc{\Hbpm}{\hat{H}^\pm}
\nmc{\Gpm}{G^\pm}
\nmc{\Gbpm}{\hat{G}^\pm}
\nmc{\Gp}{G^+}
\nmc{\Gm}{G^-}
\nmc{\Hmp}{H^\mp}
\nmc{\Gmp}{G^\mp}
\nmc{\Phin}{\Phi_n}
\nmc{\Phii}{\Phi_i}
\nmc{\Phbii}{\hat{\Phi}_i}
\nmc{\Phione}{\Phi_1}
\nmc{\Phitwo}{\Phi_2}
\nmc{\Phibone}{\hat{\Phi}_1}
\nmc{\Phibtwo}{\hat{\Phi}_2}
\nmc{\phionepm}{\phi_1^\pm}
\nmc{\phitwopm}{\phi_2^\pm}
\nmc{\phinp}{\phi_n^+}
\nmc{\phiip}{\phi_i^+}
\nmc{\phbiip}{\hat{\phi}_i^+}
\nmc{\phii}{\varphi_i}
\nmc{\phij}{\varphi_j}
\nmc{\phik}{\varphi_k}
\nmc{\phin}{\varphi_n}
\nmc{\rhoi}{\rho_i}
\nmc{\rhobi}{\hat{\rho}_i}
\nmc{\rhon}{\rho_n}
\nmc{\rhoone}{\rho_1}
\nmc{\rhotwo}{\rho_2}
\nmc{\etan}{\eta_n}
\nmc{\etai}{\eta_i}
\nmc{\etabi}{\hat{\eta}_i}
\nmc{\etaone}{\eta_1}
\nmc{\etatwo}{\eta_2}

\nmc{\vth}{\tilde v_\h}
\nmc{\tilh}{\tilde\h}
\nmc{\al}{\alpha}
\nmc{\be}{\beta}
\nmc{\vone}{v_1}
\nmc{\vtone}{\left(\vbone+\dvone\right)}
\nmc{\vtwo}{v_2}
\nmc{\vttwo}{\left(\vbtwo+\dvtwo\right)}
\nmc{\vn}{v_n}
\nmc{\vi}{v_i}
\nmc{\vtn}{\tilde v_n}
\nmc{\vti}{\tilde v_i}
\nmc{\tb}{t_\be}
\nmc{\tanb}{\tan\be}
\nmc{\ca}{\cos\al}
\nmc{\catwo}{\cos^2\al}
\nmc{\satwo}{\sin^2\al}
\nmc{\sa}{\sin\al}
\nmc{\sas}{s_\al}
\nmc{\cas}{c_\al}
\nmc{\stwoa}{\sin2\al}
\nmc{\cbe}{\cos\be}
\nmc{\ctwobe}{\cos2\be}
\nmc{\cbeptwo}{\cos^2\be}
\nmc{\sbe}{\sin\be}
\nmc{\sbeptwo}{\sin^2\be}
\nmc{\cab}{c_{\al\be}}
\newcommand{\sab}{\ensuremath{s_{\alpha\beta}}}

\nmc{\mw}{M_\mathrm{W}}
\nmc{\mmw}{\mu_\mathrm{W}}
\nmc{\sw}{s_\mathrm{w}}
\nmc{\cw}{c_\mathrm{w}}
\nmc{\mv}{M_V}
\nmc{\ms}{M_S}
\nmc{\mscale}{{M^{*}}}
\nmc{\msp}{M_{S'}}
\nmc{\mz}{M_\mathrm{Z}}
\nmc{\mmz}{\mu_\mathrm{Z}}
\nmc{\mf}{m_f}
\nmc{\mfL}{\mf^\mathrm{l}}
\nmc{\mfU}{\mf^\mathrm{u}}
\nmc{\mfD}{\mf^\mathrm{d}}
\nmc{\mfb}{m_{f,\mathrm{B}}}
\newcommand{\mt}{m_\mathrm{t}}
\newcommand{\mh}{M_{\mathrm{h}}}
\nmc{\mhl}{M_{\Hl}}
\nmc{\mhh}{M_{\Hh}}
\nmc{\mhs}{M_{\Hs}}
\nmc{\Msb}{M_{\rm sb}}
\nmc{\mhc}{M_{\Hpm}}
\nmc{\mha}{M_{\Ha}}
\nmc{\dmv}{\delta \mv}
\nmc{\dms}{\delta \ms}
\nmc{\dmf}{\delta \mf}
\nmc{\rxi}{\mathrm{R}_\xi}
\nmc{\Phibi}{\hat \Phi_{i}}
\nmc{\onePI}{\Sigma}
\newcommand{\gazt}{\onePI_{AZ}^{\mathrm{1PI},\mathrm{T}}}
\newcommand{\gwwt}{\onePI_{WW}^{\mathrm{1PI},\mathrm{T}}}
\newcommand{\gwwtbfm}{\Sigma_{WW}^{\mathrm{1PI},\mathrm{BFM},\mathrm{T}}}

\newcommand{\gaztbfm}{\Sigma_{AZ}^{\mathrm{1PI},\mathrm{BFM},\mathrm{T}}}
\newcommand{\thlhh}{t_{\Hl\Hh}}

\newcommand{\thag}{t_{\Ha\GZ}}
\nmc{\dmw}{\delta \mw}
\nmc{\dmz}{\delta \mz}
\nmc{\dmmw}{\delta \mmw}
\nmc{\dmmz}{\delta \mmz}
\newcommand{\dzhhhl}{\delta Z_{\Hh\Hl}}
\newcommand{\dzhlhh}{\delta Z_{\Hl\Hh}}
\newcommand{\dzgha}{\delta Z_{\GZ\Ha}}
\newcommand{\dzhag}{\delta Z_{\Ha\GZ}}
\newcommand{\pics}{pics}
\newcommand{\dist}{distributions}
\newcommand{\SMatrix}{$S$-matrix}
\newcommand{\SM}{SM}
\newcommand{\THDM}{2HDM}
\newcommand{\HS}{HSESM}
\newcommand{\MSSM}{MSSM}
\newcommand{\BSM}{BSM}
\newcommand{\BFM}{BFM}
\newcommand{\EFT}{EFT}
\newcommand{\FCNC}{FCNC}
\newcommand{\CMS}{CMS}
\nmc{\msbar}{{\overline{\mathrm{MS}}}}
\newcommand{\ts}{{\it FJ~Tadpole Scheme}}

\makeatletter
\newcommand{\specialnumber}[1]{%
  \def\tagform@##1{\maketag@@@{(\ignorespaces##1\unskip\@@italiccorr#1)}}%
}
\newcommand{\specialeqref}[2]{\begingroup
  \def\tagform@##1{\maketag@@@{(\ignorespaces##1\unskip\@@italiccorr#2)}}%
  \eqref{#1}\endgroup}
\makeatother

\numberwithin{equation}{section}

\begin{document}

\thispagestyle{empty}
\def\thefootnote{\fnsymbol{footnote}}
\setcounter{footnote}{1}
\null
\hfill
\\
\begin{flushright}
\today
\end{flushright}
\vskip 1.2cm
\begin{center}

{\Large \boldmath{\bf NLO electroweak corrections in extended Higgs Sectors\\
  with \RecolaTwo}
\par} \vskip 2.5em
{\large
{ Ansgar~Denner$^a$, Jean-Nicolas~Lang$^a$, Sandro Uccirati$^b$}\\[3ex]
{\normalsize \it
  $^a$Universit\"at W\"urzburg, \\
Institut f\"ur Theoretische Physik und Astrophysik, \\
D-97074 W\"urzburg, Germany}\\[1ex]
{\normalsize \it
$^b$Universit\`a di Torino e INFN, \\
  10125 Torino, Italy }
}

\par \vskip 1em
\end{center}\par
\vfill \vskip .0cm \vfill {\bf Abstract:} \par 


We present the computer code \RecolaTwo{} along with the first NLO electroweak
corrections to Higgs production in vector-boson fusion and updated results for
Higgs strahlung in the Two-Higgs-Doublet Model and Higgs-Singlet extension of
the Standard Model. A fully automated procedure for the generation of tree-level
and one-loop matrix elements in general models, including renormalization, is
presented.  We discuss the application of the Background-Field Method to the
extended models. Numerical results for NLO electroweak cross sections are
presented for different renormalization schemes in the Two-Higgs-Doublet Model
and the Higgs-Singlet extension of the Standard Model. Finally, we present
distributions for the production of a heavy Higgs boson.
\par
\vskip 1cm
\noindent
\today
\par
\null
\setcounter{page}{0}
\clearpage
\def\thefootnote{\arabic{footnote}}
\setcounter{footnote}{0}


\section{Introduction\label{sec:Introduction}}
Since the discovery of a Higgs boson at the Large Hadron Collider
(LHC)~\cite{Chatrchyan:2012xdj,Aad:2012tfa} the community is moving forward
focusing on precision. Precision is the key to probe the Standard Model (\SM)
and Beyond Standard Model (\BSM) physics and potentially allows, together with
automation, to  disprove the \SM\ or even to single out new models.  State of
the art predictions involve typically two-loop and occasionally three-loop
\QCD{} and one-loop electroweak (\EW) corrections for many processes of interest
at the LHC.  As the aim is to cover all accessible processes at the LHC and
future colliders, a lot of effort has gone into the full automation of one-loop
amplitudes.  With one-loop \QCD{} amplitudes being available since a long time,
more recently much effort has been spent on the automation of \EW{} one-loop
corrections, which are more important than ever in view of the recent progress
in multi-loop \QCD{} calculations. \SM{} \EW{} corrections are nowadays
available in various approaches, \eg \OpenLoops\cite{Kallweit:2014xda},
\MG\cite{Alwall:2014hca}, \GoSam\cite{Cullen:2014yla,Chiesa:2015mya},
\FeynArts/\FormCalc\cite{Hahn:2000kx,Gross:2014ola}, and in our fully recursive
approach \Recola\cite{Actis:2012qn,Actis:2016mpe}.  For \BSM{} physics precision
is important, and especially \EW{} corrections should not be underestimated as
they can be comparable to \QCD{} corrections in certain \BSM{} scenarios.

The automation for one-loop \BSM{} physics requires three ingredients:
First, new models need to be defined, typically in form of a
Lagrangian and followed by the computation of the Feynman rules. For
this kind of task \Feynrules \cite{Alloul:2013bka} and \SARAH
\cite{Staub:2013tta} are established tools.  Then, a systematic and
yet flexible approach to the renormalization and computation of
further ingredients is required to deal with generic models.  Finally,
the renormalized model file needs to be interfaced to a generic
one-loop matrix-element generator.  As for the automation of
renormalization, there has been progress in the \Feynrules/\FeynArts{}
approach\cite{Degrande:2014vpa}.  In this paper we present an
alternative and fully automated procedure to the renormalization and
computation of amplitudes in general models, thus, combining the
second and third step. Our approach makes use of tree-level {\it
  Universal FeynRules Output} (\UFO) model files\cite{Degrande:2011ua}
and results in renormalized one-loop model files for \RecolaTwo, a
generalized version of \Recola, allowing for the
computation of any process in the underlying theory at the one-loop
level,
\change{
 with limitations only due to available memory or CPU workload. 
}

As an application of the system, we focus on two \BSM{} Higgs-production
processes at the LHC, namely Higgs production in association with a
vector boson, usually referred to as Higgs strahlung, and Higgs production in
association with two jets, known as vector-boson fusion (VBF), in the
Two-Higgs-Doublet Model (\THDM) and the Higgs-Singlet extension of the \SM{}
(\HS).  Those processes are particularly interesting for an extended
Higgs sector, as they represent the next-to-most-dominant Higgs-production
mechanisms at the LHC.  There has been enormous progress in higher-order
calculations to Higgs strahlung and VBF in the \SM{} and \BSM{}.  For Higgs
strahlung the \QCD{} corrections are known up to NNLO for
inclusive\cite{Hamberg:1990np,Brein:2003wg,Brein:2011vx} and differential
\cite{Ferrera:2011bk,Ferrera:2014lca} cross sections.  On-shell \EW{}
corrections were computed in \citere{Ciccolini:2003jy} and followed by the
off-shell calculation in \citere{Denner:2011id}.  Higgs strahlung has also been
investigated in the \THDM{} for \QCD{} \cite{Harlander:2013mla} and
\EW\cite{Denner:2016etu} corrections.  NLO QCD corrections matched to parton
shower have been presented in \citere{Maltoni:2013sma} in an effective field
theory framework. 
For VBF, the first one-loop \QCD{} corrections were obtained in a structure
function approach\cite{Han:1992hr} followed by the first two-loop
prediction\cite{Bolzoni:2010xr,Bolzoni:2011cu} in the same framework. As for
differential results, the first one-loop \QCD{} and \EW{}
corrections were calculated in \citere{Figy:2003nv} and
\citeres{Ciccolini:2007ec,Ciccolini:2007jr}, respectively.  Since recently also the
differential two-loop\cite{Cacciari:2015jma} and three-loop\cite{Dreyer:2016oyx}
\QCD{} corrections are available.  VBF has been interfaced to parton 
showers\cite{Frixione:2013mta,Rauch:2016jxo} 
and has been subject to studies for a \mbox{100 TeV}
collider\cite{Goncalves:2017gzy}. In view of \BSM, VBF has been studied in the
MSSM\cite{Figy:2010ct}.  Higgs strahlung and VBF are nowadays available in
public codes, such as V2HV \cite{Spira2015}, MCFM \cite{Campbell2015}, HAWK2.0
\cite{Denner:2014cla} and vh@nnlo \cite{Brein:2012ne}.

This paper is organized as follows. In \refse{sec:rec2} the computer
program \RecolaTwo\ is presented as a systematic approach towards the
automated generation of one-loop processes. \RecolaTwo{} relies on
one-loop renormalized model files which are automatically generated
with the new tool \Reptil{} from nothing but Feynman rules. The
computation steps are explained in different subsections, where we
discuss the translation from \UFO{} to \RecolaTwo{} model files
(\refse{sec:MFG}), the counterterm expansion and renormalization, and
the computation of rational terms of type $R_2$ (\refse{sec:AR}).  In
\refse{sec:hawkinterface} we give details on the \HAWKTwo{} interface
with \RecolaTwo, which has been used for the phenomenology. In
\refse{sec:models} we list our conventions for the \THDM{} and the
\HS, focusing on the physical input parameters. In \refse{sec:bfm} we
discuss the application of the Background-Field Method (\BFM) in
\RecolaTwo.  We present the renormalization for extended Higgs sectors in the
\BFM{} and give details on the implementation in \Reptil. In \refse{sec:setup}
we fix the calculational setup and define the
benchmark points, which were mainly taken from the Higgs cross section
working group (HXSWG). For the numerical analysis we use different
renormalization conditions for the mixing angles, which we introduce in
\refse{sec:renormschemes}. In \refse{sec:numresults} we present the numerical
results, discussing total cross sections in view of different renormalization
schemes and distributions for heavy Higgs-boson production.  After the
conclusions in \refse{sec:conclusion}, we illustrate in \refapp{sec:colorflow}
how the colour flow is derived and provide additional information on the
derivation of a minimal basis for off-shell currents in \refapp{sec:ofc}.
Finally, in \refapp{sec:simplertadpoles} we discuss the application of on-shell
renormalization schemes combined with different tadpole counterterm schemes
focusing on the gauge dependence.

\section{{\RecolaTwo}: {\Recola{}} for general models\label{sec:rec2}}
\RecolaTwo{} is a tree-level and one-loop matrix-element provider for
general models involving scalars, fermions and vector particles. It is
based on its predecessor \Recola\cite{Actis:2012qn,Actis:2016mpe},
which uses Dyson--Schwinger equations
\cite{Dyson:1949ha,Schwinger:1951ex,Schwinger:1951hq} to compute
matrix elements in a fully numerical and recursive approach. The
implementation at tree level follows the strategy developed in
\citere{Kanaki:2000ey},
supplemented by
a special treatment of the colour algebra.  The one-loop extension,
inspired by \citere{vanHameren:2009vq}, relies on the decomposition of
one-loop amplitudes as linear combination of tensor integrals and
tensor coefficients.  The former are evaluated by means of the library
\Collier{}\cite{Denner:2016kdg}, while the latter can be computed by
making use of similar recursion relations as for tree amplitudes.  The
key point is the construction of the proper tensor structure of the
coefficients at each step of the recursive procedure, which has been
implemented in \Recola{} relying on the fact that in the Standard
Model in the 't~Hooft--Feynman gauge the combination
(vertex)$\times$(propagator) is at most linear in the momenta.
\RecolaTwo{} circumvents these and other limitations of \Recola{}.  In
the following we give an introduction to \RecolaTwo{} and its
capabilities, focusing on the generalization with respect to \Recola{}
and on the applications presented in \refse{sec:numresults}.

\change{
The generalization of \Recola{} has required to remove all SM-based 
pieces of code, replacing them
with generic structures which are able to retrieve any necessary information
from the model file. 
Furthermore, the process-generation algorithm makes use of recursive functions
dealing with different cases on equal footing.
This has produced a more compact code
as no model-dependent information has been hard-coded.  
Finally, \RecolaTwo{} just needs the Feynman rules to be provided by model files in a 
specific format to directly evaluate NLO amplitudes in the model under 
consideration by using similar recursion relations to those of the SM. 
As for \Recola{}, the key ingredients are the so-called off-shell currents
\begin{align}
w^i(P,\{n\}) =
\raisebox{-1.2em}{
\includegraphics{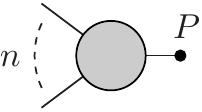}
}
\label{eq:ofc}
\end{align}
defined as the sum of all Feynman graphs which generate the 
off-shell particle $P$ combining $n$ external particles.\footnote{\change{The $n$ external particles of the sub-graph are on-shell (their wave functions 
are included, but not their propagator). 
Particle $P$ is off-shell, its wave function is not included and, for $n>1$, 
replaced by its propagator.}}
The generic index $i$ is related to the spin.  For example, in the case of a
vector field $i$ is a Lorentz index or in the case of a fermionic field $i$ is a
spinor index.  Other indices are suppressed and not relevant for the following
discussions. 

The off-shell currents \eqref{eq:ofc} are build 
recursively according to the Berends--Giele recursion relations (BGR) 
\cite{Berends:1987me}
\begin{align}
\raisebox{-4.3em}{
\includegraphics{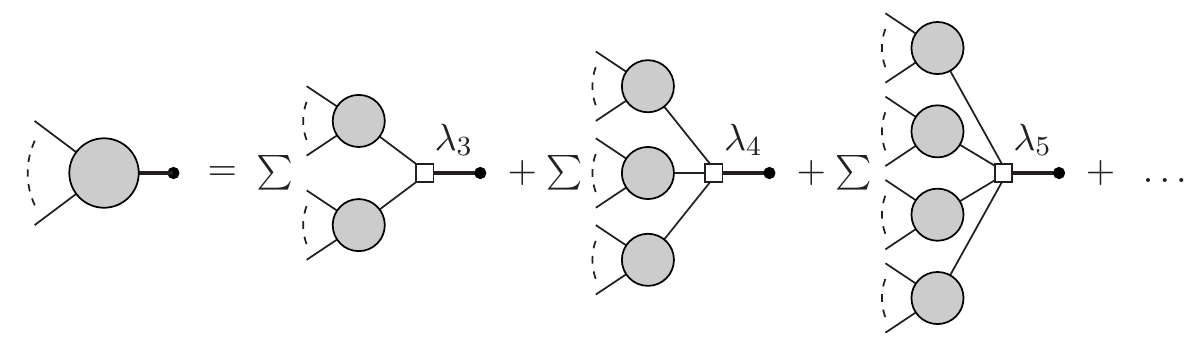}
},
\label{eq:BG}
\end{align}
which constitute a generalization of Eq.~(2.2) of
\citere{Actis:2012qn} for general models where elementary couplings
with more than four fields are present. Note that in \Recola{} \cite{Actis:2016mpe} 
the terms with $\lambda_i$, $i > 4$ are absent as
only 2-, 3-, and 4-point interaction vertices are supported.
Practically, each term on the right-hand side of the \BGR{} equation
\eqref{eq:BG} combines off-shell currents, referred to as incoming currents, and
contributes to the construction of the current on the left-hand side, referred
to as outgoing current.  An outgoing off-shell current with $n$ external
particles is calculated using the vertices of the theory connecting incoming
off-shell currents with less than $n$ external particles, which, when combined,
add up to $n$ external particles. This can be realized for tri-linear,
quadri-linear, quinti-linear, or even higher $n$-point vertices if present in
the theory.  The contribution to the outgoing current generated in each term of
equation \eqref{eq:BG} can be formally seen as the result of the action of the
$\BGR$ operator defined by
\begin{align}
  \raisebox{-3.45em}{
  \includegraphics{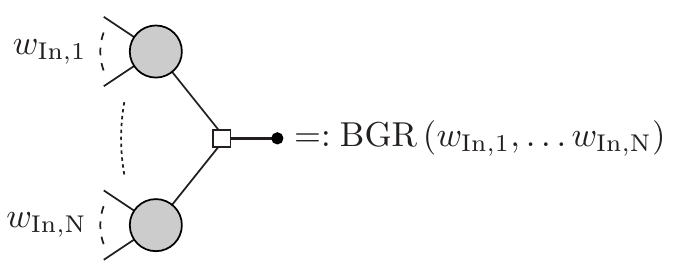}
  }
  \quad\Rightarrow\quad
  w_\mathrm{Out} = \sum
  \BGR\left(w_\mathrm{In,1},\ldots w_\mathrm{In,N}\right),
  \label{eq:BGeq}
\end{align}
with the sum running over all contributions in \eqref{eq:BG}.}

The equations \eqref{eq:BG} and \eqref{eq:BGeq} can be written in a
model-independent way as a linear combination of Lorentz structures from which
the couplings, colour structures and other relevant 
information that needs to be propagated from the left to the right
is factorized. 
\RecolaTwo{} is fully
relying on the model file to provide those rules, in addition to recursive rules
for the colour-flow and helicity-state propagation.  One could argue, that not
too many different operators are required, at least for the renormalizable
theories,
which
could have been hard-coded.  However, in view of different conventions,
different gauges and non-renormalizable theories, we decided for a flexible
system by moving this dependence to the model file.  As now the model file
provides the rules for computing off-shell currents, we can easily incorporate
the \BFM{} and \rxi-gauge for the \SM{} and \BSM{} models for \NLO{}
computations which is discussed in \refse{sec:bfm}. In addition, \RecolaTwo{}
has been generalized to deal with arbitrary $n$-point vertices,\footnote{For
reasons of optimization $n$ is restricted to $n \le 8$.} and, thus,  can compute
processes with elementary interactions between more than four fields. Dealing
with higher $n$-point vertices required to improve, among other parts of the
code, the generation of the tree graphs of the process.
\change{
The generation of those graphs is a combinatorial problem which is practically
solved in the binary representation as introduced in \citere{Caravaglios:1995cd}
(see also \citere{Kanaki:2000ey}). For elementary interactions involving an arbitrary
number of fields the method requires to compute distinct ordered
integer partitions of arbitrary size with no bitwise overlap between elements.
}

Further, \RecolaTwo{} allows for arbitrary powers of 
momenta\footnote{\change{\RecolaTwo{} has been tested with Feynman rules involving
momentum powers up to the power of $3$. Note that at one-loop order significant increase in the
rank may cause limitations due to available internal memory and CPU power.
}
}  
in Feynman rules, which is crucial for \EFT s and the \rxi-gauge at 
one-loop level.
In order to implement this important generalization, we had to generalize 
the construction of the tensor structure of loop currents (i.e. of the
coefficients of the tensor integrals), allowing the combination 
(vertex)$\times$(propagator) to contain any power of momenta. 

New theories may involve new fundamental couplings, and \RecolaTwo{} can
deal with an arbitrary number of them.\footnote{\change{This feature
has been tested with $8$ different fundamental couplings in $\Phi^8$ theory.
Note that a large number of fundamental couplings could worsen 
the performance in the process-generation and computation phase.}
}
The computation of matrix
elements is ordered according to powers of fundamental couplings, and
\RecolaTwo{} provides methods to automatically compute amplitudes and
interferences for all possible orders of these couplings. For instance, this
feature can be used to control the number of insertions of a higher dimensional
operator in a given amplitude.

Finally, \RecolaTwo{} comes with almost all features and optimizations as
provided by \Recola. \change{It is designed to be backward compatible in the sense that
a program which successfully runs with \Recola{} can be linked to \RecolaTwo{} and a \SM{} (or
\SM{} \BFM) model file and is guaranteed to run without any code adaptation.
This is realized by a dedicated \SM{} interface which has been developed on top
of the general interface to model files.
The most notable optimizations concern partial factorization in
colour-flow representation, 
the use of helicity conservation 
and the identification of fermion
loops for different fermions with equal masses.}

\subsection{\RecolaTwo{} model-file generation\label{sec:MFG}}
\RecolaTwo{} model files are generated with the tool \Reptil{} ({{\bf R}ecola's
r{\bf E}normalization {\bf P}ro\-ce\-dure {\bf T}ool at {\bf 1} {\bf L}oop}) which
is a multi-purpose tool for analytic computations at the one-loop order.
\Reptil{} is written in \PythonTwo\footnote{There is ongoing work for
\PythonThree{} compatibility.} and depends on other tools, most notably
\RecolaTwo{} for the model-independent current generation, which is used in
combination with \FORM \cite{Vermaseren:2000nd} to construct analytic vertex
functions or \SMatrix{} elements, and \Sympy \cite{10.7717/peerj-cs.103}, which
is a computer-algebra system (CAS) for Python.

\Reptil{} requires the Feynman rules in the \UFO{} format
\cite{Degrande:2011ua} which can be derived via \Feynrules
\cite{Alloul:2013bka} or \SARAH\cite{Staub:2013tta}. As there has been progress
for an automated renormalization in the \Feynrules{} framework
\cite{Degrande:2014vpa}, we stress that we do not require any results for
counterterms or rational terms. Those terms are automatically derived from the
tree-level Feynman rules in a self-contained fashion as explained in
\refse{sec:AR}.

The \RecolaTwo{} model-file generation consists of two phases. In the first phase
\Reptil{} loops over all vertices in the \UFO{} model file, disassembling
each into the vertex particles, Lorentz and colour structures, and couplings. The
colour structure is transformed to the colour-flow basis possibly rearranging
Lorentz structures and couplings. This is discussed in more detail in
\refapp{sec:colorflow}. The resulting Lorentz structures are used to derive the
\BGR{} operators in a model-independent way. For every Feynman rule \Reptil{}
tries to map the encountered  Lorentz structure onto one of those operators. If
a new structure cannot be mapped onto an existing operator a new operator is
added. In an optional second pass, the existing base of operators is minimized
(see \refapp{sec:ofc} for more details).

\begin{figure}
  \centering
  \includegraphics[width=0.79\textwidth]{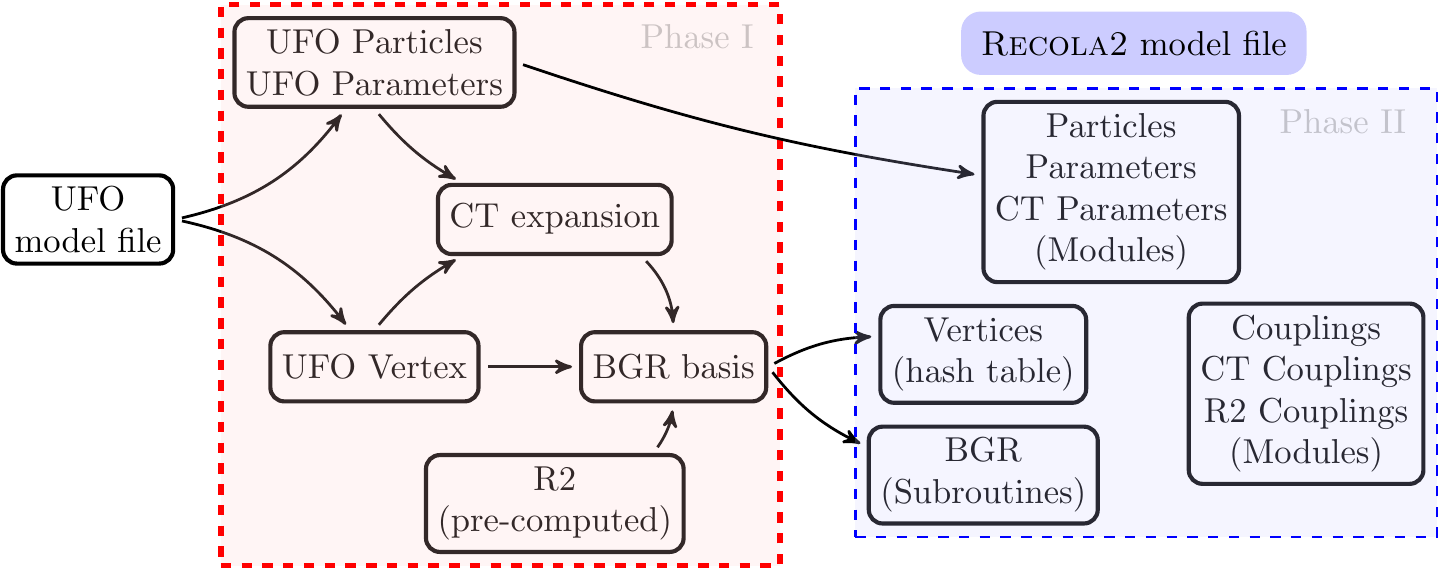}
  \caption{The \RecolaTwo{} model-file generation. \UFO{} vertices are taken as
  input and each vertex is permuted and mapped to a suited \BGR{} operator.
  Given a counterterm expansion \eqref{eq:ctexpansion}, 
  \Reptil{} can generate all counterterm vertices and
  include them in the \BGR. Once the renormalization is done and the $R_2$-terms
  are computed, the model file is derived once again,  including solutions to
  counterterm parameters and $R_2$ terms.}
  \label{fig:modelfilegen}
\end{figure}

In the second phase of the model-file generation the information is exported as
\FortranNinety{} code in form of a model-file library as depicted in
\reffi{fig:modelfilegen}.  Particle configurations are linked to the individual
contributions on the right-hand side of \eqref{eq:BG}, which differ in the
underlying \BGR{} \eqref{eq:BGeq}, colour flow, colour factors, couplings,
coupling orders or other information, via a \FortranNinety{} hash table,
allowing for a flexible and efficient access.  The actual \BGR{} are computed
and exported as \FortranNinety{} subroutines in different forms. For the
numerical evaluation tree and loop \BGR{} are used to construct tree-level and
one-loop amplitudes as it is done in \Recola. The tree \BGR{} are a special case
of the loop \BGR{}, with no loop-momentum dependence.  As a new feature in
\RecolaTwo{}, an analytic version of the \BGR{} allows to generate amplitudes as
\FORM{} code.\footnote{\Reptil{} is exporting \FortranNinety{} subroutines which
are able to write the analytic expression for the \BGR{} in \FORM{}.} 
\change{In this way the analytic expressions for the amplitudes needed in the 
renormalization conditions are derived in the same framework as the loop 
amplitudes of the computed processes, ensuring that properly defined 
renormalization schemes automatically imply UV-finite results in 
numerical computations.
In general, the UV finiteness of the theory can (and should) be verified numerically in
\RecolaTwo{} process by process by varying the scale $\mu_\mathrm{UV}$
related to the dimensional regularization of UV singularities \cite{Actis:2016mpe}.
This check also works in combination with \msbar{} subtraction schemes,
even though in this case amplitudes have an intrinsic scale dependence. To this
end, we separate the scale dependence originating from the \msbar{} subtraction
from the one of regularization.
}

Finally, \RecolaTwo{} requires particle information such as the mass, spin, and
colour of particles. This information is directly obtained from \UFO{} particle
instances and is translated to \FortranNinety{} code.
These steps conclude the tree-level model-file generation.
In the next section we discuss the counterterm generation and renormalization
and the computation of rational terms of type $R_2$.

\subsection{Counterterm expansion, renormalization and computation of
  $R_2$ terms\label{sec:AR}}
\Reptil{} supports an automated renormalization of model
files following the
standard procedure (see \eg \citere{Denner:1991}). Here we give a short summary
of all the steps, followed by details on the counterterm expansion, the
renormalization conditions, and the computation of rational terms of type $R_2$.

The starting point is a tree-level \UFO{} model file. In the first step an
independent set of parameters is identified, followed by a counterterm
expansion. The \RecolaTwo{} model file is derived, enabling the formal
counterterm expansion in \Reptil{} and leaving the values for counterterm
parameters unspecified. 
\change{Renormalization conditions are used to fix the
counterterm parameters.
\Reptil{} allows to renormalize counterterm parameters in various schemes, and
specific schemes are selected at run-time in \RecolaTwo.
}
The rational terms of type $R_2$ are constructed from
vertex functions of the underlying theory. The model file is derived once again,
including the counterterm expansion, solutions to counterterm parameters and
$R_2$ terms. The result is the desired renormalized model file, ready for
computation of processes supported by the underlying theory.

\subsubsection*{Counterterm expansion}
In the default setup, \Reptil{} defines the counterterm expansion rules of the
masses $\mv$, $\ms$, $\mf$, associated to scalars ($S$), vector bosons ($V$) and
fermions ($f$), of the not necessarily physical bosonic ($\phi$) and fermionic
fields ($\psi$), and of a set of external couplings $g_k$, according
to\footnote{We follow the conventions for the mass and field counterterms as in
\citere{Denner:1991}.}
\begin{alignat}{3}
  &\mv^2 \to \mv^2 + \dmv^2, \quad 
  &&\ms^2 \to \ms^2 + \dms^2, \quad 
  &&\mf   \to \mf + \dmf, \notag \\
  &\phi_j \to \sum_{l}\left( \delta_{jl} + \frac{1}{2}\delta Z_{jl}\right) \phi_l, \quad
  &&\psi^{\rm L}_i \to \left(1+\frac{1}{2}\delta Z_\mathrm{L,i}\right)\psi^{\rm
  L,i}, \quad
  &&\psi^{\rm R}_i \to \left(1+\frac{1}{2}\delta Z_\mathrm{R,i}\right)\psi^{\rm
  R,i}, \notag \\
  &g_k \to g_k + \delta g_k,
  \label{eq:ctexpansion}
\end{alignat}
with $\delta Z_{jl}$ being, in general, a non-diagonal matrix and L, R denoting
the left-and right-handed components of fermionic fields, which, by default, are
assumed to be diagonal.
\Reptil{} automatically deals with counterterm dependencies if the parameters,
being assigned a counterterm expansion, are declared as external parameters in
the \UFO{} format. Here, an external parameter is an independent parameter,
whereas internal parameters depend on external ones and their counterterm
expansion can be determined by the chain rule. Once all parameters have a
counterterm expansion, the most efficient way to generate counterterm vertices
of the theory is through an expansion of the bare vertices via
\eqref{eq:ctexpansion}. It is possible to add counterterm vertices by hand, or,
as a third alternative, to induce counterterm vertices from bare ones, which are
not included in the model, via counterterm expansion rules.  The latter is used
to handle $2$-point counterterms and counterterms originating from the
gauge-fixing function since both of these types have no corresponding 
tree-level Feynman rules.

\subsubsection*{Renormalization conditions}
A standard set of renormalization conditions is implemented in Python as
conditions, rather than solutions to conditions, which are 
solved upon request.
As an advantage of solving equations, the form of vertex
functions or conventions can change without breaking the system.
\Reptil{} supports on-shell, \msbar, and momentum-subtraction conditions for general
(mixing-)two-point functions. \msbar{} subtraction is implemented generically
for $n$-point functions. 
We assume standard renormalization of the physical fields and masses from 
the complex poles of Dyson-resummed propagators and their residues, while we 
allow for several choices of renormalization conditions for the gauge-fixing 
function and for unphysical fields. 
In addition, we provide standard renormalization conditions for the
\SM{} couplings, \eg the
definition of $\alpha$ in the Thomson limit (TL) and
in the $G_\mathrm{F}$ scheme, which are implemented via
self-energies\footnote{
The (gauge-dependent) vertex and box contributions of the muon decay are taken 
from the SM.
\change{In general, they depend on the model under consideration and need to be
computed explicitly, but for extended Higgs sectors they are well 
approximated by the SM ones for small muon Yukawa couplings.}
},
and the $N_\mathrm{f}$-flavour scheme for $\alpha_\mathrm{s}$ in 
\QCD\footnote{
For theories with the \SM--\QCD{} particle content a running dynamical-flavour 
scheme is supported.}, 
which is implemented as a combined \msbar{}/momentum subtraction on vertex functions.  
All conditions are implemented in a model-independent way.
Instead of the standard set of renormalization conditions already 
implemented, \Reptil{} can also handle alternative conditions properly set 
by the user. 

Setting up renormalization conditions requires a \RecolaTwo{} model file
including counterterms. The derivation of model files is done as discussed in
the previous section with enabled vertex counterterm expansion (see
\reffi{fig:modelfilegen}) and leaving the counterterm parameter unspecified.
\begin{figure}
  \centering
  \includegraphics[width=0.69\textwidth]{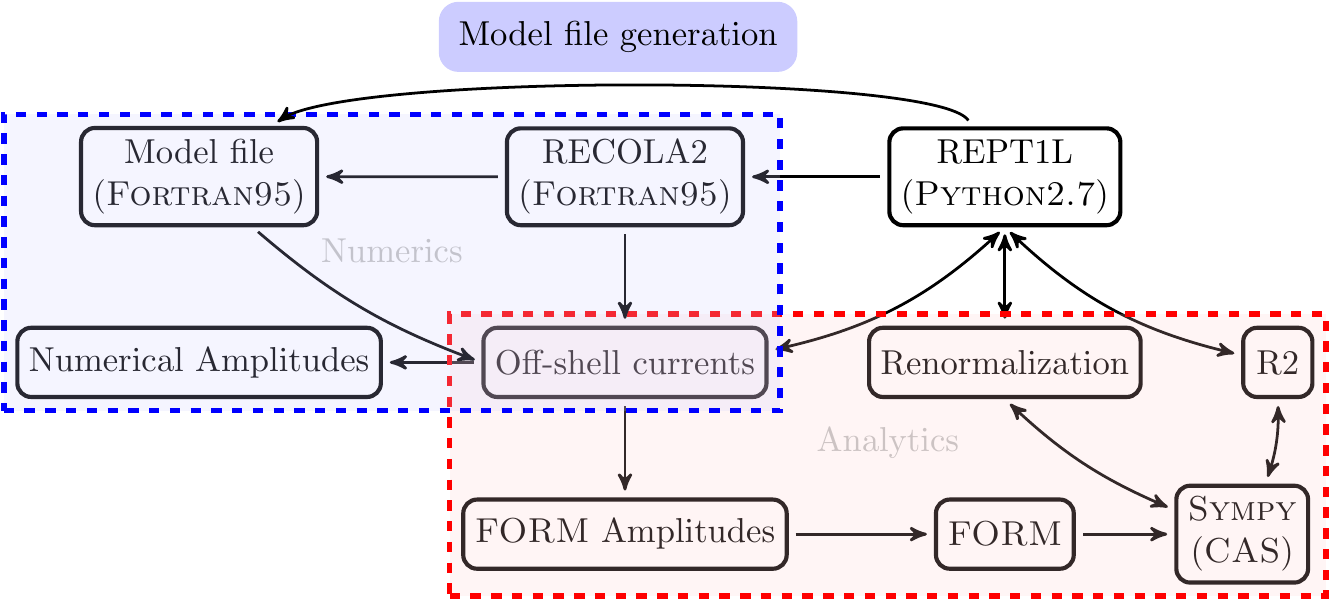}
  \caption{ The \Reptil--\RecolaTwo{} tool chain. \Reptil{} can generate
  tree-level model files which can then be used in combination with the
  \RecolaTwo{} library to generate building blocks required in the
  renormalization process.  The process generation is done via the same
  off-shell currents also used in numerical computations. The currents are
  evaluated analytically with \FORM{} and further processed with \Sympy. The
  results are then available to \Reptil{} and are used in the renormalized model
  file derivation. The red box indicates the analytic computations which uses
  the tool chain combining \RecolaTwo{}, \FORM{} and \Sympy. After the
  renormalized model-file derivation, this tool chain and \Reptil{} are no longer
  needed. The blue box, \ie \RecolaTwo{} and model files, can be used as
  stand-alone versions (pure \FortranNinety) for numerical computations.}
  \label{fig:ctandr2computation}
\end{figure}
The renormalization conditions are derived analytically as \FORM{}
code. \Reptil{} uses \RecolaTwo{} to generate the skeletons for
processes. The result is written to a \FORM{} file and evaluated,
yielding vertex functions which are parsed to Python and processed
with \Sympy{} solving the conditions for the counterterm parameters.
The procedure is visualized in
\reffi{fig:ctandr2computation}.
\change{Multiple schemes for the very same counterterm parameters can be implemented by
imposing different renormalization conditions.}
All schemes are exported to the
\RecolaTwo{} model file and, for a given parameter, a specific scheme
can be selected before the process generation phase. For instance,
this system can be used to allow the user to choose between different
\QCD{} and \EW{} renormalization schemes within the same model file.
The same system is used for dealing with singularities from light
fermions. In general, particles can be tagged as light particles,
which, when a particle is subject to on-shell renormalization, makes
\Reptil{} to regularize the associated diagonal two-point function in
three different setups, namely dimensional regularization, mass
regularization, and keeping the full mass dependence. In a \RecolaTwo{}
session a suited regularization scheme for light particles is set
automatically, depending on the choice of the mass value, unless the
regularization for a particle is explicitly required in a specific
scheme.
In the case of unstable particles, \ie massive particles with finite widths,
\Reptil{} applies, by default, the Complex-Mass Scheme (CMS) as discussed in
more detail in \refse{sec:renobfm}.

\subsubsection*{Computation of $R_2$ terms}
The computation of $R_2$ uses the methods developed in
\citeres{Ossola:2008xq,Draggiotis:2009yb,Garzelli:2009is} and follows the same
computation flow as solving renormalization conditions which is depicted in
\reffi{fig:ctandr2computation}. For renormalizable theories all
existing $R_2$ terms can
be computed. To this end, \Reptil{} can generate the skeletons 
at NLO for all vertex functions in the theory which are potentially UV divergent
by power counting. \FORM{} is used to construct each vertex function, replace
tensor integrals by their pole parts and take the limit $D\to4$. The finite
parts are identified as Feynman rules associated to the original vertices, which
are precisely the $R_2$ terms. These steps are done in Python with the help of
\Sympy.

The computation of tensor coefficients is done in conventional dimensional
regularization. Different regularizations will be supported in the future by
exchanging the responsible \FORM{}-procedure files.  In view of \EFT s, the
power counting can be disabled, and specific vertex functions can be selected.
Further, the $R_2$ extraction
rules\cite{Ossola:2008xq,Draggiotis:2009yb,Garzelli:2009is} have been
extended to higher $n$-point functions and higher rank.\footnote{$3$-and
$4$-point functions up to
rank $6$, $5$- point functions up to rank $7$, and $6$-point functions up to rank $8$.}

\section{\HAWKTwo{} interface to \RecolaTwo\label{sec:hawkinterface}}
In this section we describe the interface between \HAWKTwo{} and \RecolaTwo{}
which allows for an automated computation of \NLO{} \EW{} and \QCD{} corrections
to observables in associated Higgs production with a vector boson or two jets.
We start with the \LO{} partonic channels and virtual corrections and conclude
with the computation of the real corrections. The implementation has been
realized in a model-independent way, allowing in the future, apart from the two
presented \BSM{} models, for predictions in alternative models.

\subsection{Process definitions at \LO{} and \NLO{} with \RecolaTwo}
In the case of associated Higgs production with a vector boson, also known as
Higgs strahlung, we consider processes with an intermediate vector boson
decaying leptonically as
\begin{align}
  \Pp \Pp \to \PH \PV \to \PH l^+ l^-/\PH l^\pm \nu/ \PH \nu\nu.
  \label{eq:HS}
\end{align}
Depending on the initial-state partons, the intermediate vector boson can 
be a $\PZ$ or a $\PW$ boson. For example for the signature $\Pp \Pp \to \PH \PZ
\to \PH l^+ l^-$, neglecting bottom contributions in the PDFs, there are four
different initial-state parton combinations:
\begin{align}
  \bar \Pu \Pu, \quad
  \bar \Pd \Pd, \quad
  \bar \Pc \Pc, \quad
  \bar \Ps \Ps.
  \label{eq:quarkchannels}
\end{align}
Whenever possible, we optimize computations involving different quark
generations. For instance, in \eqref{eq:quarkchannels} the processes
involving the second generation are
not computed explicitly, but the results for the first generation are employed
instead. For the first generation of quarks the \RecolaTwo{} library is used to
generate the processes at tree and one-loop level. 

The second process class under consideration is Higgs production in
association with two hard jets
\begin{align}
  \Pp \Pp \to \PH \mathrm{j} \mathrm{j},
  \label{eq:VBF}
\end{align}
also known as VBF. There are plenty of partonic channels and,
again, we exploit optimizations with respect to the different quark generations.
For the \LO{}, virtual \NLO{} \EW, virtual \NLO{} \QCD{}, real emission \EW, and
real emission \QCD{} contributions \RecolaTwo{} generates 32 partonic
channels each, with the real kinematic channels corresponding to the Born kinematic
ones, with an additional gluon or photon.  For the gluon- and photon-induced
channels \RecolaTwo{} generates 20 channels each.

At the stage of the process definition the Higgs boson entering in \eqref{eq:HS}
or \eqref{eq:VBF} can be chosen freely\footnote{Charged Higgs bosons are not
supported by the \HAWKTwo{} Monte Carlo. Pseudo-scalar Higgs-boson production is
possible, but suppressed in the considered CP-conserving \THDM.} as long as it
is supported by the \RecolaTwo{} model file currently in use.  For instance, in
the case of the \THDM{} the Higgs flavour can be set to $\Hl$, $\Hh$ or $\Ha$
(see \refse{sec:models}), which is done in the \HAWKTwo{} input file.  In
\HAWKTwo{} the relevant parameters for process generation and computation are
set by input files. This information is
forwarded to \RecolaTwo{}, allowing to
choose specific contributions.  The selection works for individual corrections
such as \QCD{} or \EW{} either virtual or real.  For the results presented in
this work we selected the pure electroweak corrections, including photon-induced
corrections.
 
\subsection{Infrared divergences}
\change{\RecolaTwo{} provides the amplitudes for the partonic processes under
consideration as well as the colour-correlated squared matrix elements
needed for the Catani--Seymour dipole subtraction.  In order to deal
with IR singularities, an IR subtraction scheme needs to be employed.}
We adhere to the Catani--Seymour dipole subtraction
\cite{Catani:1996vz} which is used in \HAWKTwo{} and employ mass regularization
for soft and collinear divergences, \ie a small photon mass and small fermion
masses are used wherever needed. 
From the point of view of the interface,
dealing with \EW{} dipoles is a matter of replacing certain Born amplitudes with
the ones computed by \RecolaTwo.  As for the \QCD{} dipoles one needs in general
colour-correlated matrix elements.  For processes with only two partons, as it
is the case for Higgs strahlung, the colour correlation is diagonal owing to
colour conservation (see Eq.~A1 in \citere{Catani:1996vz}) and again no
colour-correlated matrix elements are required.  For VBF we compute the
colour-correlated matrix elements directly with \RecolaTwo, and use colour
conservation to minimize the number of required computations. The dipoles are
used as implemented in \HAWKTwo{} 
\change{and are not part of \RecolaTwo.}
For the
\QCD{} dipoles consider \citeres{Catani:1996vz,Catani:2002hc} and for \EW{}
dipoles see \citeres{Dittmaier:1999mb,Dittmaier:2008md}.

\section{\THDM{} and \HS{} model description\label{sec:models}}
In this section, we sketch the definition of the scalar potential of the \THDM{}
and the \HS. In both cases we restrict ourselves to a CP-conserving
$Z_2$-symmetric scalar potential, which in the case of the \THDM{} is allowed to
be softly broken. For a comprehensive introduction to the \THDM{} we refer to
\citeres{Gunion:2002,Branco:2011iw} and for the \HS{} to the original literature
\cite{Schabinger:2005ei,Patt:2006fw,Bowen:2007ia} and applications to LHC
phenomenology in
\citeres{Pruna:2013bma,Robens:2015gla,Robens:2016xkb,Chalons:2016jeu}.  For the
kinetic terms we refer to the conventions used in \citere{Pruna:2013bma}.

\subsection{Fields and potential definition}
Both models are simple extensions of the \SM, only affecting the form and fields
entering the scalar potential and 
for
the \THDM{} also the Yukawa interactions. In the case of the \THDM{} we have two
Higgs doublets, generically denoted as \Phii{} with $i=1,2$ and defined
component-wise by
\begin{align}
\Phii=\left(
  \begin{array}{c}
    \phiip\\
    \frac{1}{\sqrt{2}}\left(\vi+\rhoi+\ii\etai\right)
  \end{array}
\right),
\end{align}
with $\vi$ denoting the vevs.
Under the constraint of CP conservation plus the $Z_2$ symmetry
($\Phione\to-\Phione, \Phitwo\to\Phitwo$), the most 
\change{general, renormalizable}
potential reads
\cite{Gunion:2002}
\begin{align}
  V_{\mathrm{2HDM}}={}&m_1^2\Phione^{\dagger}\Phione+m_2^2\Phitwo^{\dagger}\Phitwo
    -m_{12}^2\left(\Phione^{\dagger}\Phitwo+\Phitwo^{\dagger}\Phione\right)\notag\\
    &+\frac{\lambda_1}{2}\left(\Phione^{\dagger}\Phione\right)^2
    +\frac{\lambda_2}{2}\left(\Phitwo^{\dagger}\Phitwo\right)^2
    +\lambda_3\left(\Phione^{\dagger}\Phione\right)\left(\Phitwo^{\dagger}\Phitwo\right)
    +\lambda_4\left(\Phione^{\dagger}\Phitwo\right)\left(\Phitwo^{\dagger}\Phione\right)\notag\\
    &+\frac{\lambda_5}{2}\left[\left(\Phione^{\dagger}\Phitwo\right)^2
    +\left(\Phitwo^{\dagger}\Phione\right)^2\right],
\label{eq:thdmpot}
\end{align}
with five real couplings $\lambda_1\ldots\lambda_5$, two real mass parameters
$m^2_1$ and $m^2_2$, and the soft $Z_2$-breaking parameter $m^2_{12}$.

The \HS{} scalar potential involves one Higgs doublet $\Phi$ and a singlet field
$S$ defined as
\begin{align}
\Phi=\left(
  \begin{array}{c}
    \phi^+\\
    \frac{1}{\sqrt{2}}\left(v+\rho_1+\ii\eta\right)
  \end{array}
\right), \qquad S = \frac{v_s + \rho_2}{\sqrt{2}}.
\end{align}
Under the same constraints, the most general,
\change{renormalizable}
potential reads
\begin{align}
  V_{\mathrm{HSESM}} = m_1^2 \Phi^\dagger \Phi + m_2^2 S^2
  +\frac{\lambda_1}{2} \left(\Phi^\dagger \Phi\right)^2
  +\frac{\lambda_2}{2} S^4
  +\lambda_3 \Phi^\dagger \Phi S^2,
  \label{eq:hspot}
\end{align}
with all parameters being real.


\change{\subsection{Parameters in the physical basis\label{sec:higgspotentials}}}
Both potentials are subject to spontaneous symmetry breaking which requires a
rotation of fields to the mass eigenstates in order to identify the physical
degrees of freedom.  For the \THDM{} there are five physical Higgs bosons $\Hl,
\Hh, \Ha, \Hpm$ and in the \HS{} there are two neutral Higgs bosons $\Hl$ and
$\Hh$, intentionally identified with the same symbols as in the \THDM.  Besides
the physical Higgs bosons, there are the three would-be Goldstone bosons $G_0$
and $G^\pm$ in the 't~Hooft--Feynman gauge.  The mass eigenstates for the
neutral Higgs-boson fields are obtained in both models by the
transformation
\begin{align}
  \left(\begin{array}{c}
    \rhoone\\
    \rhotwo
  \end{array}\right)=
  R(\al)
  \left(\begin{array}{c}
    \Hh\\
    \Hl
  \end{array}\right), \quad
  \text{with}\quad
  R(\alpha)=
  \left(\begin{array}{cc}
        \ca & -\sa\\
        \sa & \phantom{-}\ca
  \end{array}\right),
   \label{eq:neutralrotation}
\end{align}
and 
$\alpha$ being fixed such that the mass matrix  
\begin{align}
  M_{ij}:= \left.\frac{\partial^2 V}{\partial \rho_i
  \partial\rho_j}\right|_{\varphi=0},
  \label{eq:neutralmm}
\end{align}
is diagonalized via $R(-\alpha) M R(\alpha)$, with the potential $V$ being either
\eqref{eq:thdmpot} or \eqref{eq:hspot}. The solution to \eqref{eq:neutralmm} for
 symmetric $2\times 2$ matrices is generically given by (see \citere{Gunion:2002})
\begin{align}
  \stwoa = \frac{2 M_{12}}
  {\sqrt{({M}_{11} - M_{22})^2 + 4 M_{12}^2}}.
  \label{solalpha}
\end{align}
In the \THDM{} there are additional mixings between charged and pseudo-scalar
bosons and Goldstone bosons, which are diagonalized as follows
\begin{align}
  \left(\begin{array}{c}
    \phionepm\\
    \phitwopm
  \end{array}\right)={}&
  R(\beta)
  \left(\begin{array}{c}
    G^{\pm}\\
    H^{\pm}
  \end{array}\right), \qquad
  \left(\begin{array}{c}
    \etaone\\
    \etatwo
  \end{array}\right)=
  R(\be)
  \left(\begin{array}{c}
    \GZ\\
    \Ha
  \end{array}\right),\quad
  \text{with}\quad
  R(\beta)=
  \left(\begin{array}{cc}
        \cbe & -\sbe\\
        \sbe & \phantom{-}\cbe
  \end{array}\right).
  \label{eq:pseudochargedrotations}
\end{align}
The angle $\beta$ is related to the vevs according to $\tb \equiv
\tan\beta = \vtwo/\vone$ in the \THDM. For the \HS{} we define $\tb \equiv
\tan\beta = v_s/v$.
The Higgs sector is minimally coupled to the gauge bosons.  Collecting quadratic
terms and identifying the masses one obtains the well-known tree-level relations
\begin{align}
  \mw = \frac{1}{2} \g v, \qquad
  \mz = \frac{1}{2} \sqrt{\g^2 + \gy^2}\; v,\qquad
  \label{SSB}
\end{align}
where \g{} and \gy{} denote the weak isospin and hypercharge gauge
couplings, and $\mw$ and $\mz$ the \PW- and \PZ-boson masses,
respectively. For the \THDM{} we identify $v=\sqrt{\vone^2+\vtwo^2}$.
Finally one employs the minimum conditions for the scalar potential which, in both
models, read
\begin{align}
\langle \rho_i\rangle=0.
\end{align}
Then, one substitutes the potential parameters with physical parameters obtained
after spontaneous symmetry breaking and after diagonalizing the Higgs sector.
For the \THDM{} we choose the Higgs-boson masses \mhl\ (light Higgs boson),
\mhh\ (heavy Higgs boson), \mha\ (pseudoscalar Higgs boson), \mhc\ (charged
Higgs boson), the soft-$Z_2$-breaking scale \Msb{} defined via
\begin{align}
  \Msb^2=\frac{m_{12}^2}{\cbe\sbe},
  \label{eq:softz2breakingterm}
\end{align}
and the two mixing angles as $\cab:=\cos(\al-\be)$
\change{($\sab:=\sin(\al-\be)$)}
and $\tb:= \tan(\beta)$, 
which is a natural choice for studying (almost) aligned scenarios.
For the \HS{} 
we use the neutral Higgs-boson masses \mhl\ (light Higgs boson), \mhh\
(heavy Higgs boson) and the angles $\sas:=\sin(\al)$ and $\tb:= \tan
(\beta)$.
To summarize, we transform the parameters from the generic basis to the physical
one by choosing the following parameters as external ones
\begin{description}
\item [\THDM:] 
  $\lambda_1,\lambda_2,\lambda_3,\lambda_4,\lambda_5,m_1,m_2,m_{12}
   \;\to\;
   \mhl,\mhh,\mha,\mhc,\Msb,\cab,\tb,v\left(g,\mw\right),$
\item [\HS:] 
 $ \lambda_1,\lambda_2,\lambda_3,m_1,m_2
   \;\to\;
   \mhl,\mhh,\sas,\tb,v\left(g,\mw\right),$
\end{description}
where we have indicated that the vev is traded for gauge couplings and masses
according to \eqref{SSB}.


\change{\subsection{Yukawa interactions}}
The fermionic sector in the \HS{} is the same as in the \SM{}, whereas the
\THDM{} allows for a richer structure.  In the general case of the \THDM{},
fermions can couple to both $\Phione$ and $\Phitwo$, leading to
flavour-changing neutral currents (\FCNC) already at tree level.  Since \FCNC{}
\change{processes are extremely rare in nature}
they highly constrain \BSM{} models. In
order to prevent \change{tree-level}
\FCNC{}, one imposes the $Z_2$ symmetry
\begin{align}
  Z_2:\quad \Phione\to-\Phione,\quad \Phitwo\to\Phitwo,
  \label{eq:Z2symmetry}
\end{align}
as already introduced in the Higgs potential in \refse{sec:higgspotentials}.
This $Z_2$ symmetry is motivated by the Glashow--Weinberg--Paschos theorem in
\citeres{Glashow:1976nt,Paschos:1976ay}, which states that for an arbitrary
number of Higgs doublets, if all right-handed fermions couple to exactly one of
the Higgs doublets, \FCNC s are absent at tree level. This can be realized by
imposing, in addition to \eqref{eq:Z2symmetry}, a parity for right-handed
fermions under $Z_2$ symmetry. One obtains
four distinct \THDM{} Yukawa terms, \change{the so-called natural flavour-conserving models canonically described in the literature.}
\begin{description}
\item [Type I:] 
By requiring for all fermions an even parity under $Z_2$,
all have to couple to the second Higgs doublet $\Phitwo$. The corresponding
Yukawa Lagrangian reads
\begin{align}
  \mathcal{L}_\mathrm{Y}&=
  -\Gamma_\mathrm{d} \overline{Q}_\mathrm{L}\Phi_2 d_\mathrm{R}
  -\Gamma_\mathrm{u} \overline{Q}_\mathrm{L}\tilde{\Phi}_2 u_\mathrm{R}
  -\Gamma_\mathrm{l} \overline{L}_\mathrm{L}\Phi_2 l_\mathrm{R}
  +\mathrm{h.c.},
\end{align}
where $\tilde{\Phi}_2$ is the charge-conjugated Higgs doublet of
$\Phi_2$. Neglecting flavour mixing, the coefficients are directly expressed by
the fermion masses $m_\mathrm{d}$, $m_\mathrm{u}$ and $m_\mathrm{l}$, and
the mixing angle $\beta$,
\begin{align}
  \Gamma_\mathrm{d}=\frac{g\, m_\mathrm{d}}{\sqrt{2} \mw \sbe}, \qquad
  \Gamma_\mathrm{u}=\frac{g\, m_\mathrm{u}}{\sqrt{2} \mw \sbe}, \qquad
  \Gamma_\mathrm{l}=\frac{g\, m_\mathrm{l}}{\sqrt{2} \mw \sbe}.
\end{align}
Again, the vev $v$ has been substituted using Eq.~\eqref{SSB}.
\item [Type II:] 
This is the \MSSM-like scenario obtained by requiring odd parity for down-type
    quarks and leptons: $d_\mathrm{R}\to-d_\mathrm{R}$,
    $l_\mathrm{R}\to-l_\mathrm{R}$ and even parity for up-type quarks. It
    follows that the down-type quarks and leptons couple to $\Phione$, while
    up-type quarks couple to $\Phitwo$. The corresponding Yukawa Lagrangian
    reads
\begin{align}
  \mathcal{L}_\mathrm{Y}&=
  -\Gamma_\mathrm{d} \overline{Q}_\mathrm{L}\Phi_1 d_\mathrm{R}
  -\Gamma_\mathrm{u} \overline{Q}_\mathrm{L}\tilde{\Phi}_2 u_\mathrm{R}
  -\Gamma_\mathrm{l} \overline{L}_\mathrm{L}\Phi_1 l_\mathrm{R}
  +\mathrm{h.c.}.
\end{align}
Neglecting flavour mixing, the coefficients are expressed by the
fermion masses $m_\mathrm{d}$, $m_\mathrm{u}$ and $m_\mathrm{l}$, and the
mixing angle $\beta$,
\begin{align}
  \Gamma_\mathrm{d}=\frac{g\, m_\mathrm{d}}{\sqrt{2} \mw \cbe}, \qquad
  \Gamma_\mathrm{u}=\frac{g\, m_\mathrm{u}}{\sqrt{2} \mw \sbe}, \qquad
  \Gamma_\mathrm{l}=\frac{g\, m_\mathrm{l}}{\sqrt{2} \mw \cbe}.
\end{align}
\item [Type Y:] 
  This type, also referred to as
\change{{\it lepton-specific} \THDM}, is obtained by requiring odd
parity only for leptons: $l_\mathrm{R}\to-l_\mathrm{R}$.
\item [Type X:] 
  This type, also referred to as
\change{{\it flipped} \THDM}, is obtained by requiring odd parity only for
down-type quarks: $d_\mathrm{R}\to-d_\mathrm{R}$.
\end{description}
In the analysis of this paper
we focus on Type II, which is equivalent to Type I for massless
leptons and quarks, except for the top quark. We remark that exactly one
\RecolaTwo{} model file can handle all Yukawa types, and switching
between different Yukawa types is done by a simple function call.


\section{Background-Field Method for extended Higgs sectors\label{sec:bfm}}
The \BFM{} is a powerful formulation for gauge theories which renders analytic
calculations easier due to a simple structure of the Feynman rules and
additional symmetry relations. The method was originally derived by DeWitt in
\citeres{DeWitt:1964yg,DeWitt:1967ub}\footnote{See \citere{Abbott:1981ke} for a
pedagogical introduction.} and has since then been used in many applications.
\change{The additional symmetry relations emerge for gauge theories in combination with a
suited gauge-fixing term and encode the invariance of the theory
under so-called Background-Field gauge invariance.
This property is particularly
useful for the calculations of $\beta$ functions
\cite{Bornsen:2002hh} in higher orders and is also of interest in beyond flat}
space-time quantum field theory. The \BFM{} can be used to calculate \SMatrix{}
elements, as constructed in \citere{Abbott:1983zw}, which, despite having to
deal with many more Feynman rules, is in our implementation as efficient as the
conventional formalism. Further, the \BFM{},
\change{which can be viewed as a different choice of gauge,
allows for an alternative way of
computing \SMatrix{} elements and, thus, provides a powerful check of the}
consistency of the \Reptil/\RecolaTwo{} tool chain. This is particularly useful
for the validation of $R_2$ terms where mistakes are difficult to spot. In
addition, we checked a few Background-Field Ward identities.
\change{We stress that the \BFM{} can be used as a complementary method in \RecolaTwo{}
besides the usual formulation. Even though the use of the \BFM{} in practical
calculations is steered in precisely the same way as for model files in the
conventional formulation, the internal machinery is different. In particular,
the derivation of the Feynman rules and renormalization procedure requires
special attention which is discussed in the following.}

\subsection{BFM action for extended Higgs sector}
The results presented here are a simple generalization of
\citere{Denner:1994xt}, which deals with the \BFM{} applied to the \SM{} at
one-loop order. The \BFM{} splits fields in background and quantum fields and
combines the new action with a special choice for the gauge-fixing function
resulting in a manifest background-field gauge invariance for the effective
action
\change{at the quantum level. This splitting separates the classical solutions of the
field equations, represented by background fields, from the quantum excitation modes,
represented by quantum fields.}
The Feynman rules are derived as usual, treating background and
quantum fields on equal footing, which we have done with \Feynrules.
In principle, the splitting can be done for every field in the theory, however,
as we are only interested in a background-field gauge-invariant action, it is
sufficient to shift fields which enter the gauge-fixing function. Thus, we
perform
\begin{alignat}{2}
  &{W}^{a,\mu} \to \underline {W}^{a,\mu} := {W}^{a,\mu} + \hat{W}^{a,\mu}, \qquad
  &&B^\mu \to \underline B^\mu := B^\mu + \hat{B}^\mu, \notag\\
  &\Phii \to \underline \Phi_i :=\Phii + \Phibi, \qquad \qquad
      &&S \to \underline S:=S + \hat S,
  \label{eq:shiftedfields}
\end{alignat}
where $W^a$ $(\hat{W}^a)$ and $B$ $(\hat B)$ are the \SM{} quantum (background)
gauge fields in the gauge eigenbasis with $a=1,2,3$.  The index $i$ runs over
all Higgs doublets $\Phii$ in the theory under consideration, and $S$ is a
singlet field, absent in the \THDM{} or \SM{}.
While the singlet field $S$ does not appear explicitly in the gauge-fixing
function [see \eqref{eq:bfmginvgf}], the inclusion of $S$ in the splitting
\eqref{eq:shiftedfields} is necessary due to the mixing with the neutral
component of a Higgs doublet.  The components for the background- and
quantum-field doublets are defined as
\begin{align}
\Phii=\left(
  \begin{array}{c}
    \phiip\\
    \frac{1}{\sqrt{2}}\left(\rhoi+\ii\etai\right)
  \end{array}
\right),\qquad
\Phbii=\left(
  \begin{array}{c}
    \phbiip\\
    \frac{1}{\sqrt{2}}\left(\vi+\rhobi+\ii\etabi\right)
  \end{array}
\right).
\end{align}
By convention, we keep the original vev of the Higgs doublet in the Higgs
background-field doublet.  The quantum gauge-fixing term has the traditional
form. In the gauge eigenbasis it reads
\begin{align}
  \mathcal{L}_\mathrm{GF} = -\frac{1}{2}\sum_{a=1}^3 \left(F_W^a\right)^2 - \frac{1}{2}
  F_B^2,
  \label{eq:bfmginvgf}
\end{align}
with generalized gauge-fixing functions
\begin{align}
  F_W^a &= 
  \hat{D}^\mu {W}^a_\mu - \ii \frac{\g }{2}
  \sum_i
  \left[\Phibi^\dagger \sigma^a \Phii - \Phii^\dagger
  \sigma^a \Phibi \right],\notag \\
  F_B &=  \partial^\mu B_\mu + \ii \frac{\gy}{2} \sum_i \left[\Phibi^\dagger \Phii -
  \Phii^\dagger \Phibi \right],
\label{eq:bfmginvgfcomponents}
\end{align}
and $i$ running over all Higgs doublets.  The covariant derivative
$\hat D$ is similar to the usual one, but with a background-field
gauge connection instead of a quantum-field one. For a field
$\alpha^j$ in the adjoint representation it acts in the following way
\begin{align}
  \hat{D}_\mu \alpha^a =
  \partial_\mu \alpha^a + \g \epsilon^{a b c } \hat{W}^b_\mu \alpha^c,  \label{eq:bfmgt}
\end{align}
with $\epsilon^{a b c}$ being the structure constants of SU(2).
The form \eqref{eq:bfmginvgf}, \eqref{eq:bfmginvgfcomponents} is invariant under
background-field gauge transformations, which can be shown using the techniques
presented in \citere{Abbott:1981ke}, but suitably generalized in the presence of
spontaneous symmetry breaking.

The construction of the ghost term follows the standard BRST quantization
procedure. Once the symmetry transformations are defined on the fields, a valid
ghost Lagrangian, leading to a BRST invariant action, is given by
\begin{align}
  \mathcal{L}_\mathrm{ghost} =
  -\sum_{a=1}^3\bar{u}_W^a  \delta_\mathrm{B}  {F}_W^a
  - \bar{u}_B  \delta_\mathrm{B} F_B.
  \label{eq:ghostlag}
\end{align}
The fields in the gauge eigenbasis
are rotated to the physical basis in the following way
\begin{alignat}{4}
  &W_\mu^1 = \frac{ W_\mu^- + W_\mu^+}{\sqrt{2}}, \quad \notag
  &&W_\mu^2 = \frac{W_\mu^--W_\mu^+}{\ii \sqrt{2}},  \quad
  &&W_\mu^3 = \cw Z_\mu - \sw A_\mu, \quad 
  &&B_\mu   = \sw Z_\mu + \cw A_\mu, \\
  &u_W^1 = \frac{u_W^- + u_W^+}{\sqrt{2}}, \quad 
  &&u_W^2 = \frac{u_W^- - u_W^+}{\ii \sqrt{2}}, \quad
  &&u_W^3 = \cw u_Z - \sw u_A, \quad 
  &&u_B  = \sw u_Z + \cw u_A.
  \label{eq:physbase}
\end{alignat}
The BRST transformations on the gauge eigenbasis, expressed in terms
of physical fields via \eqref{eq:physbase}, read
\begin{align}
  &\delta_\mathrm{B} W^1_\mu = \underline{D}_{\mu} u_W^1 = \notag\\
  &\frac{1}{\sqrt{2}} \partial_\mu \left( u_W^- + u_W^+ \right)
  -\frac{\ii e}{\sqrt{2}} \left[  
    \left( u_W^- - u_W^+\right) \left(\underline A_\mu - \frac{\cw}{\sw}
    \underline Z_\mu\right)
    + \left( u_A - \frac{\cw}{\sw} u_Z\right) \left( \underline W_\mu^+ - \underline W_\mu^-\right)
  \right],\\
  &\delta_\mathrm{B} W^2_\mu = \underline{D}_{\mu} u_W^2 = \notag\\
  &\frac{\ii}{\sqrt{2}} \partial_\mu \left( u_W^+ - u_W^- \right)
  -\frac{e}{\sqrt{2}} \left[  
    \left( u_W^- + u_W^+\right) \left(\underline A_\mu - \frac{\cw}{\sw}
    \underline Z_\mu\right)
    - \left( u_A - \frac{\cw}{\sw} u_Z\right) \left( \underline W_\mu^+ + \underline W_\mu^-\right)
  \right],\\
  &\delta_\mathrm{B} W^3_\mu = \underline{D}_{\mu} u_W^3 = 
  \partial_\mu \left( \cw u_Z - \sw u_A \right)
  -\frac{\ii e}{\sw}   
    \left( u_W^- \underline W^+ - u_W^+ \underline W^- \right), \\
  &\delta_\mathrm{B} B_\mu = \underline{D}_{\mu} u_B = 
  \partial_\mu \left( \cw u_A + \sw u_Z \right).
\end{align}
Note that in contrast to the conventional formalism, the covariant derivatives
entering the BRST transformations use the shifted gauge fields
\eqref{eq:shiftedfields}.
For the Higgs doublets the BRST transformation rules can be defined \change{at the level}
of components as follows
\begin{align}
  \delta_\mathrm{B} \Phii := 
  \left(
  \begin{array}{c}
    \delta_\mathrm{B}\phiip\\
    \frac{1}{\sqrt{2}}\left(\delta_\mathrm{B}\rhoi+\ii\delta_\mathrm{B}\etai\right)
  \end{array}
  \right),
\end{align}
with
\begin{align}
  \delta_\mathrm{B}\phiip &= 
  \frac{\ii e}{2 \sw} (\ii \underline \eta_i +\underline \rho_i + v_i) u_W^+ 
  + \frac{\ii e \left(\cw^2 -\sw^2\right)}{2 \cw \sw} \underline \phi_i^+ u_Z
  - \ii e \underline\phi_i^+ u_A\\
  \delta_\mathrm{B}\rho_i &= \frac{e}{2\cw\sw} \underline\eta_i u_Z
  + \frac{ \ii e}{2\sw} \left( \underline \phi_i^+ u_W^- - \underline \phi_i^- u_W^+ \right)
  \\
  \delta_\mathrm{B}\eta_i &= -\frac{e}{2 \cw \sw} \left(\underline \rho_i + v_i\right)
  u_Z + \frac{e}{2\sw} \left( \underline \phi_i^+ u_W^- + \underline \phi_i^- u_W^+ \right)
  \label{eq:brsthiggsdoublet}
\end{align}
The transformations for $\delta_\mathrm{B} \rho_i$ and $\delta_\mathrm{B}\eta_i$
are fixed by taking the real and imaginary part of the BRST transformation of
the lower doublet component, respectively. In this way, if the ghost term is
formulated directly in the physical basis, as it is done in
\citere{Denner:1994xt}, the Lagrangian is guaranteed to be hermitian.


\subsection{Renormalization in the BFM\label{sec:renobfm}}
The renormalization in the \BFM{} is performed in the same fashion as in the
conventional formulation, except that only background fields need to be
renormalized.  \Reptil{} can distinguish between both types of fields by
checking the field-type attribute. A field can be assigned to be a background
and/or quantum field. In the conventional formalism, all fields play both roles
and can thus appear in tree and loop amplitudes. In the presence of pure quantum
fields, as it is the case in the \BFM, the only contributing Feynman rules to
tree and one-loop amplitudes are the ones with exactly none or two quantum
fields. 
 
Since \change{we aim at}
the computation of \SMatrix{} elements, an on-shell
renormalization of physical fields is suited. \change{However},
fixing the field
renormalization constants via on-shell conditions breaks background-field gauge invariance and,
as a consequence, some Ward identities are not fulfilled.
\change{The reason is that demanding background-field gauge invariance
requires, in particular, a uniform renormalization of all covariant derivatives in the
theory which is only possible if the field renormalization constants
of gauge fields are not independent parameters but chosen accordingly
\cite{Denner:1994xt}.
Since the theory is governed by BRST invariance, the breaking of the
background-field Ward identities does not pose a problem, especially not for  
the renormalizability of the theory and the gauge independence of observables.}
Yet, we do not break the QED
\change{background-field}
Ward identity, which relates the
fermion--fermion--photon vertex to fermionic self-energies \cite{Denner:1994xt}
\begin{align}
  k_\mu \Gamma_\mu^{\gamma \bar f f}(k,p,p^\prime) = 
  -e Q_f \left[\Gamma^{\bar f f}\left(p\right) - \Gamma^{\bar f
  f}\left(p^\prime\right)\right],
  \label{eq:qedwi}
\end{align}
and
can be used to fix the photon field renormalization constant or the
counterterm $\delta Z_e$. Requiring \eqref{eq:qedwi}
for renormalized vertex functions
yields the well-known
one-loop relation in the \BFM{}
\begin{align}
  \delta Z_e = -\frac{\delta Z_A}{2},
  \label{eq:qedwict}
\end{align}
which is consistent with the renormalization of $\alpha$ in the TL and the
photon renormalized on-shell. For LHC phenomenology the TL does not provide an
appropriate renormalization due to the difference of the scale in the underlying
processes.  A popular choice is the \Gf{} scheme
\cite{Sirlin:1980nh,Marciano:1980pb,Sirlin:1981yz} which can be defined via the
muon decay. To this end, the renormalized electric charge is related to the
experimentally measured value for the Fermi constant \Gf. At one-loop order,
neglecting pure QED corrections, finite corrections to the renormalization in
the TL are defined by
$\Delta r$ 
\begin{align}
  \delta Z_e^{\Gf} = \delta Z_e^{\mathrm{TL}} - \frac{\Delta r}{2},
\end{align}
\change{where in the \SM}
\cite{Denner:1991}
\begin{align}
  \Delta r &= \frac{\gwwt\left(0\right)- \gwwt\left(\mmw^2\right)}{\mmw^2} + \frac{2}{\cw\sw}
  \frac{\gazt(0)}{\mmz^2} + \frac{2 \delta \g}{\g} \notag\\
  &+ \frac{\g^2}{16 \pi^2}
  \left[\frac{\log\left(\cw^2\right)}{\sw^2} \left(\frac{7}{2}-2 \sw^2\right)+6\right],
  \label{eq:drFG}
\end{align}
\change{and $\Sigma^{\mathrm{1PI},\mathrm{T}}$ being an unrenormalized transverse 1PI
mixing or self-energy.
Note that all terms, except for the \PW{} self-energy, originate from vertex and
box corrections, in particular, the term $\gazt$ has just been introduced to
match the divergence structure. Equation \eqref{eq:drFG} is valid for the
conventional formulation in the 't~Hooft--Feynman gauge, but not in the \BFM{}
since mixing and self-energies, or, in general, vertex functions differ by
gauge-dependent terms in both formulations.
Since the parameter $\Delta r$ connects physical quantities it is necessarily
gauge independent, which implies that both formulations differ merely by a
reshuffling of gauge-dependent terms between the self-energy and vertex parts.}
We have determined the difference in the vertex corrections between the \BFM{}
and conventional formulation in the 't~Hooft--Feynman gauge, and, as expected,
it cancels against the difference in the \PW~self-energy.  For a
model-independent evaluation in the \BFM, the result can be expressed in the
same form as \eqref{eq:drFG}, but with a modified vertex
correction\footnote{Note that \change{$\gaztbfm(0)$}
is zero in the \BFM{} due to a Ward
identity.}
\begin{align}
  \Delta r &= 
  \frac{\gwwtbfm(0)- \gwwtbfm\left(\mmw^2\right)}{\mmw^2} + \frac{2}{\cw\sw}
  \frac{\gaztbfm(0)}{\mmz^2} + \frac{2 \delta \g}{\g} \notag \\ 
   &+ \frac{\g^2}{16 \pi^2}
  \left[\frac{\log\left(\cw^2\right)}{\sw^2} \left(-\frac{1}{2}+2 \sw^2\right)+2\right],
  \label{eq:drgdbfm}
\end{align}
which is valid only in the 't~Hooft--Feynman gauge in the \BFM.

Another subtlety concerns the renormalization within the \CMS.
\Reptil{} automatically renormalizes unstable particles in the \CMS{}
following the general prescription of
\citeres{Denner:1999gp,Denner:2005fg,Denner:2006ic}. The corresponding
on-shell
renormalization conditions require scalar integrals to be analytically
continued to complex squared momenta. This can be avoided by using an
\change{expansion around} real momentum arguments,\footnote{The expansion breaks
  down for IR-singular contributions resulting from virtual gluons or
  photons.  This can be corrected by including additional terms (see
  \citere{Denner:2005fg}) which is automatically handled in \Reptil.}
which gives rise to gauge-dependent terms of higher perturbative
orders. Thus, comparing the \BFM{} to the conventional formalism leads
to somewhat different results for finite widths. The effect can be
traced back to the difference of full self-energies in both
formulations, \eg the difference in the \PW{} self-energy is given by
\begin{align}
  \Sigma_{WW}^{\mu \nu,\mathrm{BFM}}(p)-\Sigma_{WW}^{\mu \nu}(p) =
  \frac{\g^2}{4 \pi^2} \left(\mmw^2-p^2\right) g^{\mu
  \nu}  \left[ \cw^2 B_0\left(p^2,\mmz,\mmw\right) + \sw^2
  B_0\left(p^2,0,\mmw\right)\right],
  \label{eq:cmsgaugedep}
\end{align}
with the conventions for scalar integrals as in \citere{Denner:1991}.
The gauge dependence drops out in the mass renormalization constant,
\ie ${\dmmw^{2}}^{\mathrm{BFM}} = \dmmw^2$ in the \CMS, because the
self-energy is evaluated on the complex pole, \ie for $p^2=\mmw^2$.
However, performing an expansion of the self-energy \change{around} the real
mass $\mw^2$ results in differences of the order of
$\mathcal{O}\left(\alpha^3\right)$. For a comparison of both
formulations it is useful to modify the expanded (exp) mass
counterterm to match the conventional
formalism in the following way
\begin{align}
  {\dmmw^{2}}^{\mathrm{BFM}}_{,\mathrm{exp}}& \to
  {\dmmw^2}_{,\mathrm{exp}} = \nonumber\\
  &= {\dmmw^{2}}^{\mathrm{BFM}}_{,\mathrm{exp}} - \frac{\g^2}{4 \pi^2}
  \left(\mw^2-\mmw^2\right)^2 \left[ \cw^2 B_0^\prime\left(\mw^2,\mmz,\mmw\right) +
  \sw^2 B_0^\prime \left(\mw^2,0,\mmw\right)\right],
  \label{eq:cmsbfmcorr}
\end{align}
with $B_0^\prime$ being defined as the derivative of $B_0$ with respect to
$p^2$. Note that the difference is of order $\mathcal{O}(\alpha^3)$ and
phenomenologically irrelevant.

\change{The renormalization of the tadpoles in the \BFM{} is performed analogously to the
conventional formulation.  From a theoretical point of view the renormalization of
tadpoles is not necessary, and the theory is well-defined just by including
tadpole graphs everywhere.  However, in practical calculations it is desirable
to avoid unnecessary computations of graphs with explicit tadpoles if their
contribution can be included indirectly by other means, \eg via a suited
renormalization.  The renormalization of the tadpoles has to be done with care
because a naive  treatment of the tadpole counterterms can lead to spurious
dependencies on the gauge-fixing procedure which ultimately spoil the
gauge independence of the one-loop part of \SMatrix{} elements.}
From the point of view of applicability, automation and gauge
independence, we strongly recommend the \ts{},\footnote{See original \SM{}
formulation in \citere{Fleischer:1980ub} and explicit formulas for the \THDM{}
in \citere{Krause:2016oke,Denner:2016etu}} which has been automated for arbitrary
theories \cite{Denner:2016etu}.
\change{In contrast to other schemes, the \ts{} is
purely based on the field reparametrization invariance of quantum
field theory (see \citere{Denner:2016etu}), which can be
shown to be equivalent to not renormalizing the tadpoles at all, but with the
benefit of not having to compute graphs with explicit tadpoles.}
Under the general assumption that the theory under consideration is expressed in
the physical basis without tree-level mixings and restricting to the one-loop
case, the \ts{} is equivalent to the field redefinition
\begin{align}
  \hat H_i \to \hat H_i - \frac{\delta t_{\hat H_i}}{m_{\hat H_i}^2},
\end{align}
for every physical (background-)field $\hat H_i$ that develops a vev and with
$\delta t_{\hat H_i}$ being the associated tadpole counterterm. 
By fixing
$\delta t_{\hat H_i}$ to the tadpole graphs $T_{\hat H_i}$ via
\begin{align}
 \delta t_{\hat H_i}  = - T_{\hat H_i},
  \label{eq:tadpolecond}
\end{align}
explicit tadpoles are cancelled and only tadpole counterterms to 1PI graphs
remain. \Reptil{} can automatically derive all tadpole counterterms in the \ts.
\change{In the \ts{} the value of each counterterm needs to be independent
of $\delta t_{\hat H_i}$ which can be verified analytically.\footnote{\change{The
invariance follows from Eq.~(2.26) in \citere{Denner:2016etu}.}}
Additional checks concerning the tadpole renormalization can be performed on a
process-by-process basis by including the tadpole graphs explicitly
instead of renormalizing them.
Finally, we note that \RecolaTwo{} is able to use
any tadpole counterterm scheme, but only the \ts{} is fully automated.}
%
%
\section{Setup and benchmark points\label{sec:setup}}
\subsection{Input parameters}
For the numerical analysis in the two Higgs-boson production
processes we use the following values for the \SM{} input
parameters~\cite{Olive:2016xmw}:
\begin{align}
  \Gf=1.16638\cdot10^{-5}\GeV^{-2},\quad
  \mt=173.21\GeV, \quad
  \mh=125.09\GeV, \notag \\
  \mw=80.385\GeV, \quad \Gamma_{\PW}=2.085\GeV, \quad
  \mz=91.1876\GeV, \quad \Gamma_{\PZ}=2.4952\GeV.
\end{align}
For the \THDM{} we present updated and new results for the benchmark points in
\reftas{tab:BPTHDM} and \ref{tab:BPTHDMNA} as proposed by the HXSWG
\cite{deFlorian:2016spz}.
\change{The corresponding Higgs self-couplings $\lambda_i$ are given for convenience in
\reftas{tab:BPTHDMlambda} and \ref{tab:BPTHDMNAlambda}}.
For
the \HS{} we compiled a list of benchmark points in \refta{tab:HSBP} featuring
different hierarchies and being compatible with the limits given in
\citeres{Robens:2015gla,Robens:2016xkb}.\footnote{Our conventions differ from
those of \citere{Robens:2015gla}. We identify $\cas, \tb$ in
\citere{Robens:2015gla} with $-\sas, 1/\tb$ in our conventions.}  The results
include the \SM-like and heavy Higgs-boson production for both models.  The
computations were carried out in the 't~Hooft--Feynman gauge both in the
conventional formalism and in the \BFM{}. In case of the \THDM{} the matrix
elements have undergone additional tests. Most notably, we have compared results
obtained with \RecolaTwo{} for Higgs decays into four fermions, which is closely
related to the considered processes, to an independent
calculation\cite{Altenkamp:2017ldc} based on \FeynArts/\FormCalc
\cite{Hahn:2000kx,Gross:2014ola} for various channels. 
We found agreement to more than 7 digits for 3348 out of 3500
phase-space points in the virtual amplitude, none differing by more than 5
digits.
\change{We compared off-shell two-point functions for all distinct external
states, \ie scalars, fermions, and vector bosons, against an independent
approach in \QGraf{} \cite{Nogueira:1991ex} and \QGS, which is an extension of
\GraphShot \cite{Actis}.  Against the same setup we compared Higgs decays into
scalars, fermions and vector bosons on amplitude level.  In addition, we
verified (on-shell) Slavnov--Taylor identities for two-point functions (see
Eq.~(4.16) and the following in \citere{Altenkamp:2017ldc}).}

\subsection{Cut setup}
For the analysis of Higgs strahlung we consider the case of two charged
muons in the final state, $\Pp\Pp \to \PH \Pm^+ \Pm^- + X$.
The muons are not recombined with collinear photons, and are assumed to be
perfectly isolated, treated as bare muons as described in \citere{Denner:2011id}.
We use the cuts given in \citere{Chatrchyan:2013zna}, \ie we demand the muons
to
\begin{itemize}
  \item have transverse momentum $p_{\mathrm{T},l}> 20\GeV$ for $l=\Pm^+, \Pm^-$,
  \item be central with rapidity $\left|y_l\right| < 2.4$ for $l=\Pm^+, \Pm^-$,
  \item have 
a pair invariant mass 
$m_{\mu\mu}$ of  $75\GeV < m_{\Pm\Pm} < 105\GeV$.
\end{itemize}
Further, we select boosted events with a
\begin{itemize}
  \item transverse momentum  $p_{\mathrm{T},\Pm\Pm} > 160\GeV$.
\end{itemize}
For VBF we employ the cuts as suggested by the HXSWG in
\citere{deFlorian:2016spz}, \ie we require two hard jets $\mathrm{j}_i, i=1,2$, emerging
from partons $i$ with
\begin{itemize}
  \item pseudo-rapidity $\left|\eta_i\right|< 5$.
\end{itemize}
The recombination is done in the anti-$k_T$ algorithm \cite{Cacciari:2008gp} with
jet size $D=0.4$. Further, events pass the cuts if the two hard jets have
\begin{itemize}
  \item a transverse momentum  $p_{\mathrm{T}, \mathrm{j}_i}> 19\GeV$ each,
  \item a rapidity  $\left|y_{\mathrm{j}_i}\right|< 5$ each,
  \item a rapidity difference  $\left|y_{\mathrm{j}_1} - y_{\mathrm{j}_2}\right|> 3$,
  \item an invariant mass $M_{\mathrm{j}_1 \mathrm{j}_2}> 130 \GeV$.
\end{itemize}
We present the results for hadronic cross sections at the centre-of-mass energy
of 13 TeV using the NLO PDF set NNPDF2.3 with QED corrections
\cite{Ball:2013hta}. 
\begin{table}
  \centering
  \begin{tabular}{|c|c|c|c|c|c||c|}
  \hline
  \mydstrut    & $\mhh$    & $\mha$    & $\mhc$    & $m_{12}$    & $\tb$ & $\Msb$\\
  \hline BP21A & $200\GeV$ & $500\GeV$ & $200\GeV$ & $135\GeV$   & $1.5$ & $198.7\GeV$ \\
  \hline BP21B & $200\GeV$ & $500\GeV$ & $500\GeV$ & $135\GeV$   & $1.5$ & $198.7\GeV$  \\
  \hline BP21C & $400\GeV$ & $225\GeV$ & $225\GeV$ & $0\GeV$     & $1.5$ & $0\GeV$  \\
  \hline BP21D & $400\GeV$ & $100\GeV$ & $400\GeV$ & $0\GeV$     & $1.5$ & $0\GeV$ \\
  \hline BP3A1 & $180\GeV$ & $420\GeV$ & $420\GeV$ & $70.71\GeV$ & $3$   & $129.1\GeV$\\
  \hline
  \end{tabular}
  \caption{\THDM{} benchmark points in the alignment
    limit, \ie $\sab\to-1$, $\cab\to0$, taken from
    \citere{HXSWG2016}. The parameter $\Msb$
    depends on the other parameters and is given for
    convenience. \label{tab:BPTHDM}}
\end{table}
\begin{table}
  \change{
  \centering
  \begin{tabular}{|c|c|c|c|c|c|c|}
  \hline
  \mydstrut    & $\lambda_1$    & $\lambda_2$    & $\lambda_3$    & $\lambda_4$ & $\lambda_5$\\
  \hline BP21A & $0.28$ & $0.26$ & $0.27$  & $3.45$   & $-3.47$  \\
  \hline BP21B & $0.28$ & $0.26$ & $7.19$  & $-3.47$  & $-3.47$  \\
  \hline BP21C & $6.19$ & $1.43$ & $-0.71$ & $-0.83$  & $-0.83$  \\
  \hline BP21D & $6.19$ & $1.43$ & $2.89$  & $-5.11$  & $-0.16$  \\
  \hline BP3A1 & $2.59$ & $0.29$ & $5.26$  & $-2.63$  & $-2.63$  \\
  \hline
  \end{tabular}
  \caption{\change{Higgs self-couplings for the \THDM{} benchmark points in the alignment
    limit. We omit the imaginary parts appearing in the \CMS.} \label{tab:BPTHDMlambda}}}
\end{table}
\begin{table}
\begin{tabular}{|c|c|c|c|c|c|c||c|}
\hline
  \mydstrut  & $\mhh$    & $\mha$    & $\mhc$    & $m_{12}$    & $\tb$  & $\cab$ & $\Msb$\\
\hline a-1 & $700\GeV$ & $700\GeV$ & $670\GeV$ & $424.3\GeV$ & $1.5$  & $-0.0910$ & $624.5\GeV$\\
\hline b-1 & $200\GeV$ & $383\GeV$ & $383\GeV$ & $100\GeV$   & $2.52$ & $-0.0346$ & $204.2\GeV$\\
\hline BP22A & $500\GeV$ & $500\GeV$ & $500\GeV$ & $187.08\GeV$   & $7$ & $0.28 $ & $500\GeV$\\
\hline BP3B1 & $200\GeV$ & $420\GeV$ & $420\GeV$ & $77.78\GeV$   & $3$ & $0.3 $
  & $142.0\GeV$\\
\hline BP3B2 & $200\GeV$ & $420\GeV$ & $420\GeV$ & $77.78\GeV$   & $3$ & $0.5 $
  & $142.0\GeV$\\
\hline BP43 & $263.7\GeV$ & $6.3\GeV$ & $308.3\GeV$ & $52.32\GeV$   & $1.9$ & $0.14107 $ & $81.5\GeV$\\
  \hline BP44 & $227.1\GeV$ & $24.7\GeV$  & $226.8\GeV$ & $58.37\GeV$ & $1.8$ &
  $0.14107$ & $89.6\GeV$\\
\hline BP45 & $210.2\GeV$ & $63.06\GeV$ & $333.5\GeV$ & $69.2\GeV$   & $2.4$ &
  $0.71414 $ & $116.2\GeV$\\
\hline
\end{tabular}
\caption{\THDM{} benchmark points outside the alignment limit taken from \citere{Baglio:2014nea} (a-1, b-1)
  and \citere{HXSWG2016}. The parameter $\Msb$ depends on the other parameters
  and is given for convenience. \label{tab:BPTHDMNA}}
\end{table}
\begin{table}
  \change{
  \centering
  \begin{tabular}{|c|c|c|c|c|c|c|}
  \hline
  \mydstrut    & $\lambda_1$    & $\lambda_2$    & $\lambda_3$    & $\lambda_4$ & $\lambda_5$\\
  \hline a-1 & $1.76$ & $1.97$ & $0.09$  & $-0.29$   & $-1.65$  \\
  \hline b-1 & $0.01$ & $0.26$ & $3.72$  & $-1.73$ & $-1.73$  \\
  \hline BP22A & $0.26$ & $0.26$ & $7.98$  & $0.$ & $0.$  \\
  \hline BP3B1 & $3.60$ & $0.25$ & $5.46$  & $-2.57$ & $-2.57$  \\
  \hline BP3B2 & $3.44$ & $0.27$ & $5.74$  & $-2.57$ & $-2.57$  \\
  \hline BP43 & $4.42$ & $0.43$ & $2.34$  & $-3.02$ & $0.11$  \\
  \hline BP44 & $2.85$ & $0.40$ & $1.10$  & $-1.55$ & $0.12$  \\
  \hline BP45 & $3.16$ & $0.35$ & $3.92$  & $-3.38$ & $0.16$  \\
  \hline
  \end{tabular}
  \caption{\change{Higgs self-couplings for the \THDM{} benchmark points outside the alignment
    limit. We omit the imaginary parts appearing in the \CMS.}
    \label{tab:BPTHDMNAlambda}}}
\end{table}
\begin{table}[tp]
  \centering
  \begin{tabular}{|c|c|c|c||c|c|c|}
    \hline
    \mydstrut     & $\mhh/ \GeV$    & $\tb$ & $\sas$ & \change{$\lambda_1$} & \change{$\lambda_2$}
    & \change{$\lambda_3$} \\
    \hline BP1  & $500$ & $2.2$ & $-0.979796$ & $0.41$ & $0.82$ & $-0.34$ \\
    \hline BP2  & $400$ & $1.7$ & $-0.96286$ & $0.43$ & $0.85$ & $-0.36$  \\
    \hline BP3  & $300$ & $1.3$ & $-0.950737$ & $0.38$ & $0.81$ & $-0.28$ \\
    \hline BP4  & $200$ & $0.85$ & $-0.932952$ & $0.31$ & $0.84$ & $-0.16$\\
    \hline
    \hline
  \mydstrut   & $\mhl/\GeV$     & $\tb$ & $\sas$ & $\lambda_1$ & $\lambda_2$
    & $\lambda_3$ \\
    \hline BP5  & $100$ & $0.35$  & $-0.35$ & $0.25$ & $1.44$ & $-0.09$\\
    \hline BP6  & $50$  & $0.2$ & $-0.06$ & $0.26$ & $1.05$ & $-0.06$\\
  \hline
  \end{tabular}
  \caption{\HS{} benchmark points compiled from
    \citere{Bojarski:2015kra}. In
  the upper table typical scenarios are depicted with a heavy additional scalar Higgs
  boson. In the lower table inverted scenarios are listed with $\Hh$ identified as the \SM{}
  Higgs boson and mass $\mhh=125.09\GeV$.
  \change{The Higgs self-couplings $\lambda_i$ depend on the other parameters
  and their real parts are given for convenience.}
  \label{tab:HSBP}}
\end{table}
%


\change{\subsection{Mixing angles at one-loop order\label{sec:renormschemes}}}

The prime vertices of interest in the processes studied in
\refse{sec:numresults} are the $\Hl \PV \PV$ and $\Hh \PV \PV$ vertices. Thus,
the relevant one-loop corrections require to renormalize $\alpha$ and $\beta$ in
the \THDM{} and $\alpha$, but not $\beta$, in the \HS.  We present the
counterterms for the mixing angles in an \msbar{} scheme and two
different on-shell schemes
in the following:
\begin{description}
  \item[\msbar:] The mixing angles $\alpha$, $\beta$ are renormalized using \msbar{}
    subtraction\cite{Denner:2016etu} for the vertices $\Hl \to \tau^+ \tau^-$,
    $\Ha \to \tau^+ \tau^-$, respectively, with $\beta$ only being renormalized in
    the \THDM. This is equivalent to using the identities
    \begin{align}
      \delta \alpha &= \frac{\dzhhhl^{\msbar}-\dzhlhh^{\msbar}}{4}
      =\frac{\Sigma^{\mathrm{1PI},\msbar}_{\Hh \Hl}\left(\mhh^2\right)+
             \Sigma^{\mathrm{1PI},\msbar}_{\Hh \Hl}\left(\mhl^2\right)+
      2 \thlhh^{\msbar}}{2\left(\mhh^2-\mhl^2\right)}
      , \notag\\
      \quad \delta \beta &= \frac{\dzgha^{\msbar}-\dzhag^{\msbar}}{4}
      =-\frac{\Sigma^{\mathrm{1PI},\msbar}_{\Ha \GZ}\left(0\right)+
             \Sigma^{\mathrm{1PI},\msbar}_{\Ha \GZ}\left(\mha^2\right)+
      2\thag^{\msbar} }{2\mha^2}
      \label{eq:msbarprescription}
    \end{align}
    with the relation for $\delta \alpha$ being valid in the \THDM{} and the
    \HS{} and the one for $\delta \beta$ only in the \THDM.  The origin of these
    relations can be traced back to the renormalizability of models in a
    minimal (symmetric) renormalization scheme. See \citere{Altenkamp:2017ldc}
    for the derivation of these and other UV-pole-part identities. 
    The tadpole counterterms in \eqref{eq:msbarprescription} are treated in the
    \ts{} (see Apps.~A and B in \citere{Denner:2016etu}) and using the
    renormalization condition \eqref{eq:tadpolecond} for tadpoles.
    \change{Estimating the size of higher-order contributions via the usual scale
    variations has been improved via a partial resummation including the
    renormalization-group (RG) running of parameters.\footnote{In
    \citere{Altenkamp:2017ldc} the running of  the mixing angles is investigated 
    within various \msbar{} and tadpole counterterm schemes in the \THDM.}}
    For the \THDM{} this
    requires to solve a coupled system of differential equations, 
    \begin{align}
      \frac{\partial}{\partial \log \mu^2} \alpha(\mu) &=
      f_{\alpha}(\alpha(\mu),\beta(\mu),\Msb(\mu)), \notag\\
      \frac{\partial}{\partial \log \mu^2} \beta(\mu) &=
      f_{\beta}(\alpha(\mu),\beta(\mu),\Msb(\mu)), \notag\\
      \frac{\partial}{\partial \log \mu^2} \Msb(\mu) &=
      f_{\Msb}(\alpha(\mu),\beta(\mu),\Msb(\mu)).
    \end{align}
    The functions $f_{\alpha},f_{\beta}$ and $f_{\Msb}$ can be directly read off
    the pole parts of the corresponding counterterms. The counterterm
    $\delta \Msb$ was fixed from
    the vertex $\Hh \to \Hp \Hm$ in the \msbar{} scheme. Note that
    $\delta \Msb$ does not enter the considered processes at fixed one-loop
    order. For the \HS{} we keep $\beta$ fixed, assuming no scale dependence,
    resulting in a simple differential equation for $\alpha$,
    \begin{align}
      \frac{\partial}{\partial \log \mu^2} \alpha(\mu) = 
      f_{\alpha}(\alpha(\mu)).
    \end{align}
    The (coupled) system has been 
    solved to run the parameters from the
    reference scale $\mu_0$ to $\mu=\mu_0/2$ and $\mu=2 \mu_0$. The
    results are presented in  \reftas{tab:BPrunning} and
    \ref{tab:BPrunningHS} for the 
    benchmark points of \reftas{tab:BPTHDM},  \ref{tab:BPTHDMNA}, and \ref{tab:HSBP}
    being defined at the typical scale of the process, $\mu_0=2\mhl$ if not
    stated otherwise.\footnote{Note that the running of parameters is
    independent of the scale at which they are defined.}
    The cross sections are evaluated at the scales 
    $\mu_0/2, \mu_0, 2 \mu_0$, 
    using the running
    parameters of $\cab$, $\tb$, $\Msb$ ($\sas$) at the corresponding scale as
    input parameters in the \THDM{} (\HS).
\begin{table}
  \centering
  \begin{tabular}{|c|c|c|c||c|c|c|}
    \hline
    BP  & $\tb(\mu_0/2)$ & $\cab(\mu_0/2)$ & $\Msb(\mu_0/2)/\GeV$ & $\tb(2\mu_0)$ & $\cab(2
    \mu_0)$ & $\Msb(2\mu_0)/\GeV$\\
  \hline
  \hline
  \multicolumn{1}{|c|}{BP21A}   & $1.41$ & $-0.1166$  & $192.3$
                                & $1.54$ & $0.0504$   & $197.7$ \\
  \hline
  \multicolumn{1}{|c|}{BP21B}   & $1.16$ & $-0.4163$  & $199.7$
                                & $1.51$ & $0.0293$   & $191.2$ \\
  \hline
  \multicolumn{1}{|c|}{BP21C}   & $1.40$ & $-0.0029$  & $0.0$
                                & $1.64$ & $0.0067$   & $0.0$\\
  \hline
  \multicolumn{1}{|c|}{BP21D}   & $1.37$ & $-0.0017$  & $0.0$ 
                                & $1.68$ & $-0.0119$  & $0.0$\\
  \hline
    \multicolumn{1}{|c|}{\change{BP3A1}}   & $2.34$ & $-0.0681$  & $121.6$ 
                                & $3.53$ & $0.1701$  & $133.8$\\
  \hline
  \hline
  \multicolumn{1}{|c|}{a1} & $0.86$ & $-0.3801$ & $614.1$ 
                           & $1.78$ & $-0.0202$ & $638.5$\\
  \hline
  \multicolumn{1}{|c|}{b1} & $2.36$ & $-0.1542$ & $203.6$
                           & $2.59$ & $ 0.0116$ & $203.3$\\
  \hline
  \multicolumn{1}{|c|}{BP22A} & $-$ & $-$ & $-$ 
                              & $1.52$ & $0.6538$ & $274.5$ \\
  \hline
  \multicolumn{1}{|c|}{BP3B1} & $3.15$ & $ 0.1292$ & $149.3$
                              & $2.24$ & $ 0.5972$ & $123.8$\\
  \hline
  \multicolumn{1}{|c|}{BP3B2} & $4.17$ & $ 0.2992$ & $167.9$
                              & $1.99$ & $ 0.7809$ & $119.3$ \\
  \hline
  \multicolumn{1}{|c|}{BP43}  & $1.76$ & $ 0.0997$ & $80.7$
                              & $2.08$ & $ 0.1906$ & $82.8$ \\
  \hline
  \multicolumn{1}{|c|}{BP44}  & $1.66$ & $ 0.1313$ & $88.1$ 
                              & $1.97$ & $ 0.1511$ & $91.5$ \\
  \hline
  \multicolumn{1}{|c|}{BP45}  & $2.29$ & $ 0.6504$ & $115.1$
                              & $2.53$ & $ 0.7666$ & $117.5$ \\
  \hline
  \end{tabular}
  \caption{Running values for $\tb$, $\cab$ and $\Msb$ in the \THDM{} at the
  scales $\mu_0/2$ and $2 \mu_0$. The benchmark points are defined at the central
  scale $\mu_0$ in \reftas{tab:BPTHDM} and \ref{tab:BPTHDMNA}. The results for
  the alignment-limit scenarios are in the upper part of the table whereas the
  non-alignment scenarios are in the lower part. For BP22A the running $\beta$
  reaches $\pi/2$ for a scale greater than $\mu_0/2$, thus, $\tb$ becomes
  singular.
  \change{In this particular scenario
  the steep running is caused by the Higgs self-coupling
  $\lambda_3\approx 8$ (see \refta{tab:BPTHDMNAlambda}) and can be stabilized
  by reducing its value. The running becomes stable only for values smaller than
  $\lambda_3 \lesssim 0.5$. Changing $\lambda_3$ to $0.5$ and keeping the
  values for all other $\lambda_i$ fixed has a small effect on $\mhl$ and
  $\mhh$ of the order $\mathcal{O}\left(5 \GeV\right)$, but brings the scenario
  close to the alignment limit $\cab \approx 0$.}
  \label{tab:BPrunning}}
\end{table}
\begin{table}
  \centering
  \begin{tabular}{|c|c||c|}
    \hline
  \mydstrut BP  & $\sas(\mu_0/2)$ & $\sas(2\mu_0)$ \\
    \hline
    \hline
    \multicolumn{1}{|c|}{BP1}     & $-0.9802$ & $-0.9794$  \\
    \hline
    \multicolumn{1}{|c|}{BP2}     & $-0.9646$ & $-0.9612$  \\
    \hline
    \multicolumn{1}{|c|}{BP3}     & $-0.9557$ & $-0.9455$  \\
    \hline
    \multicolumn{1}{|c|}{BP4}     & $-0.9367$ & $-0.9293$  \\
    \hline
    \multicolumn{1}{|c|}{BP5}     & $-0.2780$ & $-0.4463$  \\
    \hline
    \multicolumn{1}{|c|}{BP6}     & $-0.04647$ & $-0.08194$  \\
  \hline
  \end{tabular}
  \caption{Running values for $\sas$ in the \HS{}
  at the scales $\mu_0/2$ and $2 \mu_0$. The
  benchmark points are defined at the central scale
  $\mu_0$ in \refta{tab:HSBP}.\label{tab:BPrunningHS}}
\end{table}
    The three different predictions for $\snloew$ normalized to the
    leading-order cross section $\slo(\mu_0)$ at the central scale $\mu_0$ and
    scale-dependent relative \EW{} corrections are defined as
    \begin{align}
      \dew\left(\mu,\mu_0\right) &:=
      \frac{\snloew\left(\mu\right) - \slo\left(\mu_0\right)}{\slo(\mu_0)},
    \end{align}
    such that
    \begin{align}
      \snlo(\mu) = \Bigl(1+ \dew\left(\mu,\mu_0\right)\Bigr) \slo\left(\mu_0\right).
    \end{align}
    Note that the tree-level matrix elements only depend on the scale through
    the running of parameters, whereas the one-loop matrix elements have an
    explicit scale dependence. As a shorthand notation for the relative
    corrections in the \msbar{} scheme we use 
    \begin{align}
      \dewm &:= \dew(\mu_0, \mu_0)^{u}_{d}, \notag \\
      u &:= \dew(2 \mu_0,\mu_0)-\dew(\mu_0,\mu_0), \notag \\
      d &:= \dew\left(\frac{\mu_0}{2},\mu_0\right)-\dew(\mu_0,\mu_0)
      \label{eq:runningmsbar}
    \end{align}
    with $u$ and $d$ being
    \change{the upper and lower edges of the scale variation (see
    \eg \refta{tab:BPHlHZthdm}).}
  \item[$p^*$:]
    The renormalized mixing angles $\alpha$ and $\beta$ are defined to
    diagonalize radiatively corrected mass matrices which implies a scale and
    momentum dependence for the mixing angles. The scale dependence can
    be eliminated by a special choice for the momentum $p^2=\left(p^*\right)^2$ at
    which the mixing two-point functions, and thus the running mixing angles,
    are evaluated.
    The original idea goes back to
    \citere{Espinosa:2001xu} (see also \citere{Espinosa:2002cd}) and has been
    applied to the \HS{} in \citere{Bojarski:2015kra} and the \THDM{} in
    \citere{Krause:2016oke}. In our conventions, the counterterms are defined as
    \begin{align}
      \delta \alpha &= \frac{\Sigma^{\mathrm{1PI},\mathrm{BFM}}_{\Hh
      \Hl}\left(\frac{\mhh^2+\mhl^2}{2}\right)+\thlhh}{\mhh^2-\mhl^2}, \qquad
      \delta \beta = -\frac{\Sigma^{\mathrm{1PI},\mathrm{BFM}}_{\Ha
      \GZ}\left(\frac{\mha^2}{2}\right)+\thag}{\mha^2}.
      \label{eq:pstarprescription}
    \end{align}
    Note that for $\delta \beta$ alternatively the mixing energy with the
    charged Higgs and Goldstone boson can be used.  As the mixing energies
    are gauge-dependent an additional intrinsic prescription is required
    to fix the gauge-independent parts. We choose the \BFM{} with quantum
    gauge parameter $\xi_Q=1$, which corresponds to the gauge-fixing
    functions \eqref{eq:bfmginvgf}, \eqref{eq:bfmginvgfcomponents}. We
    remark that this is
    equivalent \cite{Denner:1994nn, Binosi:2004qe} to the
    self-energy in the pinch technique \cite{Cornwall:1981zr,Cornwall:1989gv}
    \change{and allows to extract a well-defined gauge-parameter-independent
    contribution to self-energies or, in general, vertex functions and
    hence counterterms in this scheme.}
  \item[BFM:] As an on-shell alternative to the $p^*$ scheme, the authors of
    \citere{Krause:2016oke} propose to use the on-shell scalar mixing energies
    defined within the pinch technique which has also been investigated
    in \citere{Kanemura:2017wtm}. In our framework, this corresponds to
     \begin{align}
       \delta \alpha &= \frac{\dzhhhl^{\mathrm{BFM}}-\dzhlhh^{\mathrm{BFM}}}{4}
       = \frac{\Sigma^{\mathrm{1PI},\mathrm{BFM}}_{\Hh \Hl}\left(\mhh^2\right)+
       \Sigma^{\mathrm{1PI},\mathrm{BFM}}_{\Hl \Hh}\left(\mhl^2\right)  + 2
       \thlhh}{2\left(\mhh^2-\mhl^2\right)},\\ \quad \delta \beta &=
       \frac{\dzgha^{\mathrm{BFM}}-\dzhag^{\mathrm{BFM}}}{4} =
       -\frac{\Sigma^{\mathrm{1PI},\mathrm{BFM}}_{\Ha\GZ}\left(0\right)+
       \Sigma^{\mathrm{1PI},\mathrm{BFM}}_{\Ha \GZ }\left(\mha^2\right) + 2
       \thag}{2 \mha^2},
       \label{eq:dadbbfm}
    \end{align}
    with the mixing energies evaluated in the \BFM{} with quantum gauge
    parameter $\xi_Q=1$.
\end{description}

In \citere{Krause:2016oke} it is argued that the use of the \ts{} is essential
for the consistency of on-shell schemes in combination with
\eqref{eq:tadpolecond}. There are, however, other options. A different tadpole
counterterm scheme, such as the one of \citere{Denner:1991}, yields different
values and pole parts for counterterms, \eg $\delta \alpha$ and $\delta \beta$
absorb tadpoles and become gauge dependent. Yet, the absorbed tadpoles drop out
in momentum subtraction schemes \cite{Denner:2016etu} and do not spoil the gauge
independence of the \SMatrix. In tadpole counterterm schemes other than the
\ts{} special care has to be devoted to the formulation of renormalization
conditions as they are necessarily gauge dependent.  This situation is
encountered in standard \SM{} and \MSSM{} on-shell renormalization schemes,
where certain tadpole contributions to self-energies are left out, rendering the
counterterms gauge dependent, but the \SMatrix{} remains gauge independent.
When employing gauge-fixing prescriptions in renormalization conditions,
tadpoles can be handled naively in a favoured tadpole counterterm scheme if the
same gauge is used in the 
renormalization and in the matrix-element evaluation.
This is illustrated in \refapp{sec:simplertadpoles} using the example of $\delta
\alpha$ in the $p^*$ scheme. There, we also discuss the general case
with arbitrary gauge-fixing functions, which is less trivial and
cannot be done in the naive way due to the mismatch of the gauge
prescription and the actual gauge-parameter choice. From a practical
point of view the latter is only relevant if one is interested in
verifying the gauge independence of the \SMatrix{} in tadpole
counterterm schemes other than the \ts.
We note that the use of \msbar{} schemes for the mixing angles in combination with  
alternative tadpole counterterm schemes can be made gauge independent by
including finite tadpole terms\footnote{See Eq.~(4.43) and the following ones in
\citere{Altenkamp:2017ldc},  Eq.~(43) in \citere{Freitas:2002um}, and Eq.~(5.24)
in \citere{Denner:2016etu}.} which is equivalent to the use of the
\ts.

The results for total cross sections in the \BFM{} renormalization scheme in
\refse{sec:numresults} were not computed directly, but were obtained
from the results in the $p^*$ scheme 
using 
the following formulas,
depending on the model (\THDM{} or \HS) and on the produced Higgs flavour (\Hl{}
or \Hh) as follows
\begin{description}
  \item[\THDM{} \Hl:]
    $\dewb = \dewps -2  \frac{\cab}{\sab}  \left(\delta \alpha^{p^*} - \delta
    \beta^{p^*} - \delta \alpha^{\mathrm{BFM}} + \delta \beta^{\mathrm{BFM}}\right)$
  \item[\THDM{} \Hh:]
    $\dewb = \dewps +2  \frac{\sab}{\cab}  \left(\delta \alpha^{p^*} - \delta
    \beta^{p^*} - \delta \alpha^{\mathrm{BFM}} + \delta \beta^{\mathrm{BFM}}\right)$
  \item[\HS{} \Hl:]
    $\dewb = \dewps -2  \frac{\cas}{\sas}  \left(\delta \alpha^{p^*} - \delta
    \alpha^{\mathrm{BFM}}\right)$
  \item[\HS{} \Hh:] $\dewb = \dewps +2  \frac{\sas}{\cas}  \left(\delta
    \alpha^{p^*} - \delta \alpha^{\mathrm{BFM}}\right)$
\end{description}
Note that the formulas can be applied uniquely to the observables under
consideration
\change{as these rely on the mixing-angle dependencies of the respective leading-order
couplings.}


\section{Numerical results for total cross sections\label{sec:numresults}}
In \refta{tab:BPHlHZthdm} we present updated results for the
production of a \SM-like Higgs boson in Higgs strahlung in the \THDM{}
in alignment scenarios.  Non-alignment scenarios are given in
\refta{tab:BPHlHZthdmNA}.
The corresponding \SM{} correction is $-12.4\%$.
In \refta{tab:BPHhHZthdm} we provide the
corresponding results for heavy Higgs-boson production in
non-alignment scenarios.  For the \HS{} all considered scenarios are
non-aligned. The results for light Higgs-boson production are given in
\refta{tab:BPHlHZHS}, and the ones for heavy Higgs-boson production in
\refta{tab:BPHSHZHS}. Note that for the benchmark points BP5 and BP6
with inverted hierarchy the heavy Higgs boson is \SM-like with
$\mhh=125.09\GeV$. For the benchmark points in the \THDM{} the light
Higgs boson is always identified as the \SM{} Higgs boson.  Finally,
in \refta{tab:BPVBFthdm} results for \SM-like and heavy Higgs-boson
production in VBF are presented for the \THDM{}. The \HS{} predictions
for VBF are given in \refta{tab:BPVBFHS}.
The corresponding \SM{} correction for \SM-like Higgs-boson production
in VBF amounts to $-5.5\%$.
\begin{table}
\begin{center}\renewcommand{\arraystretch}{1.2}
\begin{tabular}{|c|c|c|c|}
  \hline BP & $\slo^{\Hl}/\pba$ & 
  \mystrut $\dewm$ & $\dewps$     \\
  \hline
  \hline BP21A    & $1.65$  &   ${-11.8}^{+0.7}_{+2.3}\%$
  & $-11.8\%$ \\
  \hline BP21B    & $1.65$  &   ${-13.0}^{+1.2}_{-48} \%$  
    & $-13.0\%$\\
  \hline BP21C    & $1.65$  &   ${-13.2}^{-0.1}_{+0.1}\%$  
  & $-13.2\%$ \\
  \hline BP21D    & $1.65$  &   ${-13.6}^{-0.2}_{+0.1}\%$  
  & $-13.6\%$ \\
  \hline \change{BP3A1}    & $1.65$  &   ${-13.3}^{-6.4}_{+0.4}\%$  
  & $-13.3\%$ \\
  \hline
\end{tabular}
\caption{Relative NLO corrections
  \dew{} to \SM-like Higgs-boson production in Higgs strahlung $\Pp \Pp \to \Hl
  \mu^- \mu^+$ in alignment scenarios in the \THDM.  The results in
  the \msbar{} scheme are
  given at the central scale 
  $\mu_0=2\mh=250.18\GeV$ with
  scale uncertainties 
  estimated including the
  \change{RG running of parameters as given by
  \eqref{eq:runningmsbar}.}
  Both on-shell schemes agree
  within the integration error, and only results in the $p^*$~scheme are given.
  The \SM{} \EW{} correction is $\dew=-12.4\%$.
  \label{tab:BPHlHZthdm}}
\end{center}
\end{table}
\begin{table}
\begin{center}\renewcommand{\arraystretch}{1.2}
\begin{tabular}{|c|c|c|c|c|}
  \hline BP & $\slo^{\Hl}/\pba$ & 
  \mystrut $\dewm$ & $\dewps$ & $\dewb$  \\
  \hline
  \hline a-1      & $1.63$  &   ${-10.4}^{-1.6}_{+40.0}\%$ 
  & $-12.6\%$ & $-12.6\%$\\
  \hline b-1      & $1.64$  &   ${-12.9}^{+0.5}_{+2.5}\%$  
  & $-12.6\%$ & $-12.6\%$\\
  \hline BP22A    & $1.52$  &   ${-40.5}^{-}_{-}\%$        
  & $-15.9\%$ & $-15.9\%$\\
  \hline BP3B1    & $1.50$  &   ${-35.1}^{-16.3}_{+29.7}\%$
  & $-13.4\%$ & $-13.4\%$\\
  \hline BP3B2    & $1.23$  &   ${-66.6}^{-}_{-}\%$        
  & $-13.6\%$ & $-13.6\%$\\
  \hline BP43     & $1.61$  &   ${-15.0}^{-0.67}_{+1.2}\%$ 
  & $-12.6\%$ & $-12.6\%$\\
  \hline BP44     & $1.61$  &   ${-11.2}^{-}_{-3.4}\%$     
  & $-12.6\%$ & $-12.6\%$\\
  \hline BP45     & $0.806$ &   ${-31.3}^{+4.3}_{-6.7}\%$  
  & $-13.0\%$ & $-13.0\%$\\
  \hline
\end{tabular}
\end{center}
\caption{
  Relative NLO corrections \dew{} to \SM-like Higgs-boson production in Higgs
  strahlung $\Pp \Pp \to \Hl \mu^- \mu^+$ in non-alignment scenarios in the \THDM.
  \change{The results in the \msbar{} scheme are given at the central scale
  $\mu_0=2\mh=250.18\GeV$ with scale uncertainties estimated including the
  RG running of parameters as given by \eqref{eq:runningmsbar}.
  The scale uncertainties are large, and for}
  some points (BP22A,
  BP3B2, BP44) the running is unstable and yields corrections beyond $100\%$,
  which is indicated as ``$-$''.\label{tab:BPHlHZthdmNA}}
\end{table}
\begin{table}
\begin{center}
\begin{tabular}{|c|c|c|c|}
  \hline \mystrut 
   BP  & $\slo^{\Hh}/\fba$ & $\dewps$ & $\dewb$ \\
  \hline
  \hline BP22A& $6.43$  & $-12.5\%$ & $-12.9\%$\\
  \hline BP3B1& $79.4$  & $-17.5\%$ & $-17.4\%$\\
  \hline BP3B2& $220.4$ & $-16.2\%$ & $-16.1\%$\\
  \hline BP43& $10.1$   & $-2.67\%$ & $-2.74\%$\\
  \hline BP44& $13.9$   & $-8.35\%$  & $-8.39\%$\\
  \hline BP45& $411.6$  & $-13.8\%$ & $-13.8\%$\\
\hline
\end{tabular}
\end{center}
\caption{
  Relative NLO corrections \dew{} to heavy Higgs-boson $\Hh$ production in Higgs
  strahlung $\Pp \Pp \to \Hh \mu^- \mu^+$ in the \THDM{}. No results for the
  \msbar{} scheme are presented due to large scale uncertainties exceeding
  $100\%$.\label{tab:BPHhHZthdm}}
\end{table}
\begin{table}
\begin{center}\renewcommand{\arraystretch}{1.2}
\begin{tabular}{|c|c|c|c|c|}
  \hline \mystrut
  BP & $\slo^{\Hl}/\fba$ & $\dewm$ & $\dewps$ & $\dewb$ \\
  \hline
  \hline BP1 & 1580  & $-11.1 \%$                 & $-12.3 \%$ & $-12.4 \%$ \\
  \hline BP2 & 1526  & $-10.5 \%$                 & $-12.2 \%$ & $-12.3 \%$ \\
  \hline BP3 & 1486  & $-10.2 \%$                 & $-12.3 \%$ & $-12.3 \%$ \\
  \hline BP4 & 1432  & $-9.2_{-0.3}^{+0.1} \%$    & $-12.4 \%$ & $-12.4 \%$ \\
  \hline BP5 & 242.0 & $-$                        & $-11.7 \%$ & $-11.7 \%$ \\
  \hline BP6 & 9.4   & $+1.65_{-48.1}^{-11.1}\%$  & $-8.86 \%$ & $-10.4 \%$ \\
  \hline
\end{tabular}
\caption{Relative NLO corrections
  \dew{} to light Higgs-boson \Hl{} production in Higgs strahlung $\Pp \Pp \to
  \Hl \mu^- \mu^+$ in the \HS.
  \change{The scale uncertainties in the \msbar{} scheme are estimated including the
  RG running of parameters  as given by \eqref{eq:runningmsbar}.}
  The central scale for BP1--4 is $\mu_0 = 2\mh =250.18 \GeV$. For BP6 we set
  the scale to $\mu_0 = 130 \GeV$. For BP5 the \msbar{} scheme is unstable. 
  The scale uncertainties for BP1--3 are smaller than the given accuracy.
  \label{tab:BPHlHZHS}}
\end{center}
\end{table}
\begin{table}
\begin{center}\renewcommand{\arraystretch}{1.2}
\begin{tabular}{|c|c|c|c|c|}
  \hline \mystrut
  BP & $\slo^{\Hh}/\fba$ & ${\rule{0ex}{2.5ex} \dewm}$ & $\dewps$ & $\dewb$ \\
  \hline
  \hline BP1 & 3.28  & $-53.7_{+1.0}^{-0.7} \%$ & $-20.3 \%$ & $-20.5 \%$ \\
  \hline BP2 & 12.3  & $-47.4_{+1.7}^{-1.5}\%$  & $-20.0 \%$ & $-20.3 \%$ \\
  \hline BP3 & 36.0  & $-40.8_{+0.5}^{-0.4} \%$ & $-16.8 \%$ & $-16.9 \%$ \\
  \hline BP4 & 114.0 & $-36.8_{+1.3}^{-1.2}\%$  & $-16.0 \%$ & $-15.1 \%$ \\
  \hline BP5 & 1444  & $-12.6_{+ 0.0}^{+4.7} \%$  & $-12.5 \%$ & $-12.5 \%$ \\
  \hline BP6 & 1640  & $-12.3_{-0.1}^{+0.4}\%$  & $-12.5 \%$ & $-12.6 \%$ \\
  \hline
\end{tabular}
\caption{Relative NLO corrections
  \dew{} to heavy Higgs-boson \Hh{} production in Higgs strahlung $\Pp \Pp \to
  \Hh \mu^- \mu^+$ in the \HS.
  \change{The scale uncertainties in the \msbar{} scheme are estimated including the
  RG running of parameters  as given by \eqref{eq:runningmsbar}.}
  For the BP1--4 the central scales are $580\GeV,
  480 \GeV, 380\GeV$, and $280\GeV$, respectively. For BP5 and BP6 the central
  scale is $\mu_0 = 2 \mh=250.18 \GeV$.  \label{tab:BPHSHZHS}}
\end{center}
\end{table}
\begin{table}
\begin{center}
\begin{tabular}{|c|c|c|c|}
  \hline \mystrut
  BP     & ${\rule{0ex}{2.5ex} \slo^{\Hl}/\pba}$ & $\dewps$ & $\dewb$\\
  \hline
  \hline BP21A &  $2.20$ & $-5.3\%$ & $-5.3\%$ \\
  \hline a1    &  $2.18$ & $-5.9\%$ & $-5.9\%$ \\
  \hline b1    &  $2.19$ & $-6.0\%$  & $-6.0\%$\\
  \hline BP22A &  $2.02$ & $-9.6\%$ & $-9.6\%$ \\
  \hline BP3B1 &  $2.00$ & $-7.3\%$ & $-7.3\%$ \\
  \hline BP3B2 &  $1.65$ & $-7.8\%$ & $-7.8\%$ \\
  \hline BP43  &  $2.15$ & $-6.1\%$ & $-6.1\%$ \\
  \hline BP44  &  $2.15$ & $-6.0\%$ & $-6.0\%$ \\
  \hline BP45  &  $1.08$ & $-6.4\%$ & $-6.4\%$\\
  \hline
  \hline\mystrut
         BP     & ${\rule{0ex}{2.5ex}\slo^{\Hh}/\fba}$ & $\dewps$ & $\dewb$\\
  \hline
  \hline BP22A  &  $26.3$ & $+1.5\%$ & $+1.1\%$ \\
  \hline BP3B1  &  $126.2$ & $-6.1\%$ & $-6.0\%$ \\
  \hline BP3B2  &  $350.5$ & $-5.6\%$ & $-5.5\%$ \\
  \hline BP43   &  $19.7$ & $-8.4\%$ & $-8.5\%$ \\
  \hline \change{BP44}  &  $23.0$ & $-3.5\%$ & $-4.6\%$ \\
  \hline BP45   &  $637.1$ & $-5.6\%$ & $-5.6\%$ \\
  \hline
\end{tabular}
\caption{Relative NLO corrections
  $\dew$ to Higgs-boson production in VBF $\Pp \Pp \to \Hl/\Hh \mathrm{j}
  \mathrm{j}$ in the
  \THDM. 
  {The \SM-like Higgs production is in the upper table, indicated as $\slo^{\Hl}$
  whereas the heavy one is the lower table, indicated as $\slo^{\Hh}$.} The \SM{}
  \EW{} correction to $\slo^{\Hl}$ is $\dew=-5.5\%$.  \label{tab:BPVBFthdm}}
\end{center}
\end{table}
\begin{table}
\begin{center}
\begin{tabular}{|c|c|c|c|}
  \hline \mystrut
         BP     & ${\rule{0ex}{2.5ex} \slo^{\Hl}/\fba}$ & $\dewps$ & $\dewb$\\
  \hline
  \hline BP1      &  $2108$  & $-5.6\%$  & $-5.6\%$\\
  \hline BP2      &  $2035$  & $-5.6\%$  & $-5.7\%$\\
  \hline BP3      &  $1984$  & $-5.5\%$  & $-5.6\%$\\
  \hline BP4      &  $1911$  & $-5.6\%$  & $-5.6\%$\\
  \hline BP5      &  $315.6$  & $-5.7\%$  & $-5.7\%$\\
  \hline BP6      &  $12.8$  & $-3.8\%$  & $-5.3\%$\\
  \hline
  \hline \mystrut
         BP     & ${\rule{0ex}{2.5ex} \slo^{\Hh}/\fba}$ & $\dewps$ & $\dewb$\\
  \hline
  \hline BP3    &  $79.2$  & $-4.6\%$  & $-4.7\%$ \\
  \hline BP4    &  $181.7$ & $-4.4\%$  & $-4.5\%$\\
  \hline BP5    &  $1927$  & $-5.6\%$  & $-5.6\%$\\
  \hline BP6    &  $2188$  & $-5.5\%$  & $-5.6\%$\\
  \hline
\end{tabular}
\caption{Relative NLO corrections
  $\dew$ to Higgs-boson production in VBF $\Pp \Pp \to \Hl/\Hh \mathrm{j}
  \mathrm{j}$ in the \HS.
  {
  The light Higgs production is in the upper table, indicated as $\slo^{\Hl}$ whereas
  the heavy one is in the lower table, indicated as
  $\slo^{\Hh}$.\label{tab:BPVBFHS}}}
\end{center}
\end{table}


\subsection{Discussion of the numerical results}
In the following, we compare cross sections in different
renormalization schemes and models for Higgs-boson production in Higgs
strahlung.  For VBF the picture is
similar and not 
discussed in detail. In particular, for the
\msbar{} scheme we collect some observations concerning large corrections.
An analysis of the exact origin of these contributions would go beyond the scope
of this paper. 

\subsubsection*{\msbar{} scheme}
We start with the \msbar{} scheme and \SM{}-like Higgs production in the
alignment limit of 
the \THDM{} in \refta{tab:BPHlHZthdm}. In a fixed-order
calculation no scale dependence appears in the \msbar{} scheme, because
the relevant counterterms $\dzhhhl$, $\delta \alpha$ and
$\delta \beta$ entering the vertices $\Hl \PZ \PZ$ and $\Hl \PW \PW$ are
screened by the factor $\cab/\sab=0$ in the alignment limit. 
For the same reason, the
on-shell schemes agree with the \msbar{} scheme at the central value. Yet, with
the running of parameters, a small scale dependence is visible.
For BP21B the
correction is unstable for smaller scales, 
signalling a potential problem
\change{with the benchmark point (in fact, this scenario is close to the
non-perturbative limit, see \refta{tab:BPTHDMlambda}.) or with the \msbar{} scheme.}
In non-alignment scenarios the \msbar{} results for the \THDM{} in
\refta{tab:BPHlHZthdmNA} are almost all unstable and suffer from large scale
dependencies,\footnote{\change{The Higgs self-couplings are all within the
conventional tree-level perturbativity band, \ie $\lambda_i \le 4 \pi$, but typically
one or two are of the order $\lambda_i = \mathcal{O}\left(5\right)$ (see
\reftas{tab:BPTHDMlambda} and \ref{tab:BPTHDMNAlambda}).}
}
\change{which are reflected in 
the running parameters $\cab$ and $\tb$ in}
\refta{tab:BPrunning}. 
For heavy Higgs-boson production in the \THDM{} (\refta{tab:BPHhHZthdm}) no
predictions in the \msbar{} scheme are presented as
\change{these scale uncertainties are}
even more enhanced due to ratios $\sab/\cab$ entering the
predictions.

The situation for the \msbar{} renormalization in the \HS{} for light
(\refta{tab:BPHlHZHS}) and heavy (\refta{tab:BPHSHZHS}) Higgs-boson
production is clearly more stable \change{for the considered benchmark
scenarios (see \refta{tab:HSBP} for the $\lambda_i$
values)}. This is reflected in a reasonable
running of the parameter $\sas$ in \refta{tab:BPrunningHS}, except for
BP5 and arguably for BP6. Due to the smaller running, we obtain
\change{results in the expected ballpark, with no artificially large corrections},
even for the heavy Higgs-boson production near the
alignment limit, where potential problems coming from the mixing energy
would be enhanced
by uncancelled finite parts.
\change{In the \HS{} large scale uncertainties are observed}
in the
\msbar{} scheme 
for light Higgs-boson production in BP6 and in particular for
almost degenerate neutral Higgs bosons in BP5.
Further, one observes that the \msbar{} scheme leads to larger
deviations from the \SM{} corrections, which, however, do
not come with large scale
uncertainties for the well-behaved benchmark points.

In the \HS{} the main source for large corrections are the top-quark
contributions to the neutral scalar mixing energy, which is not
subtracted in the \msbar{} scheme. This particular effect is enhanced
for degenerate neutral Higgs bosons owing to the denominator structure
in \eqref{eq:msbarprescription} which is not cancelled against the one
coming from the on-shell off-diagonal field renormalization constants.
Besides the top-quark contributions it is possible to induce moderate
contributions coming from the Higgs potential by tuning $\lambda_3$.
This requires,
however, large $\mhh^2-\mhl^2$ with not too small $\sas$, and 
it is not straightforward to tune the parameters in order to
exceed the top-quark contribution without getting close to the
non-perturbative limit $\left|\lambda_i\right| \sim 4 \pi$.  
In
the \THDM, the reason for the large corrections in the \msbar{} scheme
is more difficult to grasp, especially because in view of our
observables we have to deal with the renormalization of $\beta$ which
is known to cause difficulties in the
MSSM\cite{Freitas:2002um}.\footnote{Note that $\beta$ in the \HS{}
  suffers the same problems, but does not enter our fixed-order
  calculations and the observables we consider.  For this reason
  $\beta$ has been decoupled from the running of $\alpha$ in the \HS{}
  in order to avoid problems related to its renormalization.}  The
problem with $\beta$ can be traced back to large contributions in the
tadpoles.  
For
$\alpha$, the largest contributions cannot be
explained by tadpoles nor by the top-quark contribution in the neutral
scalar mixing energy. Here, we observe that the large contributions to
the neutral scalar mixing energy are mediated through the charged and
pseudo-scalar Higgs boson, which, eventually, exceed all other
contributions. 
\change{Since these large contributions are only found in the off-diagonal
LSZ-factors they remain uncancelled in the \msbar{} scheme.}

\subsubsection*{On-shell schemes}
For the considered on-shell schemes none of the observed problems of
the \msbar{} scheme is encountered because the large contributions in
the mixing energy and the tadpoles are subtracted via $\delta \alpha$
and $\delta \beta$, \ie all terms involving the poles
$1/(\mhh^2-\mhl^2)$ and $1/\mha^2$ cancel in \SMatrix{} elements.
Further, in view of the size of the corrections the on-shell methods
perform much better in the sense that the \SM-like Higgs-boson
production processes (see \reftas{tab:BPHlHZthdm},
\ref{tab:BPHlHZthdmNA} for the \THDM{} and \reftas{tab:BPHlHZHS},
\ref{tab:BPHSHZHS} for the \HS) yield corrections which are close to
the \SM{} correction.
In heavy Higgs-boson production (see \refta{tab:BPHhHZthdm} for
the \THDM{} and
\refta{tab:BPHSHZHS} for the \HS)  
the results for on-shell
renormalization schemes remain stable even for aligned\footnote{For heavy-Higgs
production one expects large one-loop corrections in almost aligned scenarios
(\eg benchmark point b1) because in the exact alignment the LO vanishes. In that
case the cross section should be computed including squared one-loop amplitudes,
making the one-loop computation effectively a LO approximation.} or degenerate
scenarios.  The difference between the $p^*$ and \BFM{} schemes is tiny. It
seems to us that the schemes are too similar for their difference to provide a
qualitative estimate of higher orders.  The difference between these schemes
just results from the momentum dependence of the neutral scalar mixing energy,
which turns out to be small and starts at the order
$\mathcal{O}(\mhl^2-\mhh^2)$. Note also, that the 
large contributions in the neutral \change{scalar} mixing energy
were observed to have almost no momentum dependence.  For VBF the computation
has only been carried out in the on-shell schemes.  For the \THDM{}
(\refta{tab:BPVBFthdm}) and \HS{} (\refta{tab:BPVBFHS}) the \SM-like scenarios
almost coincide with the \SM{} predictions. 

\subsection{Distributions}
We present distributions for the transverse momentum $p_{\mathrm{T}, \Hh}$ and
rapidity $y_{\Hh}$ of heavy Higgs bosons in Higgs strahlung and VBF.  In
addition, we show distributions in the rapidity $y_{\mu^-}$ of the muon $\mu^-$
in Higgs strahlung and in the rapidity $y_{j_1}$ of the hardest jet $j_1$ in
VBF.  We selected a typical subset of all benchmark points, namely the benchmark
points BP3B1, BP43 and BP45 in the \THDM{} and BP3 in the \HS.  All results are
given in the $p^*$ renormalization scheme for $\alpha$ and $\beta$. We do not
show any \SM-like Higgs-production scenarios in the \THDM{} or \HS{} as no 
shape distortions are observed compared to the \SM{} and basically only the
normalization of the distributions is affected. Our results are thus consistent
with the observation made in \SM{} \EFT{} matched to the full model for the
\HS{} and
\THDM{} in \citere{Brehmer:2015rna}, where it is stated that for small mixing
angles, near the alignment, new operators do not play a significant role.

\begin{figure}
  \centering
  \begin{subfigure}[b]{0.49\textwidth}
    \includegraphics[width=0.89\textwidth]{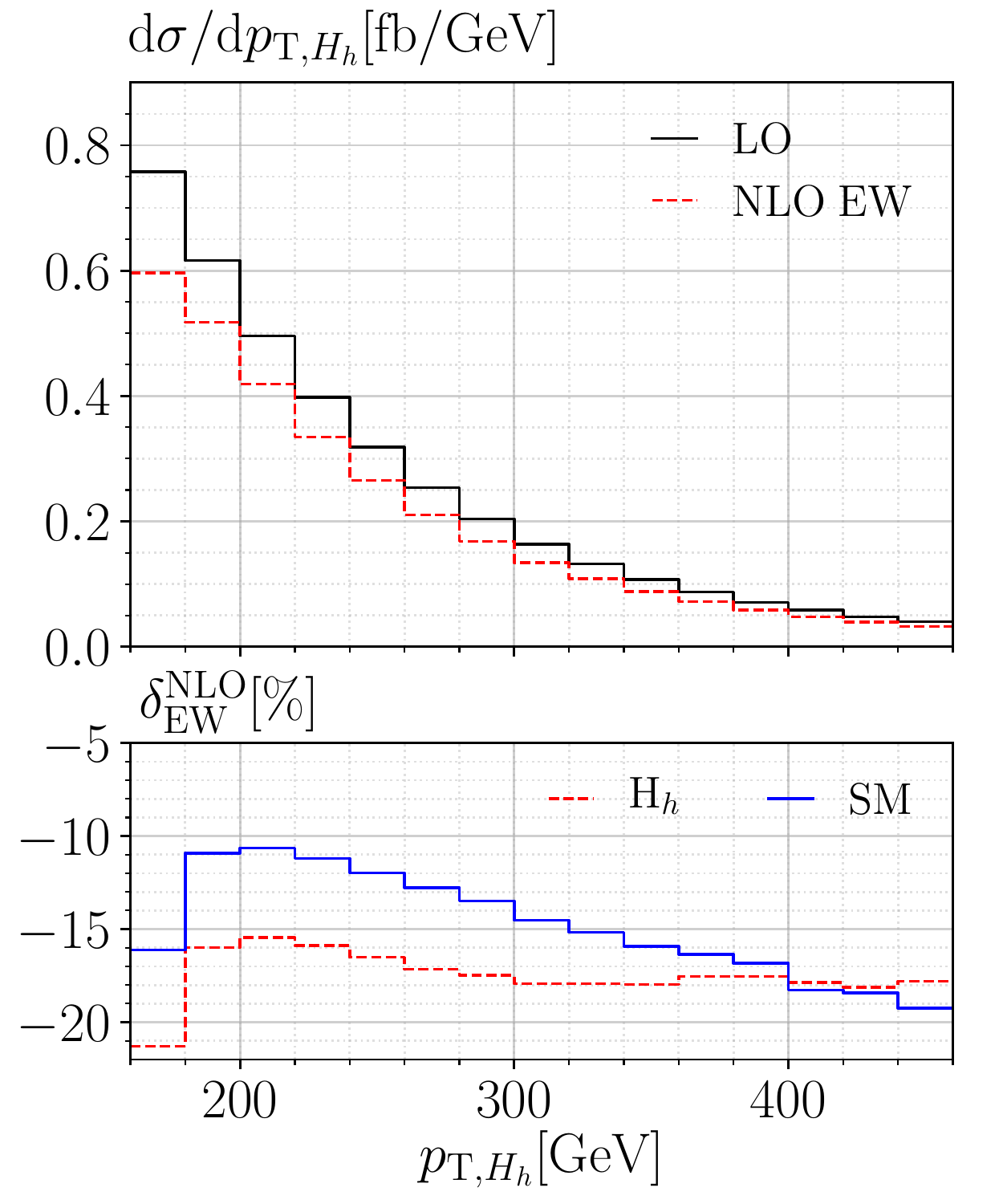}
    \caption{BP3B1 (\THDM)\\\phantom{.}}
  \end{subfigure}
  \begin{subfigure}[b]{0.49\textwidth}
    \includegraphics[width=0.89\textwidth]{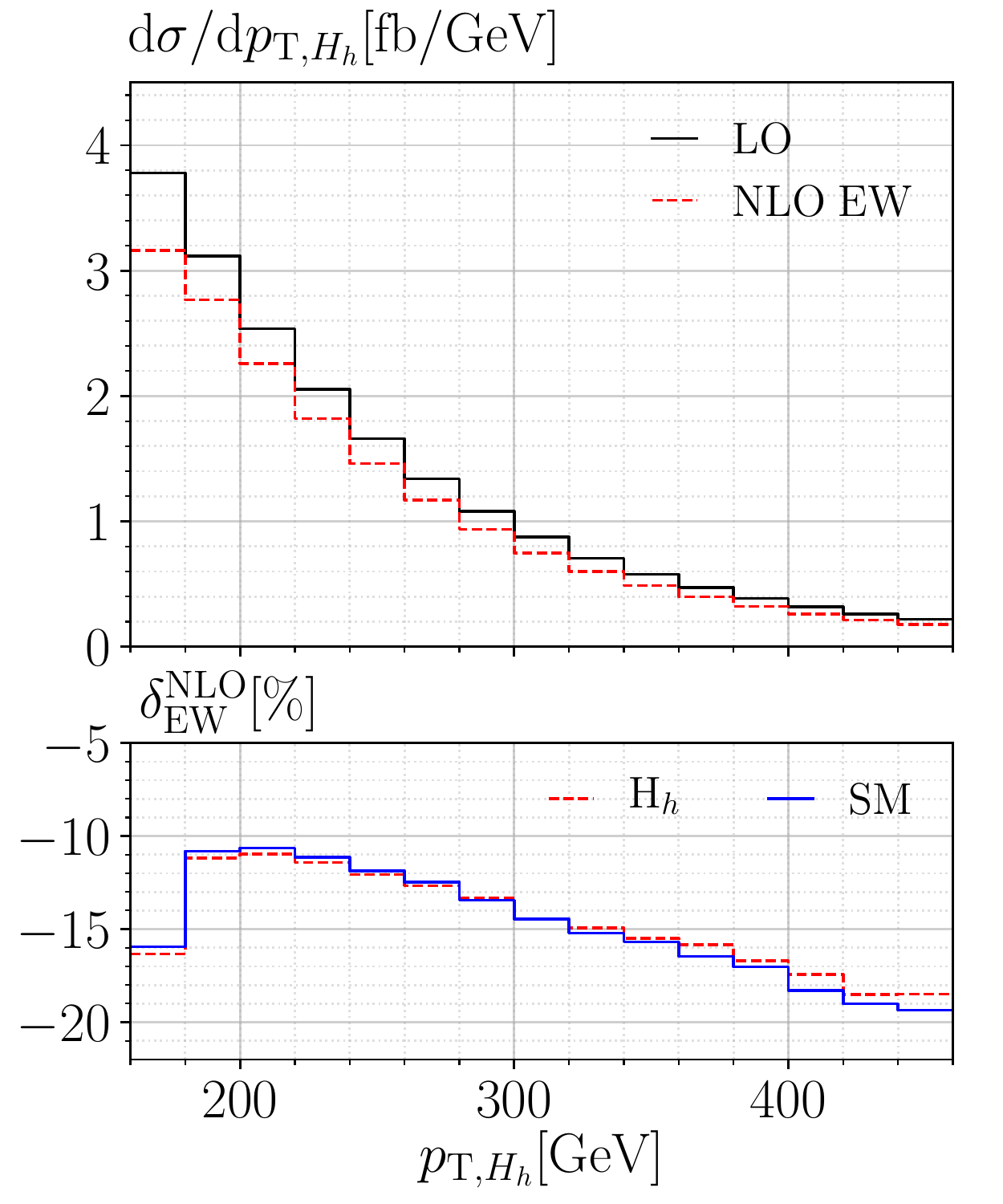}
    \caption{BP45 (\THDM)\\\phantom{.}}
  \end{subfigure}
  \begin{subfigure}[b]{0.49\textwidth}
    \includegraphics[width=0.89\textwidth]{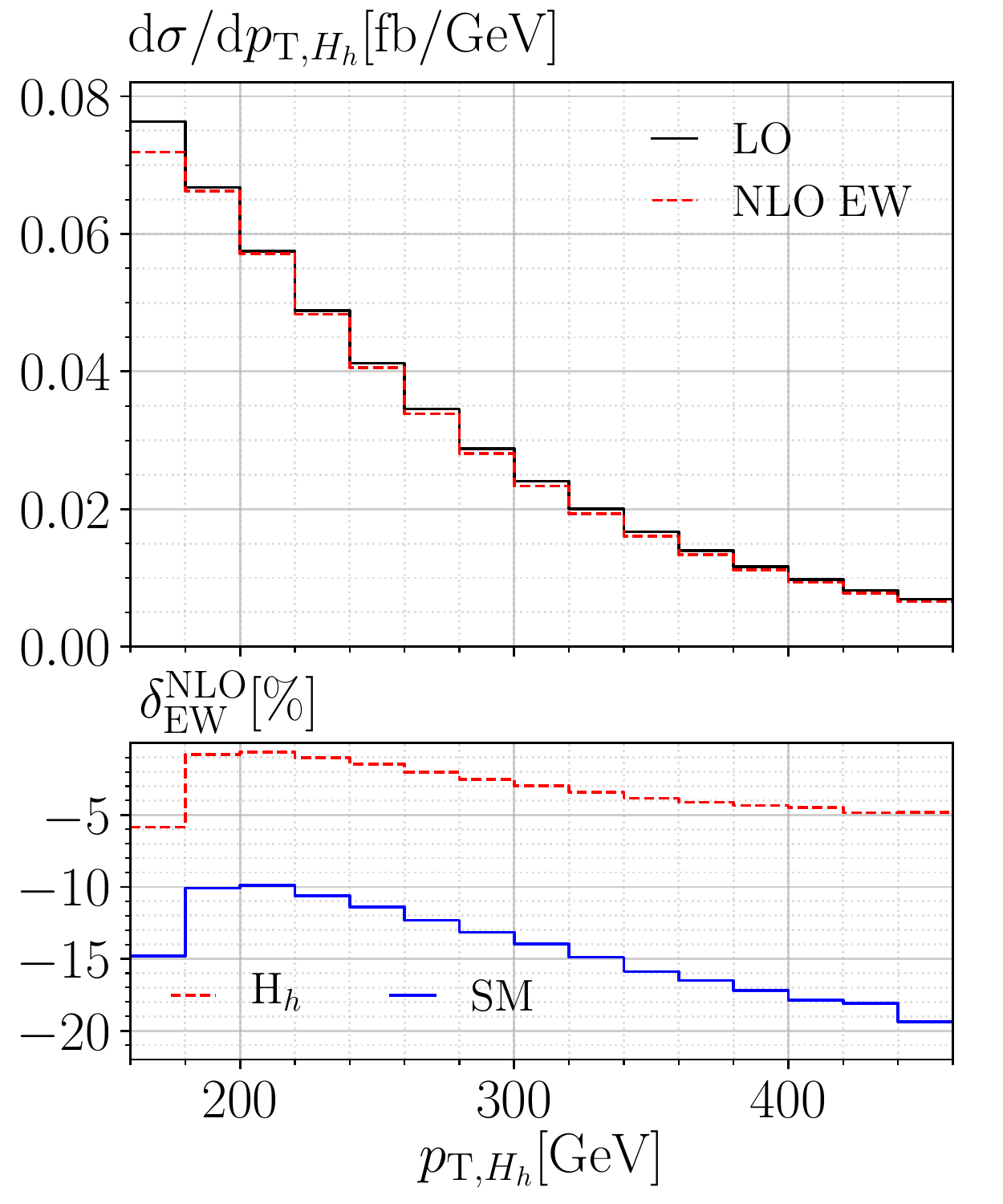}
    \caption{BP43 (\THDM)}
  \end{subfigure}
  \begin{subfigure}[b]{0.49\textwidth}
    \includegraphics[width=0.89\textwidth]{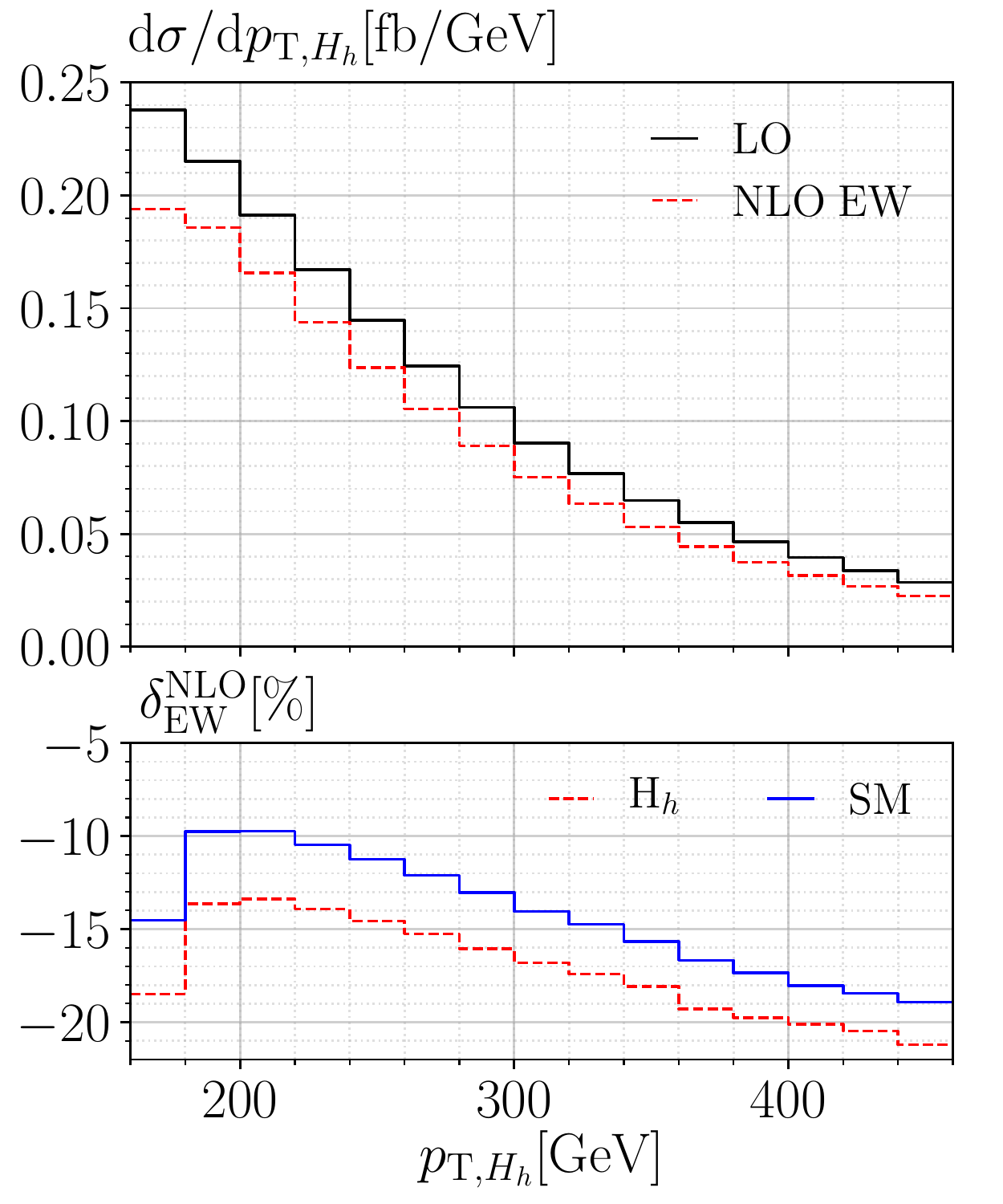}
    \caption{BP3 (\HS)}
  \end{subfigure}
  \caption{
    Distributions in the transverse momentum of the Higgs boson
    $p_{\mathrm{T},\Hh}$ for heavy Higgs production in Higgs strahlung for the
    benchmark points BP3B1 in (a), BP45 in (b) and BP43 in (c) in the \THDM, and
    BP3 in (d) in the \HS.}
  \label{fig:pth}
\end{figure}

\begin{figure}
  \centering
  \begin{subfigure}[b]{0.49\textwidth}
    \includegraphics[width=0.89\textwidth]{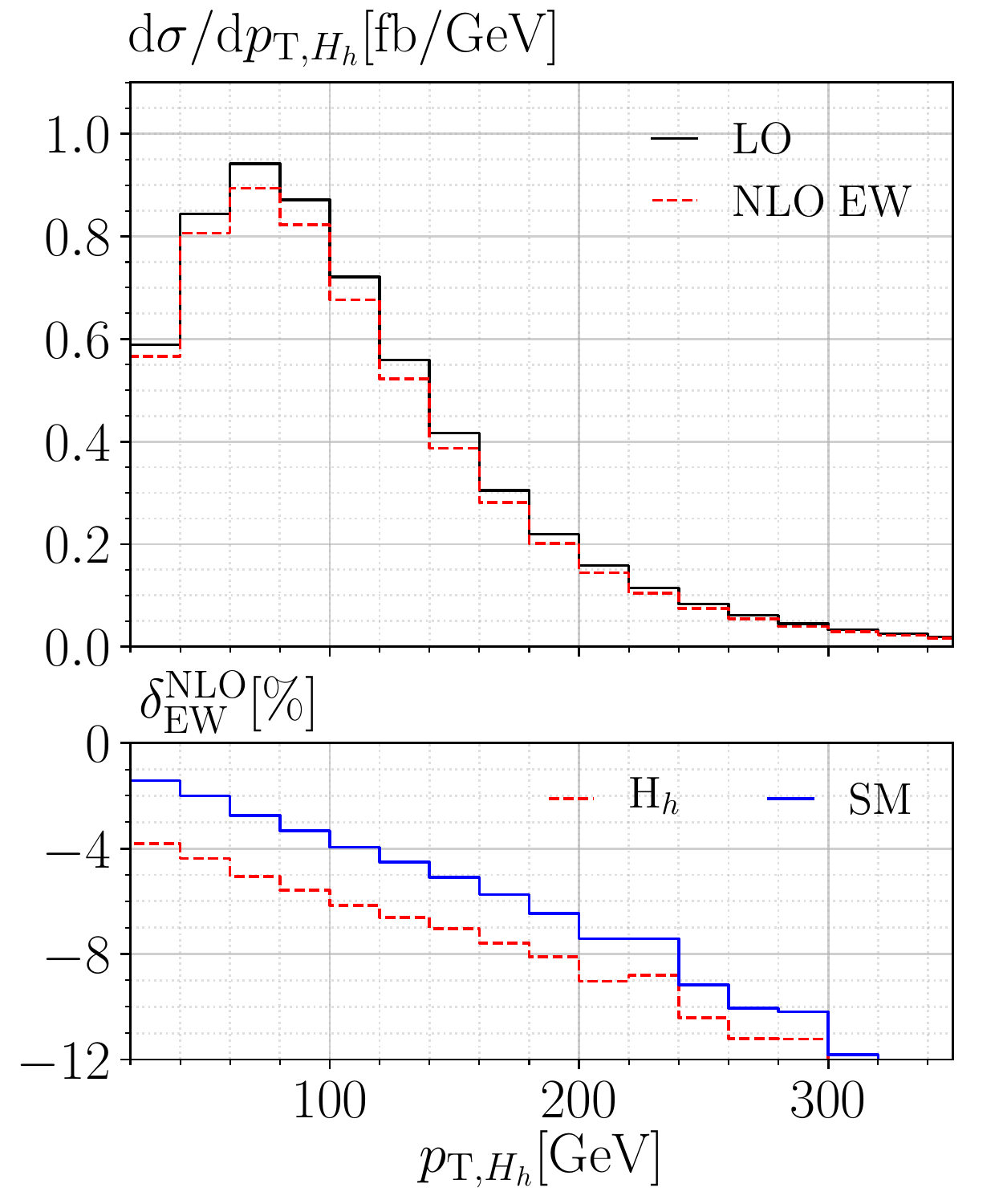}
    \caption{BP3B1 (\THDM)\\\phantom{.}}
  \end{subfigure}
  \begin{subfigure}[b]{0.49\textwidth}
    \includegraphics[width=0.89\textwidth]{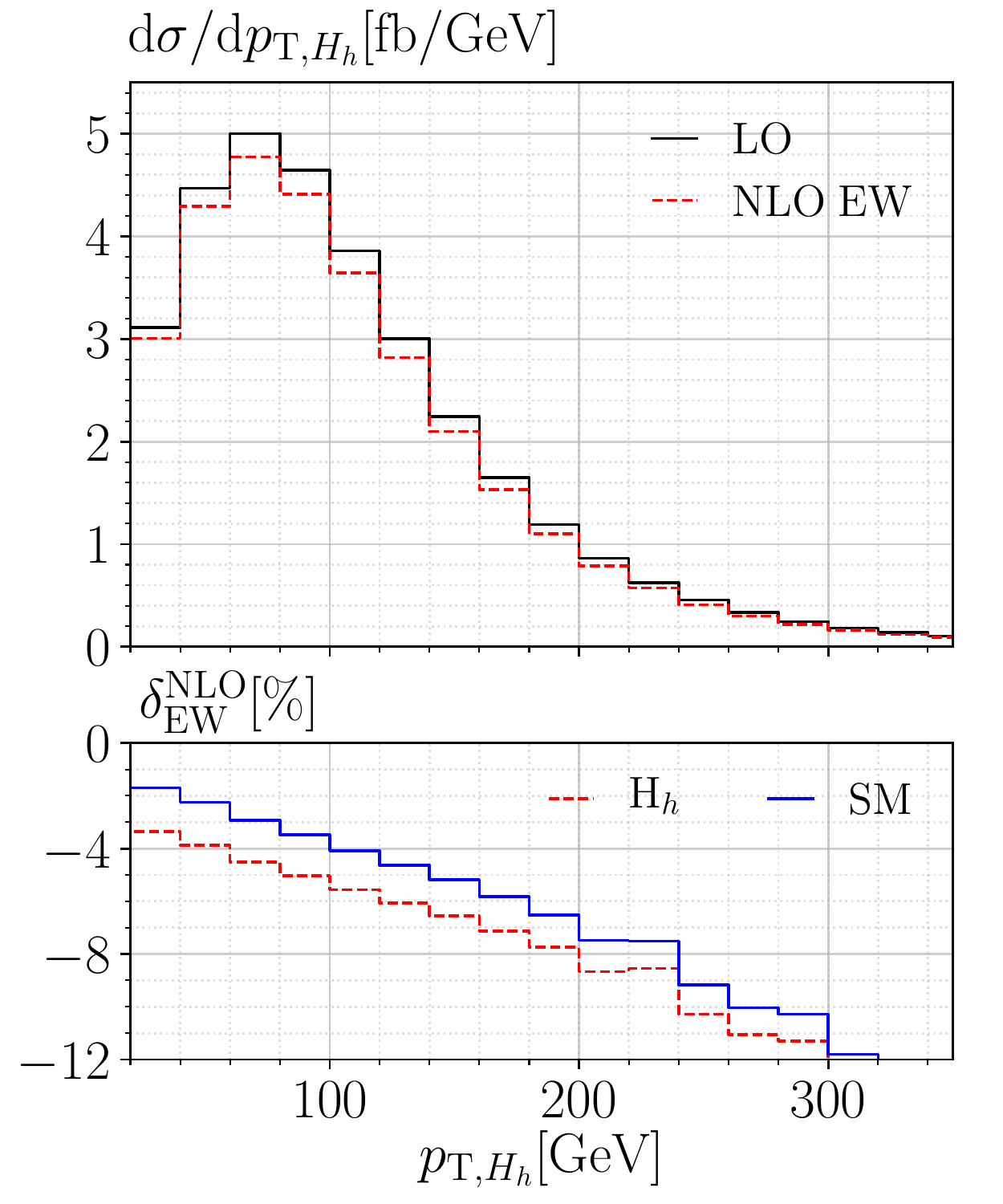}
    \caption{BP45 (\THDM)\\\phantom{.}}
  \end{subfigure}
  \begin{subfigure}[b]{0.49\textwidth}
    \includegraphics[width=0.89\textwidth]{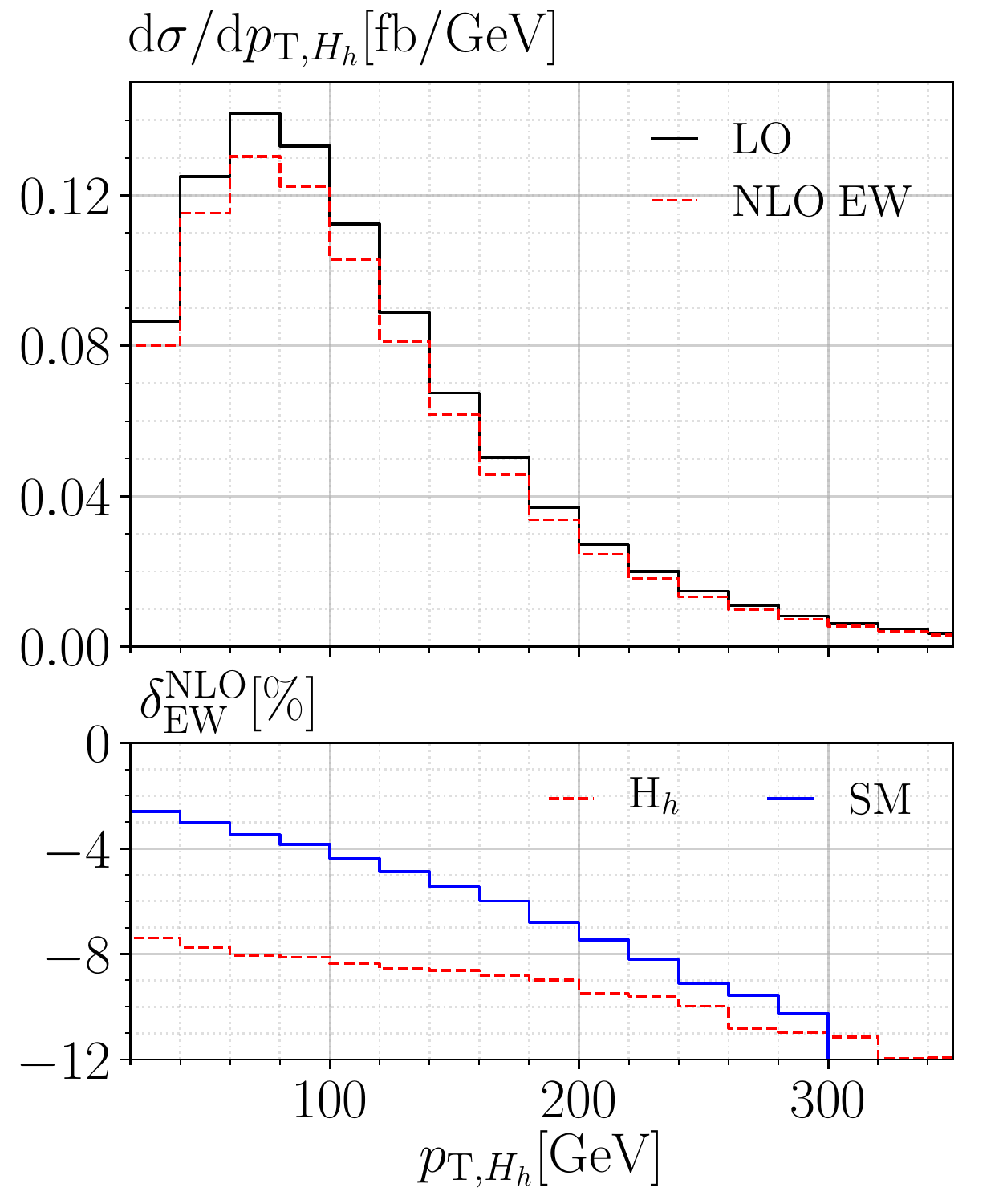}
    \caption{BP43 (\THDM)}
  \end{subfigure}
  \begin{subfigure}[b]{0.49\textwidth}
    \includegraphics[width=0.89\textwidth]{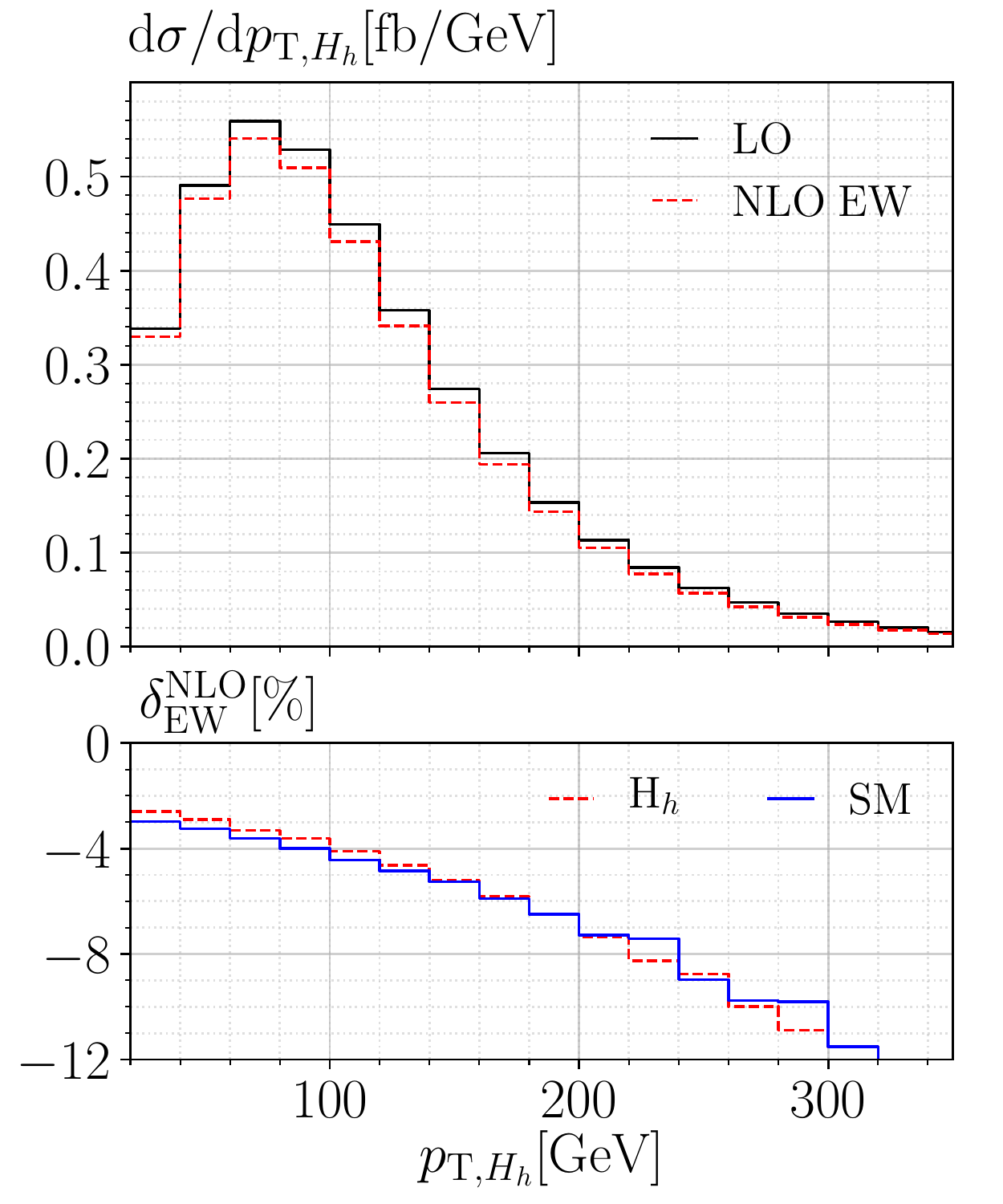}
    \caption{BP3 (\HS)}
  \end{subfigure}
  \caption{
    Distributions in the transverse momentum of the Higgs boson
    $p_{\mathrm{T},\Hh}$ for heavy Higgs production in VBF for the benchmark
    points BP3B1 in (a), BP45 in (b) and BP43 in (c) in the \THDM, and BP3 in
    (d) in the \HS.}
  \label{fig:pthVBF}
\end{figure}

\begin{figure}
  \centering
  \begin{subfigure}[b]{0.49\textwidth}
    \includegraphics[width=0.89\textwidth]{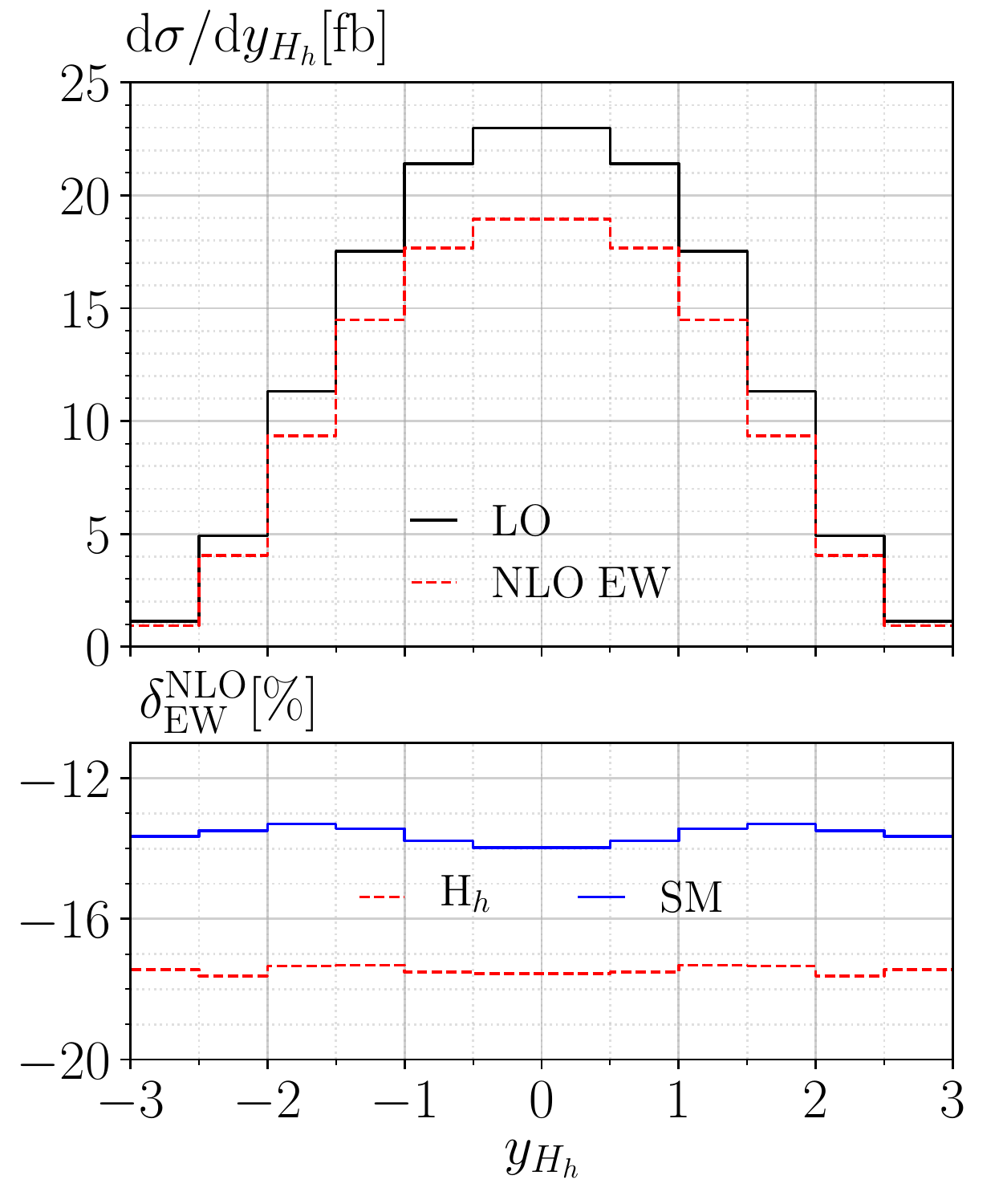}
    \caption{BP3B1 (\THDM)\\\phantom{.}}
  \end{subfigure}
  \begin{subfigure}[b]{0.49\textwidth}
    \includegraphics[width=0.89\textwidth]{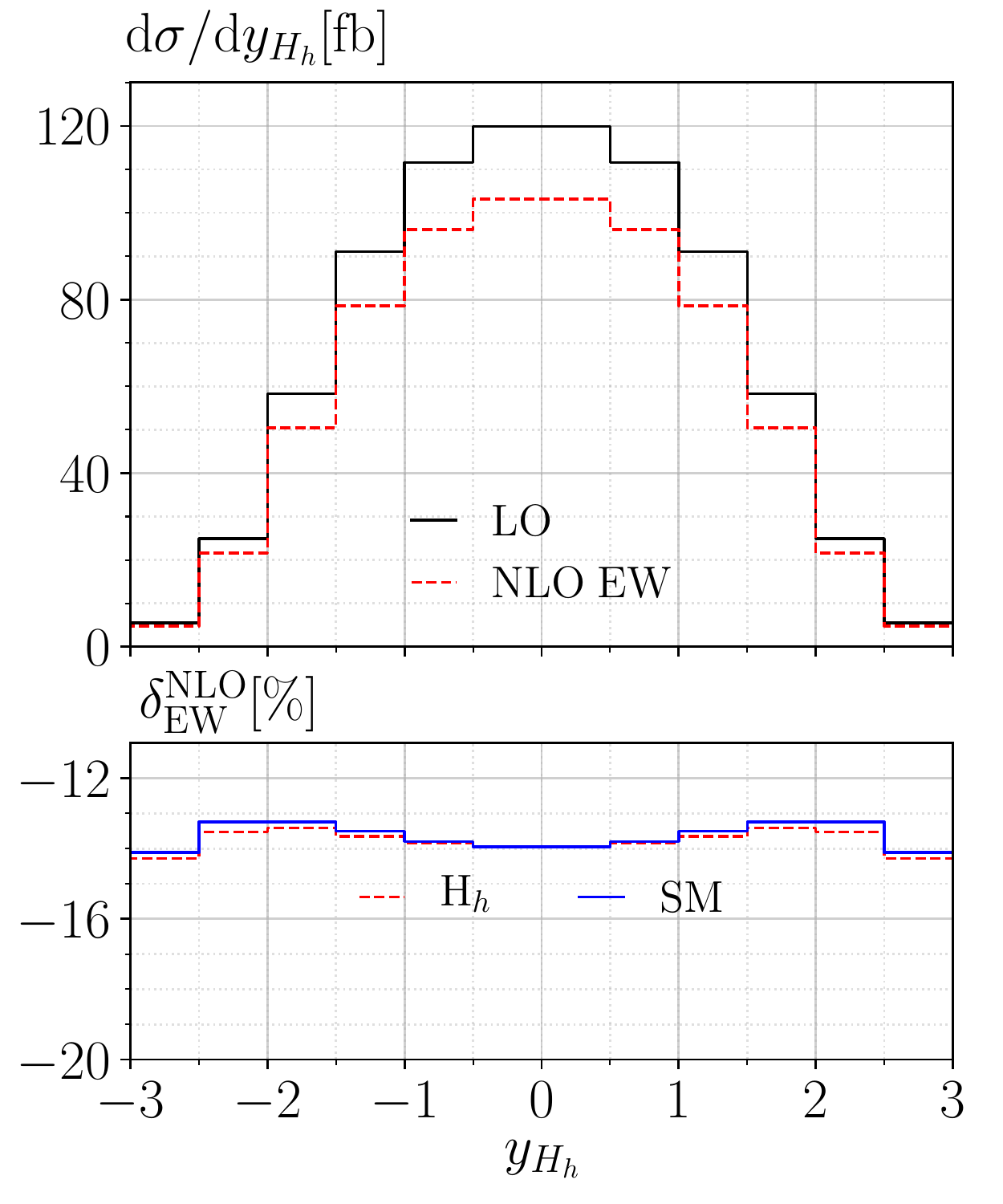}
    \caption{BP45 (\THDM)\\\phantom{.}}
  \end{subfigure}
  \begin{subfigure}[b]{0.49\textwidth}
    \includegraphics[width=0.89\textwidth]{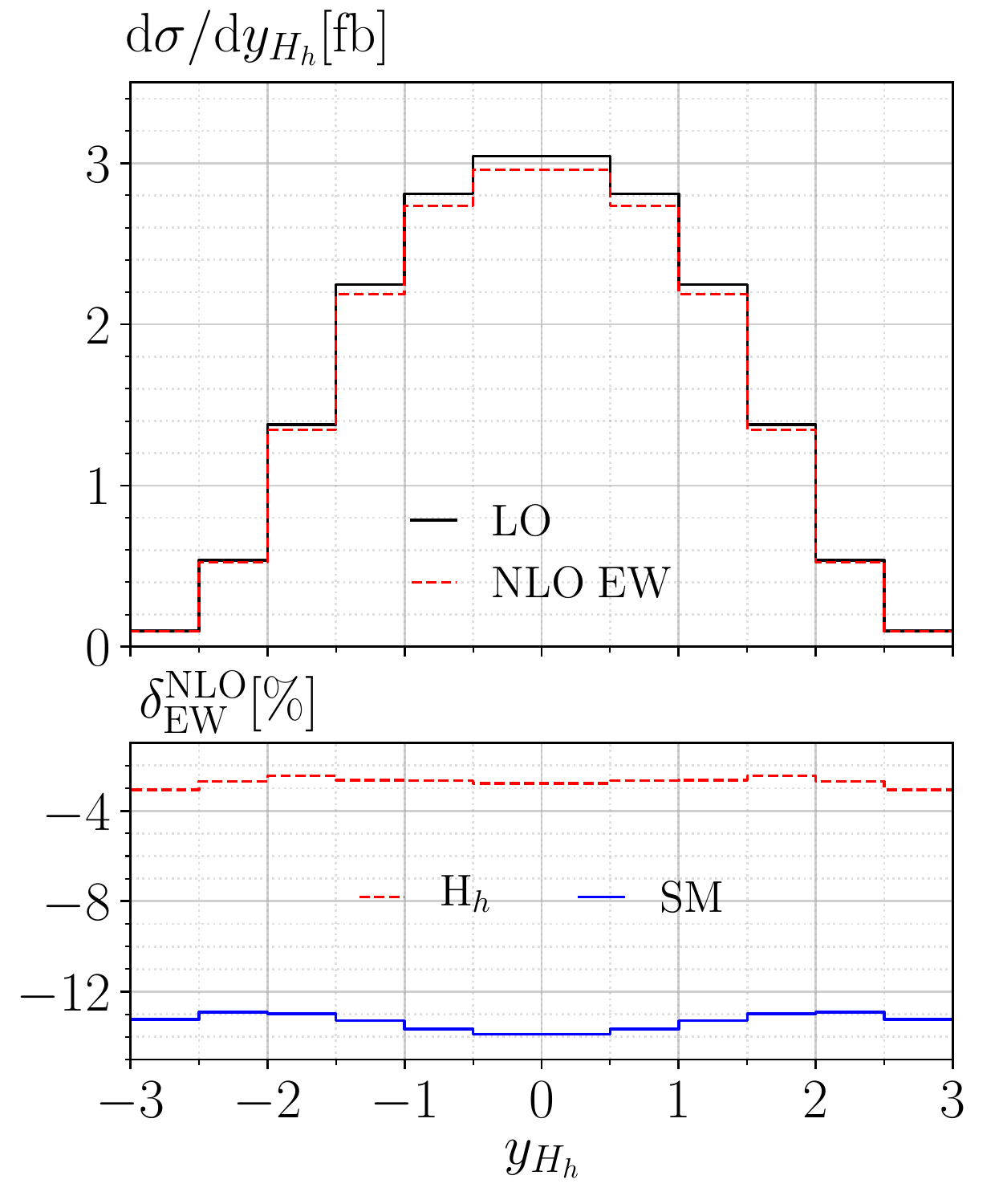}
    \caption{BP43 (\THDM)}
  \end{subfigure}
   \begin{subfigure}[b]{0.49\textwidth}
    \includegraphics[width=0.89\textwidth]{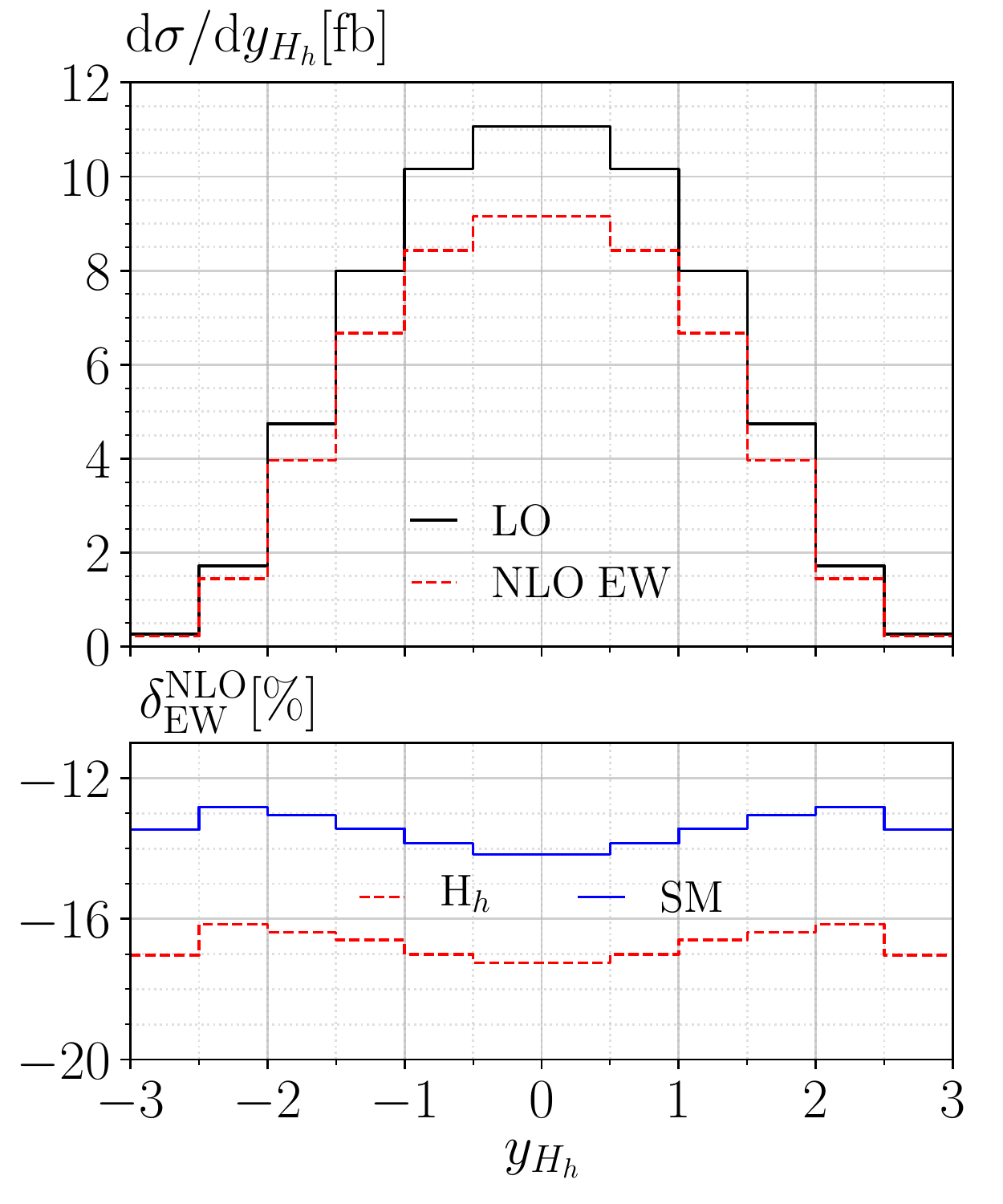}
    \caption{BP3 (\HS)}
  \end{subfigure}
  \caption{
    Distributions in the rapidity of the Higgs boson
    $y_{\Hh}$ for heavy Higgs production in Higgs strahlung for the
    benchmark points BP3B1 in (a), BP45 in (b) and BP43 in (c) in the \THDM, and
    BP3 in (d) in the \HS.}
  \label{fig:yh}
\end{figure}

\begin{figure}
  \centering
  \begin{subfigure}[b]{0.49\textwidth}
    \includegraphics[width=0.89\textwidth]{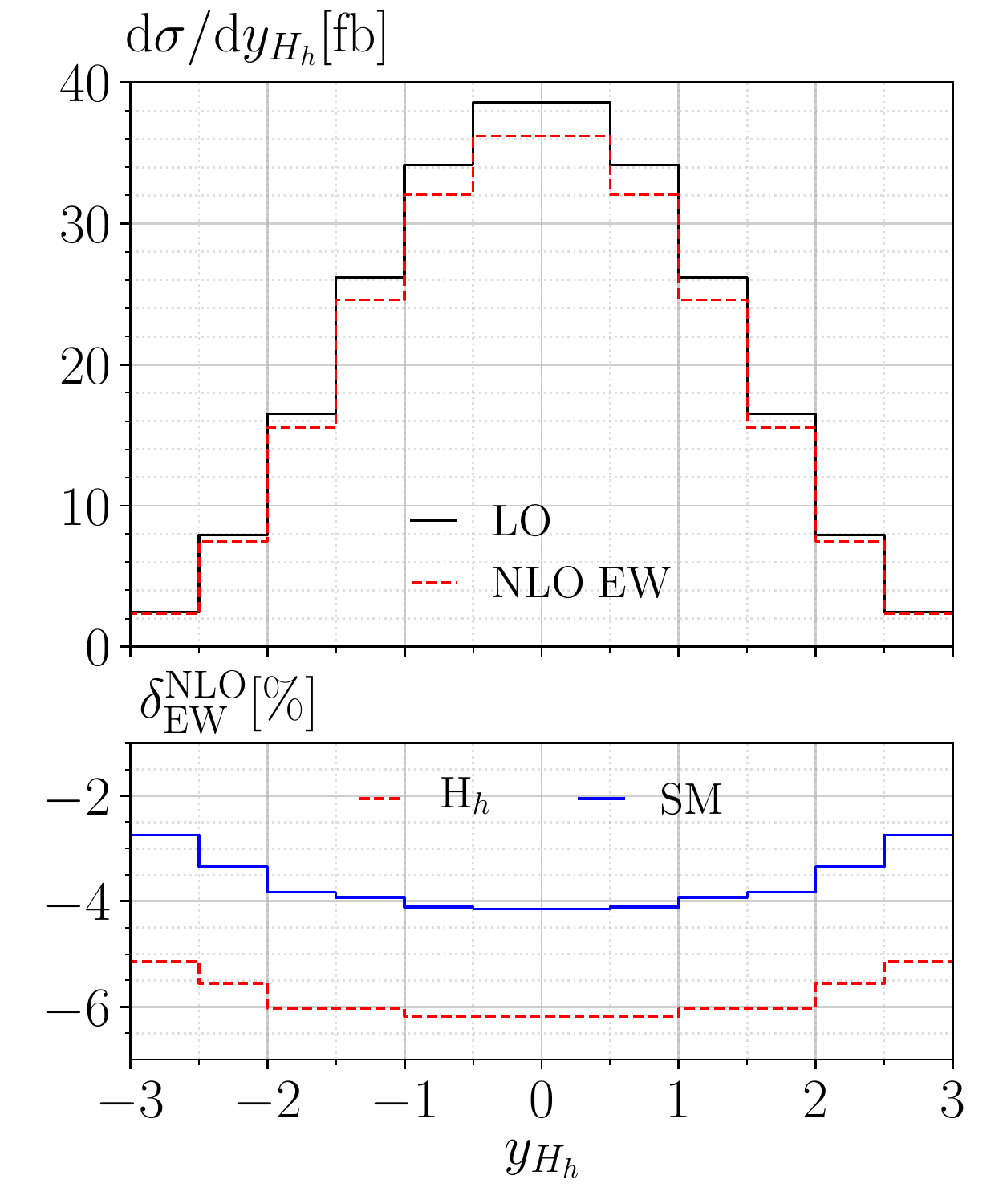}
    \caption{BP3B1 (\THDM)\\\phantom{.}}
  \end{subfigure}
  \begin{subfigure}[b]{0.49\textwidth}
    \includegraphics[width=0.89\textwidth]{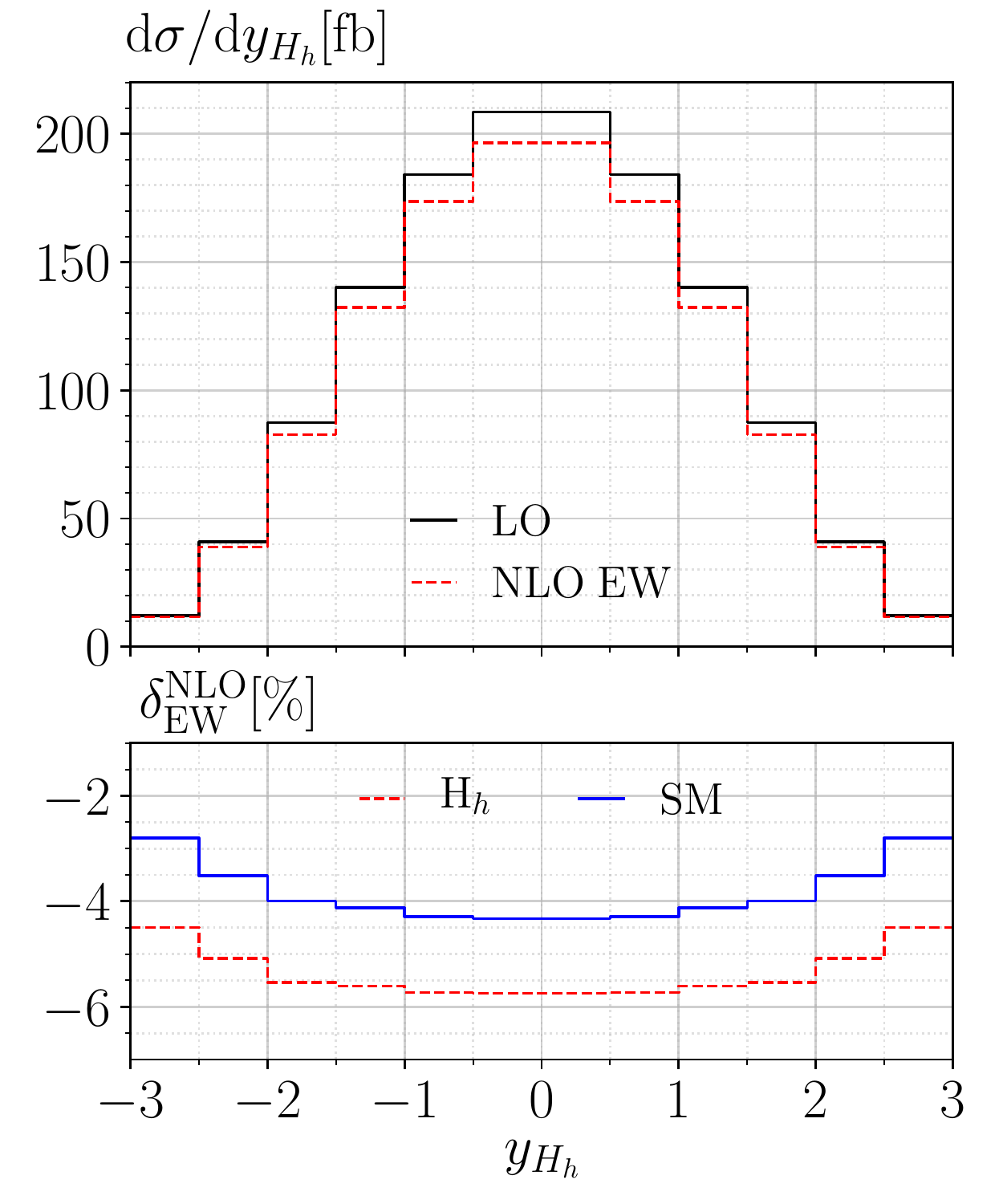}
    \caption{BP45 (\THDM)\\\phantom{.}}
  \end{subfigure}
  \begin{subfigure}[b]{0.49\textwidth}
    \includegraphics[width=0.89\textwidth]{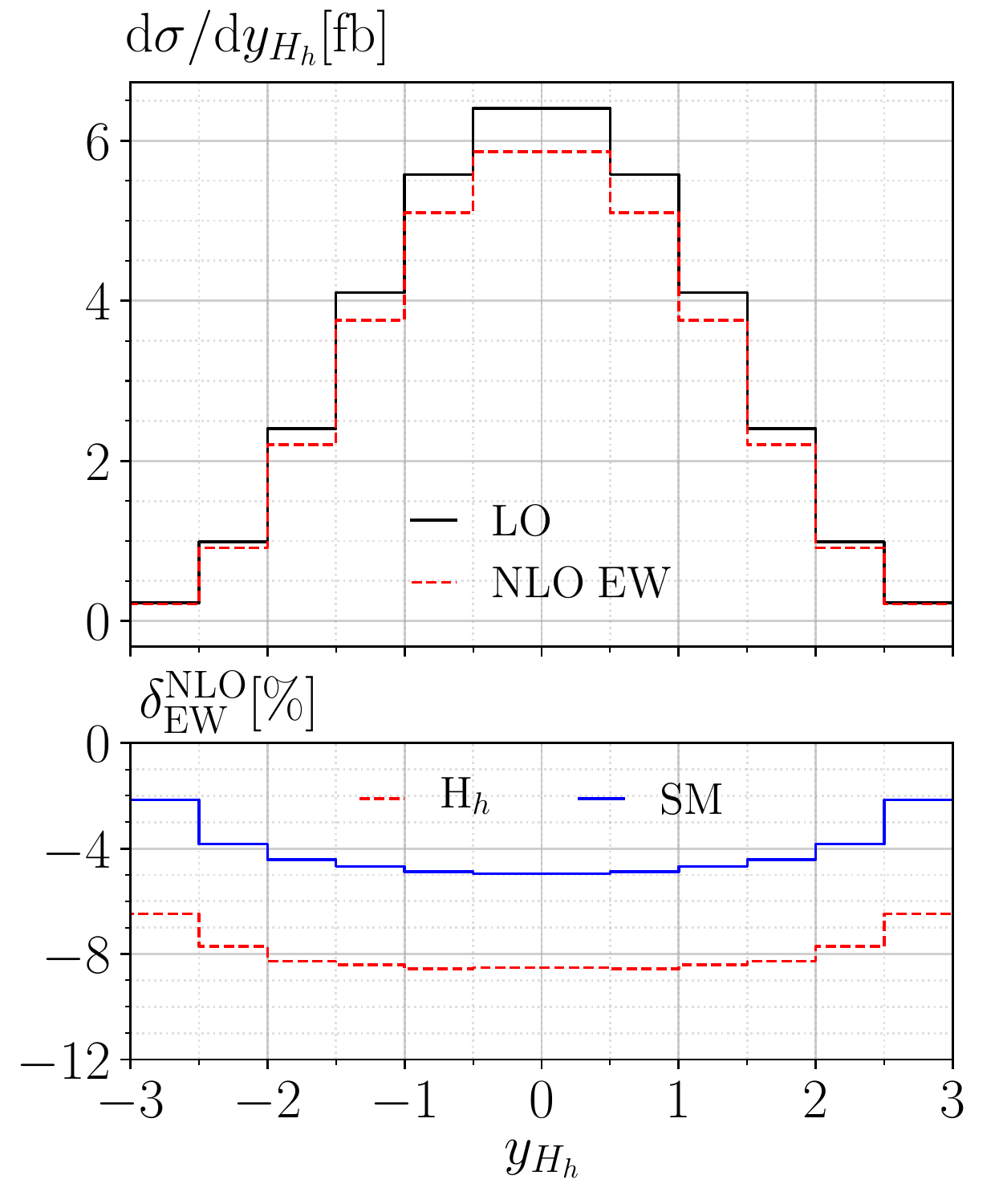}
    \caption{BP43 (\THDM)}
  \end{subfigure}
  \begin{subfigure}[b]{0.49\textwidth}
    \includegraphics[width=0.89\textwidth]{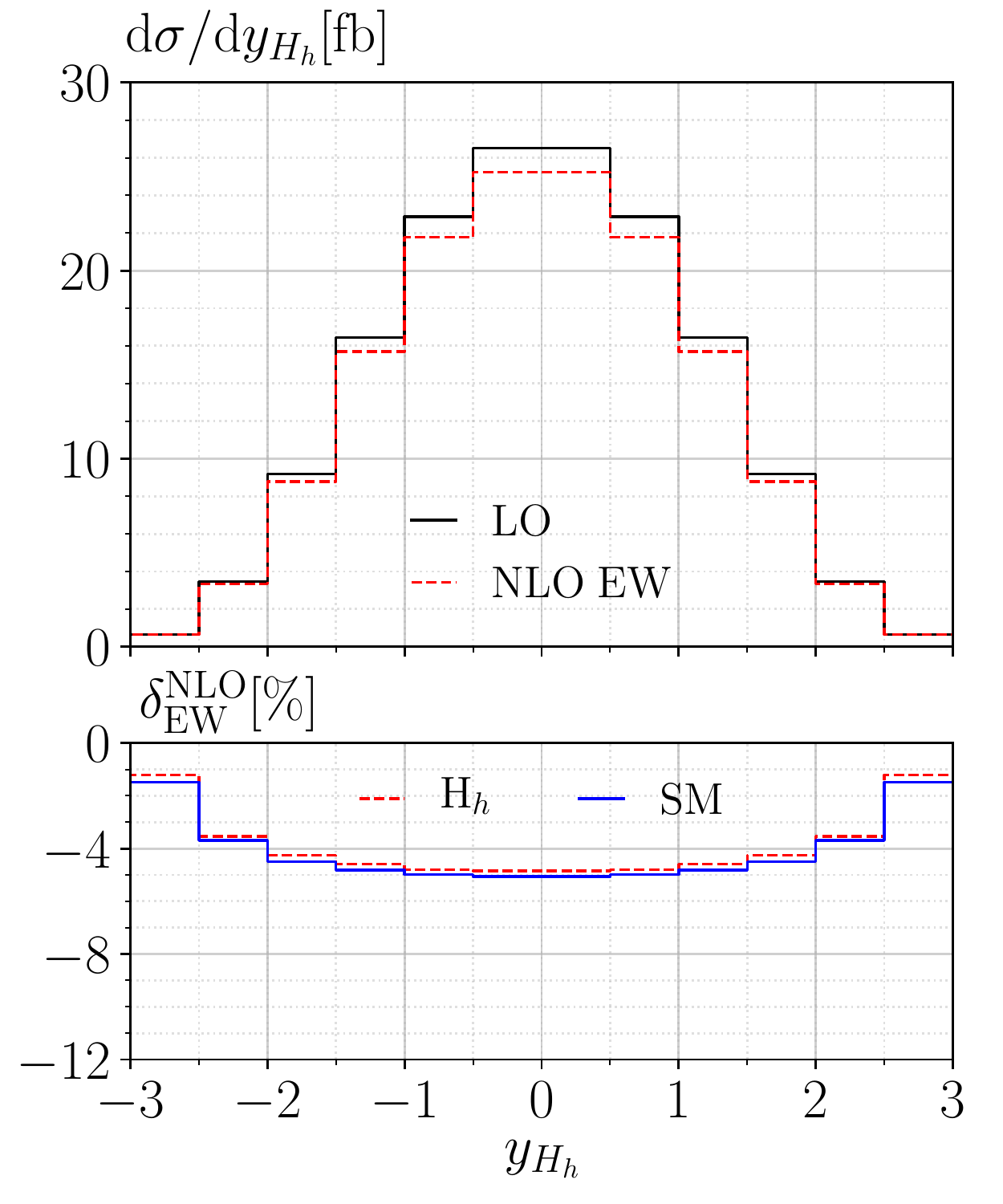}
    \caption{BP3 (\HS)}
  \end{subfigure}
  \caption{
    Distributions in the rapidity of the Higgs boson
    $y_{\Hh}$ for heavy Higgs production in VBF for the
    benchmark points BP3B1 in (a), BP45 in (b) and BP43 in (c) in the \THDM, and
    BP3 in (d) in the \HS.}
  \label{fig:yhVBF}
\end{figure}

\begin{figure}
  \centering
  \begin{subfigure}[b]{0.49\textwidth}
    \includegraphics[width=0.89\textwidth]{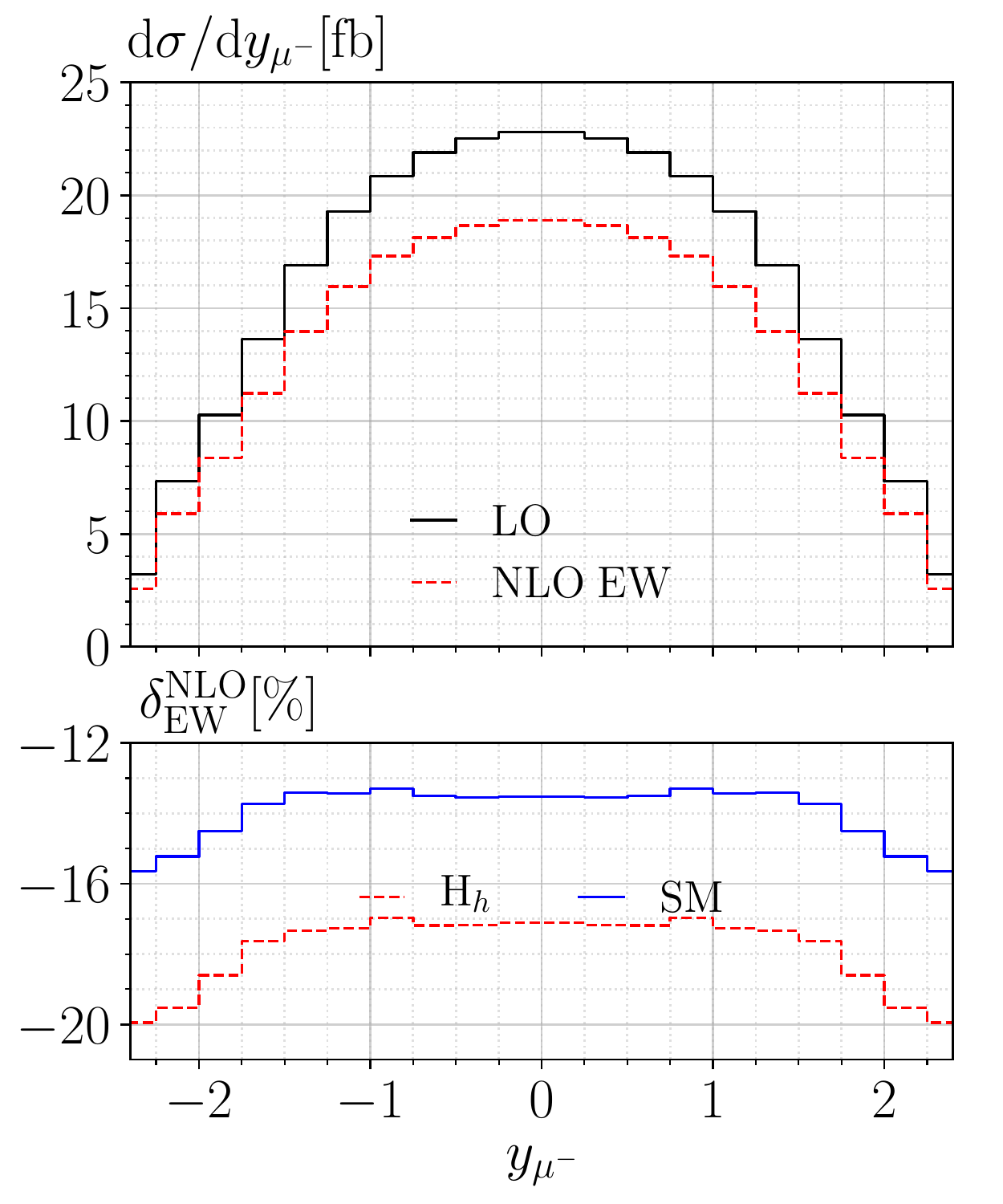}
    \caption{BP3B1 (\THDM)\\\phantom{.}}
  \end{subfigure}
  \begin{subfigure}[b]{0.49\textwidth}
    \includegraphics[width=0.89\textwidth]{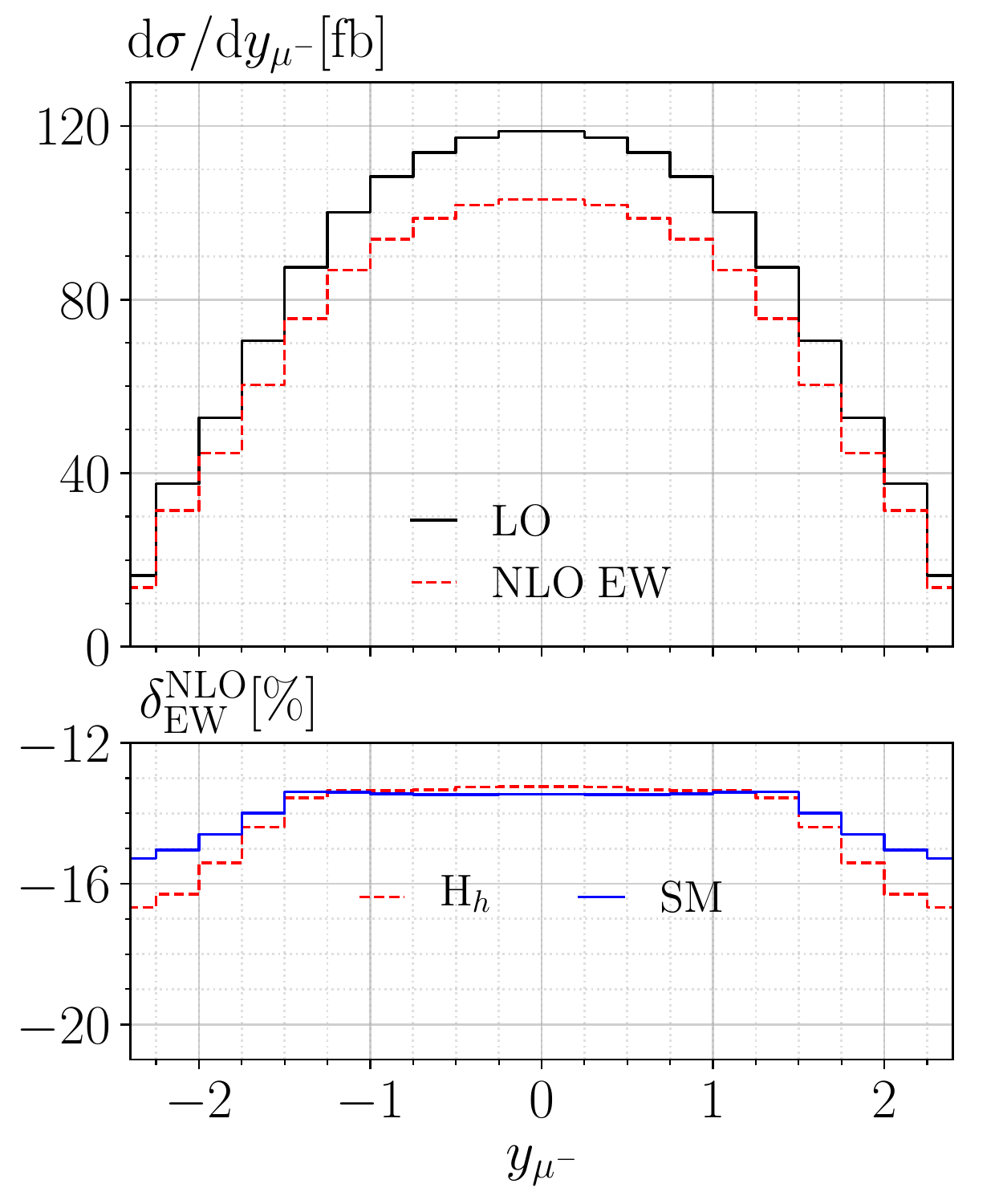}
    \caption{BP45 (\THDM)\\\phantom{.}}
  \end{subfigure}
  \begin{subfigure}[b]{0.49\textwidth}
    \includegraphics[width=0.89\textwidth]{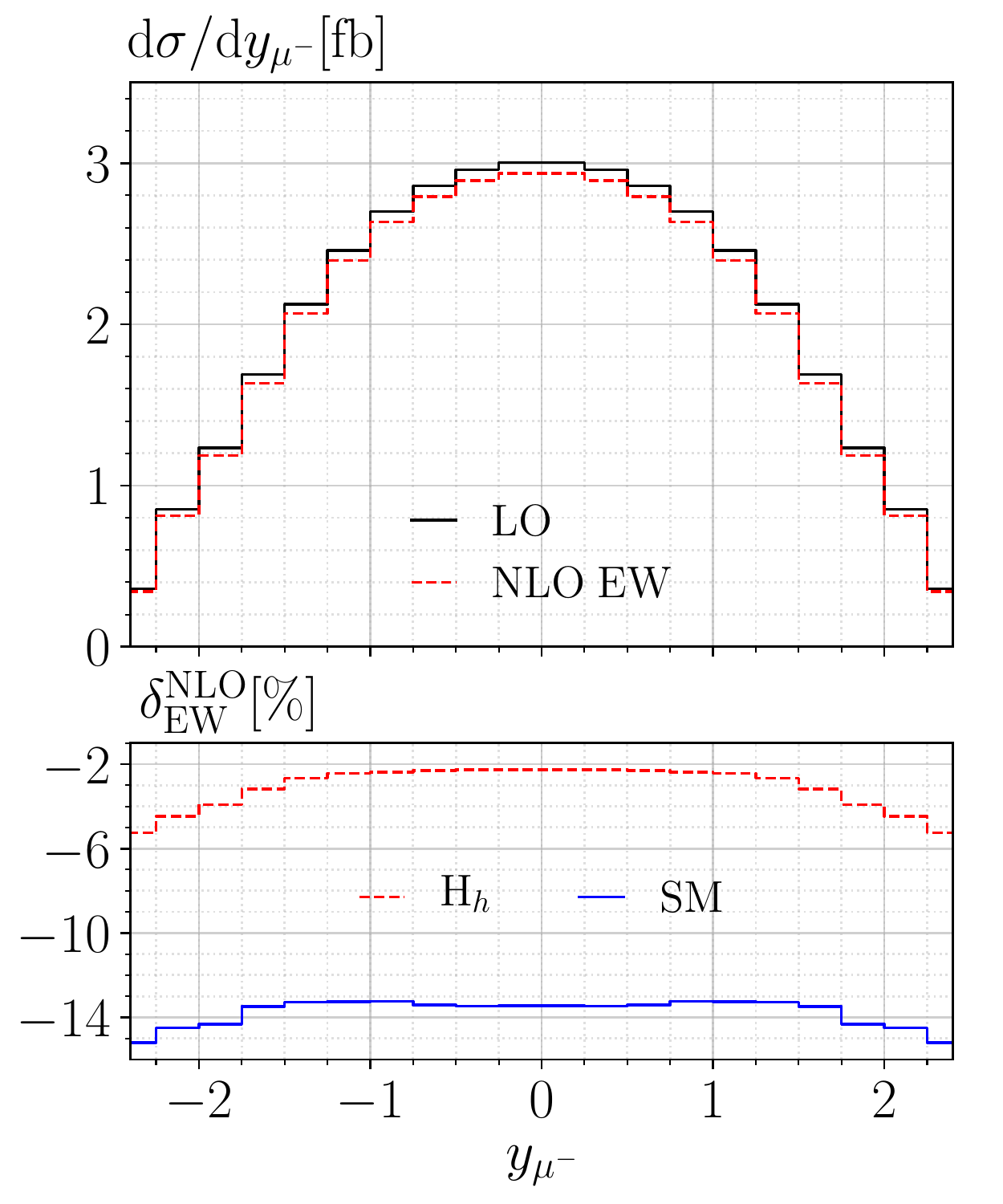}
    \caption{BP43 (\THDM)}
  \end{subfigure}
  \begin{subfigure}[b]{0.49\textwidth}
    \includegraphics[width=0.89\textwidth]{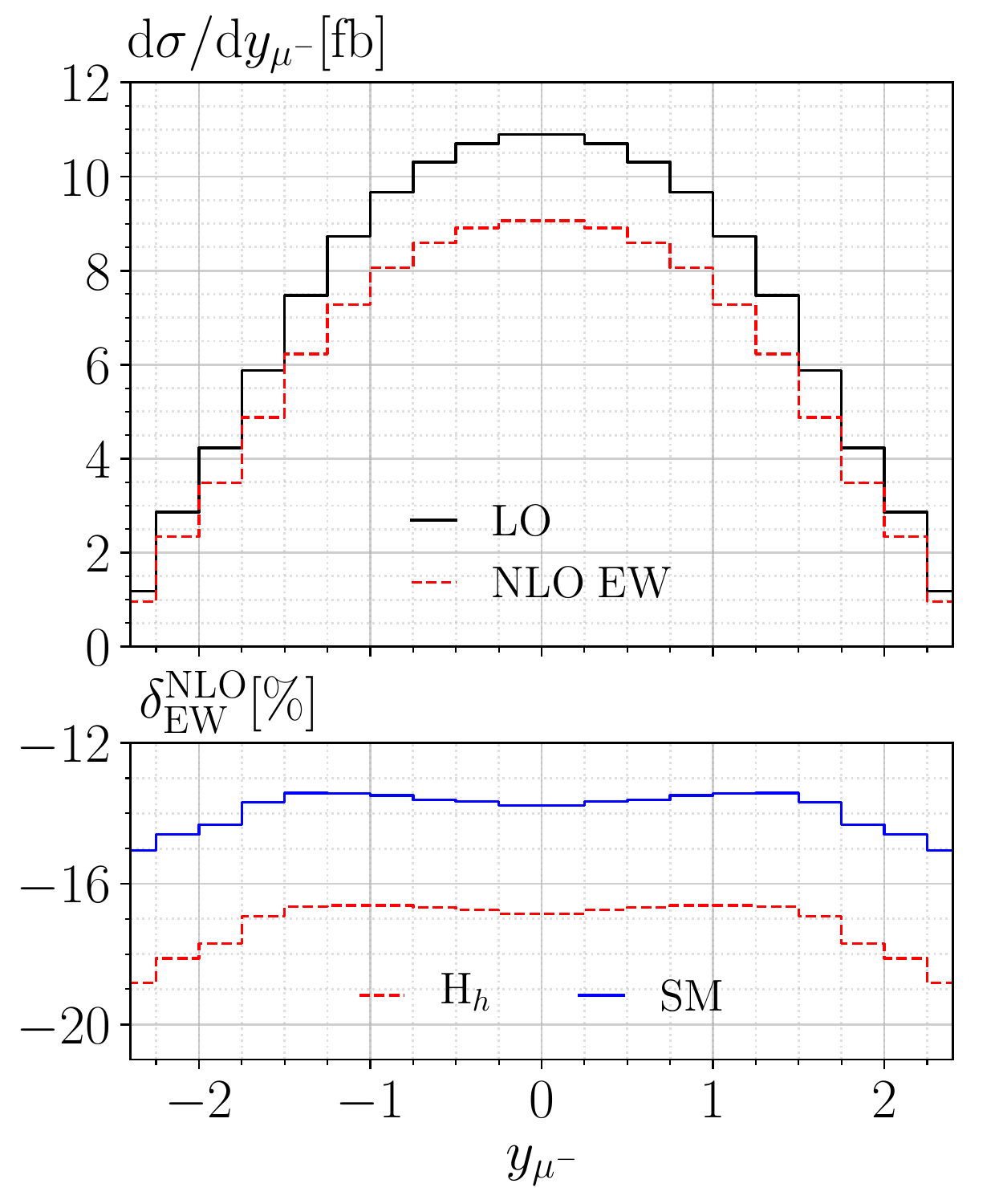}
    \caption{BP3 (\HS)}
  \end{subfigure}
  \caption{
    Distributions in the rapidity of the muon
    $y_{\Pm^-}$ for heavy Higgs production in Higgs strahlung for the
    benchmark points BP3B1 in (a), BP45 in (b) and BP43 in (c) in the \THDM, and
    BP3 in (d) in the \HS.}
  \label{fig:ylm}
\end{figure}

\begin{figure}
  \centering
  \begin{subfigure}[b]{0.49\textwidth}
    \includegraphics[width=0.89\textwidth]{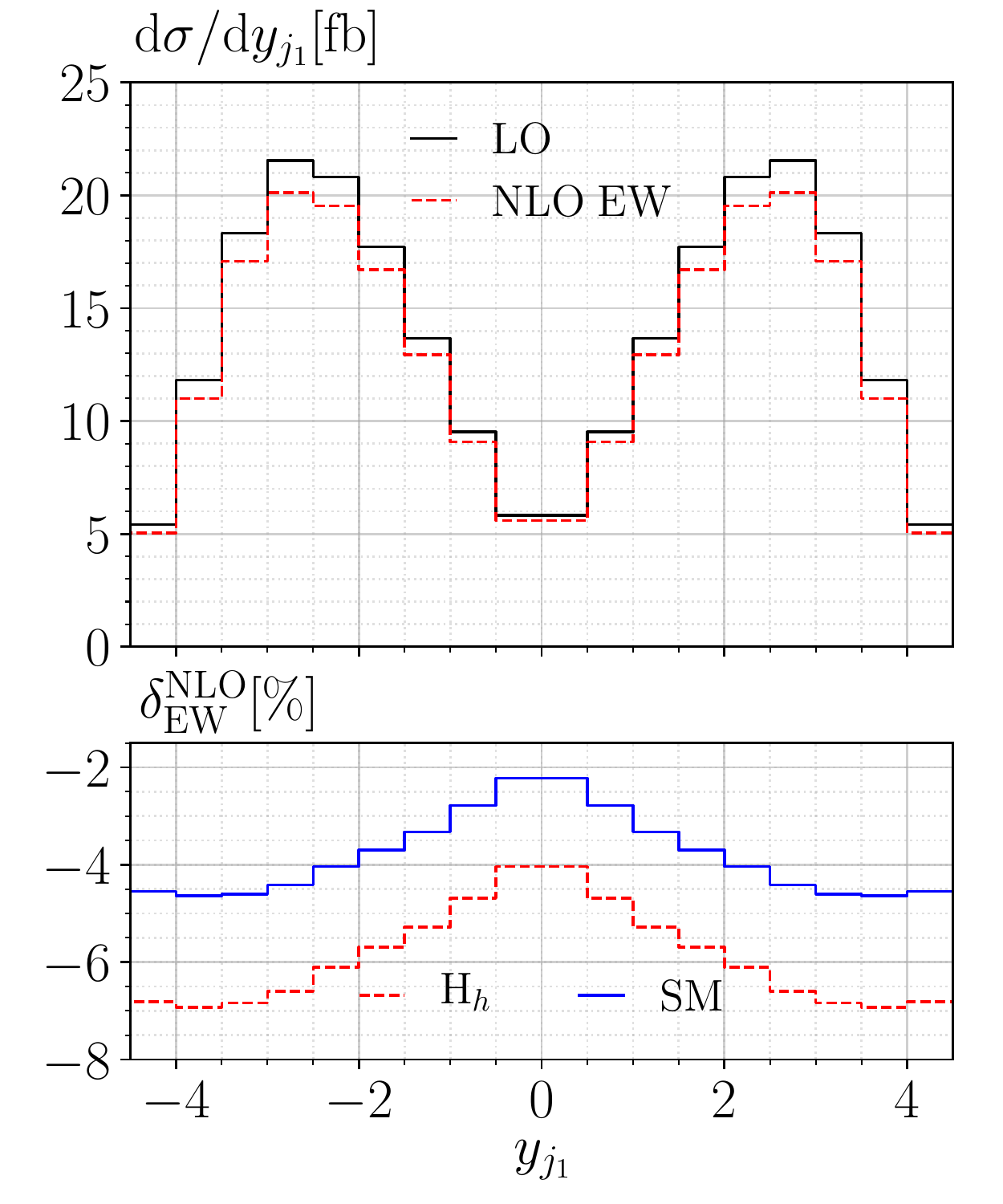}
    \caption{BP3B1 (\THDM)\\\phantom{.}}
  \end{subfigure}
  \begin{subfigure}[b]{0.49\textwidth}
    \includegraphics[width=0.89\textwidth]{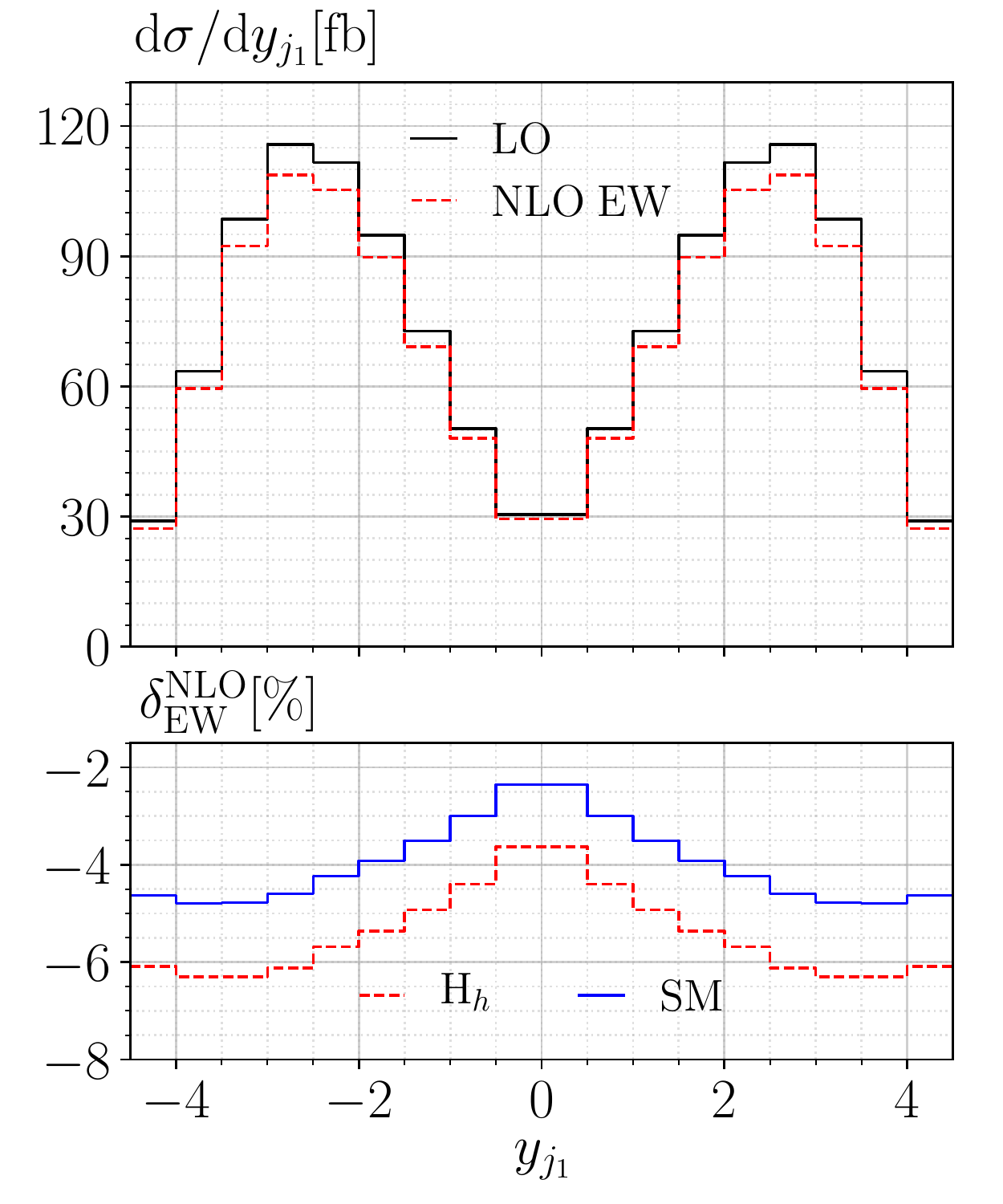}
    \caption{BP45 (\THDM)\\\phantom{.}}
  \end{subfigure}
  \begin{subfigure}[b]{0.49\textwidth}
    \includegraphics[width=0.89\textwidth]{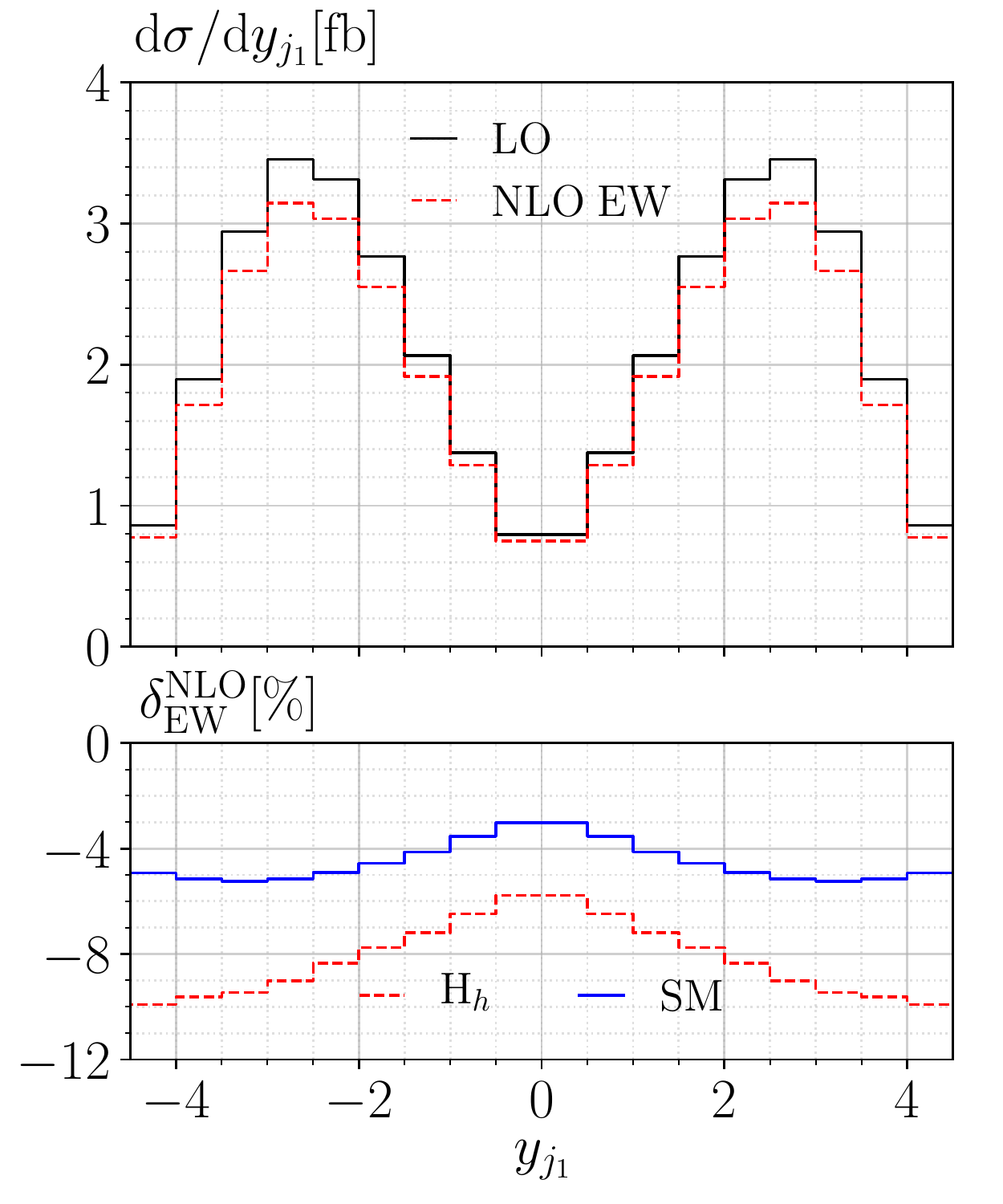}
    \caption{BP43 (\THDM)}
  \end{subfigure}
  \begin{subfigure}[b]{0.49\textwidth}
    \includegraphics[width=0.89\textwidth]{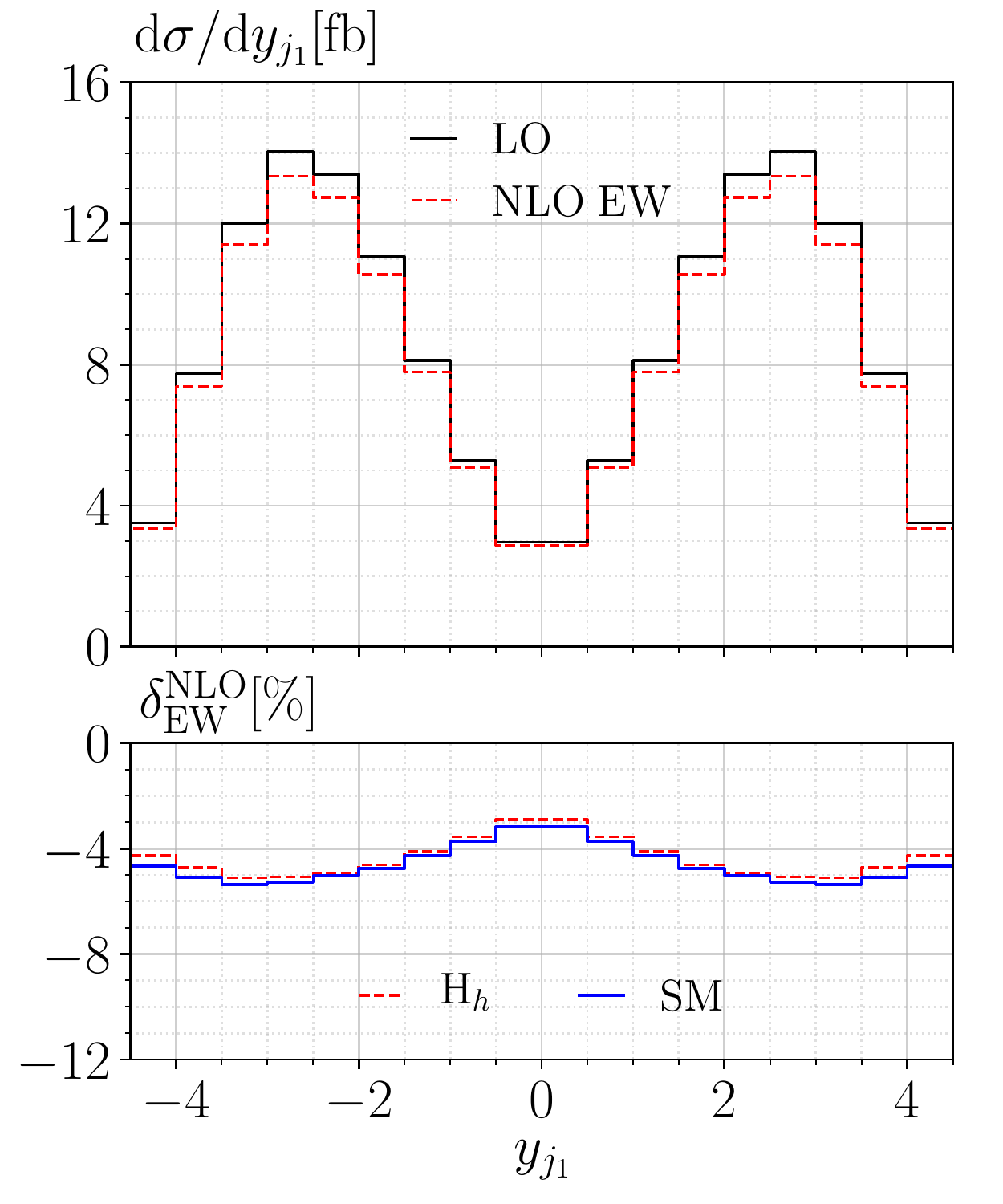}
    \caption{BP3 (\HS)}
  \end{subfigure}
  \caption{
    Distributions in the rapidity of hardest jet 
    $y_{j_1}$ for heavy Higgs production in VBF for the
    benchmark points BP3B1 in (a), BP45 in (b) and BP43 in (c) in the \THDM, and
    BP3 in (d) in the \HS.}
  \label{fig:yj1VBF}
\end{figure}

The results for $p_{\mathrm{T}, \Hh}$ distributions in Higgs strahlung and VBF
are shown in \reffi{fig:pth} and \reffi{fig:pthVBF}, respectively, the ones for
$y_{\Hh}$ in Higgs strahlung and VBF in \reffi{fig:yh} and \reffi{fig:yhVBF},
and those for $y_{\mu^-}$ and $y_{j_1}$ in \reffi{fig:ylm} and
\reffi{fig:yj1VBF}.\footnote{All rapidity distributions were symmetrized.}
In the upper plots we show the LO and NLO EW differential cross section. In the
lower plots the 
relative EW corrections $\dew$ are depicted. In order to
isolate the 
genuine effects  of the underlying model 
from the kinematic ones, we have computed the pure \SM{} corrections
with the \SM{} Higgs-boson mass set to the heavy Higgs-boson mass
$\mhh$ denoted as ``SM'' in the lower panels.  
The corresponding \SM{} total EW
cross sections are listed in \refta{tab:BPdistSM}.
\begin{table}
\begin{center}
\begin{tabular}{|c|c|c|c|}
  \hline  \mystrut BP & $\slo^{h}/\fba$ & $\dew$ \\
  \hline
  \hline BP3B1 & 877   & $-13.7 \%$ \\
  \hline BP45  & 802   & $-13.7 \%$ \\
  \hline BP43  & 501   & $-13.5 \%$ \\
  \hline BP3   & 366   & $-13.7 \%$ \\
  \hline
  \hline  \mystrut BP & $\slo^{h}/\fba$ & $\dew$ \\
  \hline
  \hline BP3B1 & 1402  & $-4.0 \%$ \\
  \hline BP45  & 1324  & $-4.1 \%$ \\
  \hline BP43  & 991   & $-4.7 \%$ \\
  \hline BP3   & 823   & $-4.8 \%$ \\
  \hline
\end{tabular}
\caption{Relative NLO corrections
  \dew{} to Higgs-boson production in Higgs strahlung $\Pp \Pp \to
  \Hl \mu^- \mu^+$ in the upper table and VBF $\Pp \Pp \to \Hl/\Hh \mathrm{j}
  \mathrm{j}$ in the lower table in the \SM. The Higgs-boson mass $\mh$ is set to the
  \change{heavy Higgs-boson mass $\mhh$}
  in the corresponding benchmark point.
  \label{tab:BPdistSM}}
\end{center}
\end{table}
 

In the following we focus on shape-distortion effects relative to the \SM{}
results.  Starting with the distributions in Higgs strahlung, we observe quite
large effects in the $p_{\mathrm{T}, \Hh}$ distribution in \reffi{fig:pth} for
BP3B1 and BP43 in the \THDM, small effects for BP3 in the \HS{}, and no effect
in BP45 in the \THDM, which perfectly reproduces the \SM{} result.  The
situation changes for the distributions in the rapidities $y_{\Hh}$ and
$y_{\mu^-}$ in \reffis{fig:yh} and \ref{fig:ylm}. Here, the largest deviations
from the SM are observed for BP43, where the relative EW corrections to the
$y_{\Hh}$ distribution in the \THDM{} are flatter than in the \SM.  For the
$y_{\mu^-}$ curve the
 opposite tendency is observed, \ie the
\SM{} correction is flatter.
For BP3B1, BP45, and BP3 shape distortions relative to the \SM{} appear
at large rapidities, which are less important due to low statistics
in those regions.
Switching to the distributions for VBF in \reffis{fig:pthVBF}, \ref{fig:yhVBF},
and \ref{fig:yj1VBF}, we observe a stronger trend towards \SM{}-like results.
The largest differences are observed for BP43 in the $p_{\mathrm{T}, \Hh}$ and
$y_{j_1}$ distributions. For BP3B1 the effects for the same distributions are
smaller but significant. For BP3 the shape distortion in the $p_{\mathrm{T},
\Hh}$ distribution for VBF is not larger than the one for Higgs strahlung. In
general in the considered benchmark points for the \HS{} the effects in VBF, but
also in Higgs strahlung, are tiny compared to the ones observed in the \THDM.

\change{The reason for the rather mild effects in the \HS{} is due to the
similiarity of the $\PH \PV \PV$ vertices to the SM ones.
In particular, in the \HS{} all couplings of the light and heavy Higgs~boson to
gauge bosons or fermions are \SM-like, but modulated with $(-\sas)$ and $\cas$,
respectively.  In the relative corrections these factors drop out, and the only
difference due to the presence of an additional light Higgs~boson and modified
Higgs-boson couplings is small in the benchmark points under consideration (all
$\lambda_i \lesssim 1$).
Remarkably, even the corrections to the $\PH \PV \PV$ vertices
involving Higgs self-couplings
(and thus all corrections) scale as the corresponding tree level with
either $(-\sas)$ or $\cas$, respectively, in the (anti-)alignment limit.
Furthermore, all mixing effects between $\Hl$ and $\Hh$ vanish in this limit.
For these reasons the corrections cannot become enhanced with respect to the LO unless tree-level
perturbativity is violated. In fact, in the \HS{} the one-loop corrected $\Hh
\PV \PV$ vertices are exactly zero in the 
alignment limit.
Note that these arguments apply to the whole phase-space region, thus, no significant
shape-distortion effects are expected for the processes under consideration in
the \HS.

The \THDM{}, on the other hand, exhibits non-decoupling effects in the alignment
limit $\cab\to 0$, where the underlying vertices for heavy Higgs-boson
production become loop induced.  The largest corrections in BP43 are due to the
non-decoupling term in the top Yukawa coupling\footnote{\change{The effect of
the top contribution has been studied for on-shell heavy Higgs-boson decay.}}
proportional to $\sab$. In this case the Yukawa coupling is of the same size as
the corresponding \SM{} one, but with a different sign and further enhanced with
respect to the LO by a factor of $1/\cab$, leading to a non-\SM-like
bosonic--fermionic interplay. Furthermore, the corrections in the \THDM{} are
very sensitive to the presence of new particles, especially the pseudo-scalar
Higgs~boson in the case of BP43.
In general, the contributions involving Higgs self-couplings can be large since
non-decoupling terms remain in the alignment limit
giving rise to enhanced corrections with respect to the LO.}


\section{Conclusion\label{sec:conclusion}}
We reported progress towards fully automated one-loop computations in BSM
models. The presented code \RecolaTwo{} allows one to compute \QCD{} and \EW{}
corrections for extensions of the \SM{} for arbitrary processes.
\RecolaTwo{} can produce NLO corrections in general models, which requires the
model file for each BSM model built in a specific format containing the
ordinary, counterterm and $R_2$ Feynman rules. The model-file generation and the
renormalization of general quantum-field-theoretic models is performed with the
new tool \Reptil{} in a fully automated way, relying on nothing but the Feynman
rules of the model in the \UFO{} format. Once the renormalization conditions for
the model are established, \Reptil{} performs the renormalization, computes the
$R_2$ rational terms and builds the one-loop renormalized model files in the
\RecolaTwo{} format. 
\change{We introduced the Background-Field Method 
as a complementary method in \RecolaTwo, which is useful for practical
calculations and serves as a powerful validation method.
We described the renormalization procedure in the
Background-Field Method which is handled in \RecolaTwo{} on equal footing with
the usual formulation.

In summary, we realized the following generalizations with respect to \Recola: 
\begin{itemize}
  \item We developed a true model-independent amplitude provider, featuring a
    dynamic process generation in memory without the need for intermediate compilation.
  \item A generic interface has been developed supporting all methods available
    in \Recola{}, but generalized to fit in the model-file
    approach. This includes the computation of amplitudes and squared amplitudes, the selection of
    specific polarizations and resonances, and the computation of interferences
    with different powers in new fundamental couplings.
    Furthermore, we provide spin- and colour-correlated squared matrix
    elements required in the
    Catani--Seymour dipole formalism. The latter methods are restricted to singlet,
    triplet and octet states of SU(3).
  \item \RecolaTwo{} is limited to scalars, Dirac fermions
        and vector bosons. In the near future we will allow for Majorana fermions.
  \item We support Feynman rules with a general polynomial momentum dependence and
    allow for elementary interactions between more than four fields. Due to
    internal optimizations the number of fields per elementary
    interaction is restricted to at most 8. 
  \item We generalized \RecolaTwo{} to support the \BFM{} as a
    complementary method. Furthermore, the \rxi{}-gauge can be used
    for massive vector bosons or, alternatively, non-linear gauges can
    be implemented.   
  \item With \Reptil{} we have formed the basis for a fully automated
    generation of renormalized model files for \RecolaTwo. We provide
    a simple framework for the implementation of custom
    renormalization conditions.  Presently available model files for
    \RecolaTwo\ include the $Z_2$-symmetric Two-Higgs-Doublet Model
    with all types of Yukawa interactions and the Higgs-Singlet
    extension of the Standard Model as well as models files with
    anomalous triple vector-boson and Higgs--vector-boson couplings.
\end{itemize}
The considered simple models do by far not exhaust the range of
applicability of \RecolaTwo{} and \Reptil, and further models will be
implemented in the future.}

As an application of the new tools we present first results for NLO
electroweak corrections to vector-boson fusion and updated results for
Higgs strahlung in the Two-Higgs-Doublet Model and the Higgs-Singlet
extension of the Standard Model.  We compared Higgs-production cross
sections for different renormalization schemes in both models.
\change{
We analysed the scale dependence in an \msbar{} renormalization scheme for the mixing
angles, which has been improved  including the renormalization-group running of parameters.
We found unnaturally large corrections and scale uncertainties at one-loop order for the \msbar{}
scheme, while the considered on-shell schemes remain well-behaved.
These enhanced contributions can be  
related to uncancelled finite parts in the \msbar{} scheme and}
should be investigated in more detail in
the future, since a proper estimation of higher-order uncertainties,
as it can be done based on scale variation in \msbar{} schemes, is
highly desirable.  For the on-shell schemes, our results for the
electroweak corrections to \SM-like Higgs-boson production are almost
not distinguishable from the corresponding \SM{} corrections for all considered
benchmark points.  Finally, we presented distributions for the production of
heavy Higgs bosons.  Here, interesting shape-distortion effects for the
electroweak corrections at the level of several percent are observed in the
\THDM.


\section{Acknowledgements\label{Acknowledgements}}
We thank L.~Altenkamp for an independent check of $\Hl \to 4f$ in the \THDM{}
which is closely related to the Higgs-production processes considered
in this work. We thank
L.~Jenniches and C.~Sturm for valuable discussions and an independent check of
the running of $\tb$ and $\cab$. A.~D. and J.-N.~L. acknowledge support from the
German Research Foundation (DFG) via grants DE~623/4-1 and DE~623/5-1. The work
of J.-N.~L. is supported by the Studienstiftung des deutschen Volkes.
The work of S.U.\ was supported in part by the European Commission through
the ``HiggsTools'' Initial Training Network PITN-GA-2012-316704.

\appendix

\begin{appendices}
\section{Colour-flow vertices}
\label{sec:colorflow}
In \RecolaTwo{} the colour flow  is constructed recursively. For a given
off-shell current  the outgoing colour configuration is determined from the
incoming ones and the possible colour flows associated to the
interaction vertex. 
As the \UFO{} format does not incorporate the colour flow, we need to translate
between the two representations. We implemented a dynamical system for
computing the colour flow from the generators and structure constants, rather
than substituting for known results. In the conventions of
\citere{Actis:2016mpe} the colour flow associated to a given colour structure
\begin{align}
  C_{a_1, \ldots,}{}^{i_1, \ldots}_{j_1, \ldots}
  \label{eq:colstruc}
\end{align}
is obtained by multiplying
  \eqref{eq:colstruc} with the normalized generator
  ${(\Delta_{a_p})^{i_p}}_{j_p}$
for each index $a_p$ corresponding to an open index in
adjoint representation.
The indices $i_p$ and $j_p$ refer to the colour and anti-colour indices,
  respectively. The generators $(\Delta_a)^i_j$ and structure constants $f_{abc}$
  define the SU(3) Lie algebra\footnote{The $\Delta_a$ generators are related to the
  conventional ones $T_a$, as used \eg in \Feynrules, via 
  ${(\Delta_{a})^{i}}_{j}=\sqrt{2} {(T_{a})^{i}}_{j}$ with
  $\mathrm{Tr}\left\{ T_a T_b\right\} = \delta_{ab}/2$ and
  $\left[ T_a, T_b\right] = \ii f_{abc} T_c$. 
  Note that the structure
  constants $\tilde f_{abc}$ in \citere{Actis:2016mpe} are related to the ones
  in this paper via $\tilde f_{abc} = \sqrt{2} f_{abc}$. } 
\begin{align}
  \left[ \Delta_a, \Delta_b\right] = \ii \sqrt{2} f_{abc} \Delta_c,
  \quad \mathrm{Tr}\left\{ \Delta_a \Delta_b\right\} = \delta_{ab},
  \quad \Delta_a = \frac{\lambda_a}{\sqrt{2}},
  \label{eq:defsu3}
\end{align}
with $\lambda_a$ being the Gell-Mann matrices.  The computation then consists of
  eliminating the structure constants and the generators by solving
  \eqref{eq:defsu3} for the structure constants and using the (Fierz)
  completeness relation for the generators as follows
\begin{align}
  f_{abc} &=  -\frac{1}{\sqrt{2}} \ii \mathrm{Tr}\left\{ \Delta_a \left[
    \Delta_b, \Delta_c\right] \right\} \notag,\\
  \sum_a {\left(\Delta_a\right)^{i_1}}_{j_1} {\left(\Delta_a\right)^{i_2}}_{j_2} &=
  \delta^{i_1}_{j_2} \delta^{i_2}_{j_1}-\frac{1}{3}\delta^{i_1}_{j_1} \delta^{i_2}_{j_2}.
  \label{eq:su3normalization}
\end{align}
Performing all contractions yields a sum of Kronecker deltas which
represent the individual colour flows.
For instance, the quartic gluon vertex of the \SM{} reads
\begin{align}
  g_s^2\sum_k \left(f_{ka_1a_2}f_{ka_3a_4} L_1^{\mu_1\mu_2\mu_3\mu_4}+
  f_{ka_1a_3}f_{ka_2a_4} L_2^{\mu_1\mu_2\mu_3\mu_4}+
  f_{ka_1a_4}f_{ka_2a_3}L_3^{\mu_1\mu_2\mu_3\mu_4}\right),
\label{eq:fourgluonvertex}
\end{align}
with $L_1, L_2, L_3$ being Lorentz structures which, for the following
discussion, are left unspecified.
Focusing on the colour structure $\delta^{i_1}_{j_2}
\delta^{i_2}_{j_3}\delta^{i_3}_{j_4} \delta^{i_4}_{j_1}$, 
we obtain for the two relevant contributions
\begin{align}
  \sum_{k,a_1,a_2,a_3,a_4}
  {\left(\Delta_{a_1}\right)^{i_1}}_{j_1}
  {\left(\Delta_{a_2}\right)^{i_2}}_{j_2}
  {\left(\Delta_{a_3}\right)^{i_3}}_{j_3}
  {\left(\Delta_{a_4}\right)^{i_4}}_{j_4}
  f_{ka_1a_2}f_{ka_3a_4} 
  = \frac{1}{2}\left(
  -
  \delta^{i_1}_{j_2} \delta^{i_2}_{j_3}\delta^{i_3}_{j_4} \delta^{i_4}_{j_1}
  + \ldots
\right),
  \notag \\
  \sum_{k,a_1,a_2,a_3,a_4}
  {\left(\Delta_{a_1}\right)^{i_1}}_{j_1}
  {\left(\Delta_{a_2}\right)^{i_2}}_{j_2}
  {\left(\Delta_{a_3}\right)^{i_3}}_{j_3}
  {\left(\Delta_{a_4}\right)^{i_4}}_{j_4}
  f_{ka_1a_4}f_{ka_2a_3}
  = \frac{1}{2}\left(
  +
  \delta^{i_1}_{j_2} \delta^{i_2}_{j_3}\delta^{i_3}_{j_4} \delta^{i_4}_{j_1}
  +\ldots
\right).
\end{align}
Combining this result with \eqref{eq:fourgluonvertex}, results in the contribution
\begin{align}
  \delta^{i_1}_{j_2} \delta^{i_2}_{j_3} \delta^{i_3}_{j_4}\delta^{i_4}_{j_1} \times
  \frac{g_s^2}{2} \left(L_3 - L_1 \right).
\end{align}
Thus, diagonalizing the vertex in colour-flow basis requires, in general, to
redefine Lorentz structures and couplings.

\section{Off-shell currents}
\label{sec:ofc}
For a given Lorentz structure and a definite colour-flow state the \BGR{} is
derived from the Feynman rules by selecting one of the particles as the outgoing
one, multiplying with 
the corresponding 
propagator and the incoming currents of the other
particles.  Since the structure of currents depends on the outgoing particle,
one needs to derive the \BGR{} for all distinct outgoing particles.  Consider
for instance the \QED{} vertex $\Pe^+ \Pe^- \gamma$. \Reptil{} constructs three
different recursion relations
\begin{align}
  w_\al  = \mathrm{i} e \sum_{\be,\delta,\mu} D^{e^-}_{\al\be}\left(\gamma^\mu\right)_{\be\delta}
  \times w_\mu \times w_{\delta}, \notag\\
  \bar w_\beta  = \mathrm{i} e \sum_{\al,\delta,\mu}D^{e^+}_{\al\be}
  \left(\gamma^\mu\right)_{\delta\al} \times w_\mu \times \bar w_{\delta},
  \notag\\
  w_\mu  = \mathrm{i} e \sum_{\al,\be,\nu} D^{\gamma}_{\mu\nu}
  \left(\gamma^\nu\right)_{\al\be} \times \bar w_\al \times w_\be,
  \label{eq:bgrqed}
\end{align}
with $w_i, \bar w_j, w_\mu$ being either incoming or outgoing off-shell
currents, depending on whether they are on the right- or left-hand side of
\eqref{eq:bgrqed}. For many Feynman rules, the underlying \BGR{} are formally
the same if the couplings  or masses of the particles are not further specified.
Assuming that the colour flow has been factorized as explained in
\refapp{sec:colorflow}, all fermion--fermion--vector rules, \eg $\PZ \Pe\to \Pe$
or $\gamma \Pe \to \Pe$, can be mapped onto the same structures realizing that
$\gamma^\mu\omega^+$ and $\gamma^\mu\omega^-$ form a suitable basis,
\begin{align}
  w_\al &= \sum_{\be,\delta,\mu} D_{\al\be}^f\left(c_1 \gamma^\mu \omega^+ + c_2 \gamma^\mu
  \omega^-\right)_{\be\delta} \times w_\mu \times w_\delta,\notag\\
  \bar w_\be &= \sum_{\al,\delta,\mu} D_{\al\be}^{\bar f}\left(c_1 \gamma^\mu \omega^+ + c_2
  \gamma^\mu \omega^-\right)_{\delta\al} \times w_\mu \times \bar w_\delta,\notag\\
  w_\mu &= \sum_{\al,\be,\nu} D_{\mu \nu}^{V} \left(c_1 \gamma^\nu \omega^+ +
  c_2 \gamma^\nu \omega^-\right)_{\al\be} \times \bar w_\al \times w_\be,
\end{align}
with $D_{\al\be}^f, D_{\al\be}^{\bar f}, D_{\mu \nu}^{V}$ denoting generic
propagators for fermions, anti-fermions and vector bosons, respectively.
\Reptil{} has the ability to derive a minimal basis, dynamically, \ie depending
on the operators of the theory, without relying on the Lorentz basis in the
\UFO{} format.  This is done in two steps.  In the first step, all distinct
\BGR{} in the underlying theory are registered. In the next step the \BGR{} are
merged recursively until a minimal basis is obtained.  The size of the \BGR{}
can be controlled by a parameter for the maximal number of allowed distinct
generic couplings, and it is possible to allow for vanishing couplings to
improve the merging. If a merge yields a \BGR{} size larger than allowed, the
merging is not accepted.  Finally, all vertices are mapped to the minimal basis.

\section{Translation of $\delta \alpha$ to other tadpole
  schemes\label{sec:simplertadpoles}} In this section we discuss the
translation of on-shell renormalization conditions from one tadpole
counterterm scheme to others using the example of $\delta \alpha$
renormalized in the $p^*$ scheme \eqref{eq:pstarprescription}.  Once,
we treat the tadpole counterterms in the \ts, with the tadpole
renormalization \eqref{eq:tadpolecond}. As an alternative tadpole
counterterm scheme, we consider the one of \citere{Denner:1991},
denoted as scheme S1, which is commonly used in the
\SM{}.\footnote{For the corresponding tadpole counterterms in the
  \THDM{} see ``Scheme I'' in \citere{Denner:2016etu}.}  In S1 the
tadpole counterterm to the neutral scalar mixing energy is zero. Then,
$\delta \alpha_{\mathrm{S1}}$ in the $p^*$ scheme is consistently
defined by absorbing all tadpole contributions as
\begin{align}
  \delta \alpha_{\mathrm{S1}} &= \delta \alpha -\frac{\thlhh }{\mhh^2-\mhl^2}
  \label{eq:alphaDS}
\end{align}
with $\delta \alpha$ and $\thlhh$ being
evaluated in the \ts{} together with the tadpole
renormalization \eqref{eq:tadpolecond}.   Note that \eqref{eq:alphaDS} holds in
any gauge, but only the tadpole $\thlhh$ is affected by the gauge choice, as
$\delta \alpha$ is {\it defined} as gauge-parameter independent. For the 't
Hooft--Feynman gauge \eqref{eq:alphaDS} reduces to
\begin{align}
  \delta \alpha_{S1}
  = \frac{\Sigma^{\mathrm{1PI},\mathrm{BFM}}_{\Hh
  \Hl}\left(\frac{\mhh^2+\mhl^2}{2}\right)}{\mhh^2-\mhl^2}.
  \label{eq:dapsFeynmangauge}
\end{align}
In order to verify the gauge independence of the \SMatrix{} one has to use the
gauge dependence originating exclusively from the absorbed tadpoles.  Equation
\eqref{eq:alphaDS} implies for the gauge-parameter dependence of $\delta
\alpha_{S1}$ 
\begin{align}
  \frac{\partial \delta \alpha_{S1}}{\partial \xi} = 
  -\frac{1}{\mhh^2-\mhl^2} \frac{\partial \thlhh}{\partial \xi},
  \label{eq:gaugedepalphaDS}
\end{align}
with $\xi$ generically denoting a parametrization of a gauge choice, not
necessarily \rxi{} gauge. We stress that the gauge dependence of
\eqref{eq:gaugedepalphaDS} is not equivalent to the gauge dependence of the
mixing energy \eqref{eq:dapsFeynmangauge}. Finally, when studying the gauge
dependence of \SMatrix{} elements, the gauge dependence
\eqref{eq:gaugedepalphaDS} necessarily cancels against the tadpole counterterm
gauge dependence absorbed in other on-shell renormalized counterterms, \eg mass
counterterms and $\delta \tb$ (and $\delta \Msb^2$ in the \THDM).
\end{appendices}


\bibliographystyle{JHEPmod}
\bibliography{recolatwobib}

\end{document}